\documentclass[useAMS,usenatbib]{mn2e}
\usepackage{journal_abbreviations}
\usepackage{graphicx, amsmath, amssymb}
\usepackage{enumerate}
\usepackage{float}
\usepackage[T1]{fontenc}
\usepackage{aecompl}
\usepackage{comment}

\usepackage{mathrsfs}
\usepackage{graphics}

\title[The three phases of galaxy formation]{The three phases of galaxy formation}
\author[Clauwens et al.]{Bart Clauwens$^{1,2}$\thanks{E-mail: clauwens@strw.leidenuniv.nl}, Joop Schaye$^{1}$, Marijn Franx$^{1}$, Richard G. Bower$^{3}$\\
$^{1}$Leiden Observatory, Leiden University, PO Box 9513, 2300 RA Leiden, The Netherlands\\$^{2}$Instituut-Lorentz for Theoretical Physics, Leiden University, 2333 CA Leiden, The Netherlands\\$^{3}$Institute for Computational Cosmology, Department of Physics, University of Durham, South Road, Durham, DH1 3LE, UK}

\begin{document}

\date{Accepted, 2018 May 9. Received, 2018 May 3; in original form, 2017 October 31.}

\pagerange{\pageref{firstpage}--\pageref{lastpage}}

\maketitle

\label{firstpage}

\begin{abstract}

We investigate the origin of the Hubble sequence by analysing the evolution of the kinematic morphologies of central galaxies in the EAGLE cosmological simulation. By separating each galaxy into disc and spheroidal stellar components and tracing their evolution along the merger tree, we find that the morphology of galaxies follows a common evolutionary trend. We distinguish three phases of galaxy formation. These phases are determined primarily by mass, rather than redshift. For $M_{*}\lesssim 10^{9.5} {\rm M_{\odot}}$ galaxies grow in a disorganised way, resulting in a morphology that is dominated by random stellar motions. This phase is dominated by in-situ star formation, partly triggered by mergers. In the mass range $10^{9.5}{\rm M_{\odot}}\lesssim M_{*} \lesssim10^{10.5}{\rm M_{\odot}}$ galaxies evolve towards a disc-dominated morphology, driven by in-situ star formation. The central spheroid (i.e. the bulge) at $z=0$ consists mostly of stars that formed in-situ, yet the formation of the bulge is to a large degree associated with mergers. Finally, at $M_{*}\gtrsim 10^{10.5}{\rm M_{\odot}}$ growth through in-situ star formation slows down considerably and galaxies transform towards a more spheroidal morphology. This transformation is driven more by the buildup of spheroids than by the destruction of discs. Spheroid formation in these galaxies happens mostly by accretion at large radii of stars formed ex-situ (i.e. the halo rather than the bulge). 
\end{abstract}

\begin{keywords}
galaxies: structure -- galaxies: formation -- galaxies: evolution -- galaxies: kinematics and dynamics -- galaxies: bulges
\end{keywords}

\section{Introduction}
\label{SectionIntroduction}

Low-redshift galaxies have a wide range of morphologies, ranging from pure stellar discs, to discs with increasingly massive central stellar bulges, to elliptical galaxies. This morphological diversity is traditionally classified according to the Hubble sequence. We can decompose most galaxies into a rotationally supported stellar disc and a spheroid, which is supported to a large degree by random and more radial stellar orbits. This decomposition is motivated by the fact that classical bulges are very similar to elliptical galaxies without an accompanying disc, suggesting a similar formation mechanism. The main difference is that there is an offset between their mass-size relations \citep[e.g.][]{Gadotti08}.

Galaxy morphology is tightly linked to other galaxy properties. More massive galaxies are generally less disky and, at a fixed mass, star forming galaxies tend to be disc-dominated while quiescent galaxies are typically bulge-dominated \citep[e.g.][]{Gadotti08,Bluck14,Whitaker15}. Above $10^{10}{\rm M_{\odot}}$ the stellar mass in the low-redshift Universe is roughly equally divided between ellipticals, classical bulges and discs \citep{Gadotti08}. There is good evidence that high-redshift galaxies are built from these same morphological components with a qualitatively similar dependency on star formation and mass. \citet{Tacchella15} find that most massive galaxies at $z\approx2$ have fully grown and quenched bulges in their cores and \citet{Dokkum14} state that: `the presence of a dense core is a non-negotiable requirement for stopping star formation in massive galaxies'. The likely progenitors of massive quenched bulges are compact star forming galaxies at high redshifts as observed by CANDELS, 3D-HST \citep{Barro13,Barro14} and ALMA \citep{Barro16}.

Observationally a distinction is made between classical bulges and pseudobulges \citep{Kormendy93,Wyse97}. Classical bulges are dispersion dominated while pseudobulges (which can be disky, boxy/peanut shaped or nuclear bars) are rotationally dominated. Our focus will be on the dispersion dominated classical bulges, which account for a factor $>4$ more in mass \citep{Gadotti08}.

There are many possible scenarios for bulge formation. Here we will briefly summarise the main ideas. Pseudobulges can form through secular processes \citep[e.g.][]{Kormendy04} such as bar formation, followed by a buckling instability that transforms the bar into a peanut shaped pseudobulge \citep[e.g.][]{Raha91,Pohlen03,Guedes12,Perez17}. Classical bulges can form from diverse processes such as the collapse of primordial gas clouds \citep{Eggen62}, disc instabilities \citep[e.g.][]{DeLucia11}, clump migration to the galaxy centre in violently unstable gas rich discs at high redshift \citep[e.g.][]{Noguchi98,Bournaud07,Elmegreen08,Dekel09,Bournaud11,Perez13,Ceverino15}, gas funneling to the centre in marginally unstable discs at high redshift \citep{Krumholz17}, misallligned accretion \citep{Sales12,Aumer13} and mergers \citep[e.g.][]{Aguerri00,Bournaud11,Aumer13,Hopkins10,DeLucia11,Ceverino15}. 

Mergers can influence bulge growth and overall morphological changes in diverse ways. \citet{Hernquist89} finds that tidal effects during mergers may induce instabilities that can funnel a large amount of gas into the central region of a galaxy, thereby inducing a starburst which creates a spheroidal component. In order to prevent too much bulge formation, stellar feedback is needed to remove low angular momentum gas, also during merger induced starbursts \citep[e.g.][]{Governato09,Governato09b,Brook11,Brook11b,Christensen15,Zjupa16}. This may not be sufficient and AGN feedback might be needed for a further supression. Discs can be destroyed by a major merger, but they can also regrow afterwards \citep[e.g.][]{Governato09,DeLucia11,Sparre16}. For massive galaxies AGN feedback may be needed to prevent disc regrowth in order to form realistic ellipticals \citep[e.g.][]{Genel15,Dubois16,Sparre16}. Generally, gas-poor (dry) mergers are thought to spin down galaxies, while gas rich (wet) mergers spin them up \citep[e.g.][]{Naab13,Lagos17}, although \citet{Penoyre17} find that in the Illustris simulation this distinction has little influence. Finally, the time at which the merger takes place also matters. Late mergers are thought to give rise to a diffuse halo \citep{Brook11,Pillepich14}.

Mergers are the prime suspect for transforming disc galaxies into galaxies with large bulges and elliptical galaxies. However, this is not a settled matter. \citet{Lofthouse16} conclude from observations at $z\approx2$ that major mergers are not the dominant mechanism for spheroid creation, because only one in five blue spheroids at this redshift shows morphological disturbances. \citet{Sales12} argue that in the GIMIC simulation \citep{Crain09}, speroid formation does not rely on mergers, because it takes place even when most stars form in-situ, as opposed to having been accreted after forming ex-situ (i.e. in a galaxy other than the main progenitor). Furthermore, \citet{RodriguezGomez17} state that in the Illustris simulation mergers play no role in morphology below $10^{11}\rm{M_{\odot}}$, because accreted stellar fractions and mean merger gas fractions are indistinguishable between spheroidal and disc-dominated galaxies.

There are different ways to determine the morphology or bulge-to-total ratio ($B/T$) of a galaxy from observations.
Usually the $B/T$ ratio is  determined photometrically, based on a decomposition of the light profile into a disc and a bulge component. The disc and bulge components are then generally assumed to have fixed Sérsic indices of $n=1$ and $n=4$ respectively \citep[e.g.][]{Bluck14}, but sometimes these indices are allowed to vary \citep[e.g.][]{Gadotti08,Sachdeva17}. The bulge can also be determined kinematically as a non-rotationally supported component. When similar methods are applied to galaxy simulations, in general 2D-photometric bulge determination leads to lower $B/T$ ratios than kinematic bulge determination \citep{Scannapieco10} and these differences can be large. In the Illustris cosmological simulation the median B/T difference between both methods becomes larger than 0.5 for galaxy masses below $10^{10.6}{\rm M_{\odot}}$ \citep{Bottrell17}, thus classifying galaxies as disky based on their light profile even when the kinematics show no ordered rotation.

In this work we investigate the evolution of kinematic morphologies (thus derived from stellar motions) of galaxies in the EAGLE cosmological simulation \citep{Schaye15,Crain15}, with emphasis on the central bulge component. \citet{Oser10} emphasized the two-phase nature of the formation of massive galaxies, whose inner regions are formed first and in-situ, while the stars in the outer parts are mainly formed ex-situ and were accreted later. Here, we investigate the provenance of in-situ/ex-situ stars in different kinematic galaxy components and we try to determine to what extent mergers are responsible for the morphological transformations of EAGLE galaxies. This will lead to a three-phase picture of galaxy formation, where low-mass galaxies are kinematically hot (i.e. spheroidal/puffy) even though most of their stars are formed in-situ, intermediate-mass galaxies also grow mostly through in-situ star formation but are kinematically cold (i.e. disky), and the growth of massive galaxies is dominated by accretion of stars formed ex-situ, making them more spheroidal. Within this three phase picture, the first phase is most speculative, since low-mass galaxies are closer to the resolution limit and since historically hydrodynamic simulations have produced galaxies that are too small and kinematically hot due to overcooling. However, recent findings for the VELA simulation \citep{Zolotov15} and the FIRE-2 simulation \citep{ElBadry17} hint at a similar early phase in galaxy formation.

Although EAGLE lacks the resolution to confidently reproduce the smallest observed bulges, it has overcome the largest hurdle: the overcooling problem. Overcooling would produce too massive and dense central stellar concentrations at high redshift, akin to bulges. EAGLE does well in this regard. It approximately reproduces the observed evolution of the galaxy stellar mass function \citep{Furlong15} and galaxy sizes, with passive galaxies being smaller at fixed mass \citep{Furlong15b}. Conclusions about the origin of galaxy morphology drawn from simulations that do not match the evolution of the mass function and the size-mass relation could be misleading, since the physical processes that determine a galaxy's stellar mass and size are also thought to determine its morphology. The galaxies in EAGLE also agree relatively well with the observed passive fraction as a function of mass \citep{Schaye15,Trayford17}. Furthermore, the galaxies have representative rotation curves \citep{Schaller15}. It is thus a useful cosmological simulation to study the origin of morphology changes and bulge formation.

We build on earlier work related to the angular momentum of EAGLE galaxies. \citet{Zavala15} find that $z=0$ galaxy morphology is correlated with a loss of angular momentum at late times, both in the stellar component and in the inner dark matter component, due to mergers. \citet{Lagos16} find that galaxies with low angular momentum can be either the result of merger activity or of early star formation quenching in the absence of mergers. \citet{Lagos17} find that dry mergers tend to reduce the total stellar angular momentum while wet mergers tend to increase it, with a dependency on the alignment of the spin vectors of the merger pair. Finally, \citet{Correa17} show that the kinematic morphology of EAGLE galaxies is closely related to mass and colour, with blue cloud galaxies having predominantly a disky structure and red sequence galaxies a spheroidal morphology.

We will shortly introduce the EAGLE simulation in section \ref{SectionEAGLE}. Section  \ref{SectionKinematicMorphology} describes our method for determining the kinematic morphology of a galaxy. We apply this to determine the morphological evolution of EAGLE galaxies in section \ref{SectionMorphologyEvolution}. Section \ref{SectionOriginBulge} focuses on the origin of stars in the stellar bulge and halo. Section \ref{SectionMergers} investigates the effects of mergers and in-situ star formation on the overall morphology of galaxies, while section \ref{SectionMergers2} isolates the contribution of mergers on bulge and spheroid formation. For a summary of our main conclusions and a discussion of the three phases of galaxy formation, see section \ref{SectionConclusions}.

\section{The EAGLE simulation}
\label{SectionEAGLE}

Our results are based on the $\rm (100 \, Mpc)^3$ sized reference run (Ref-L100N1504) of the EAGLE hydrodynamical simulation \citep{Schaye15,McAlpine16}. The simulation includes radiative cooling and heating \citep{Wiersma09a}, star formation \citep{Schaye08}, stellar mass loss \citep{Wiersma09b}, stochastic stellar feedback \citep{DallaVecchia12} (which depends on the local density and metallicity in order to prevent the overproduction of bulge-like dense stellar cores at high redshift due to numerical radiative losses) and stochastic feedback from active galactic nuclei. Gas is allowed to form stars if the density exceeds the metallicity-dependent density threshold derived by \citet{Schaye04} for the transition from the warm, atomic to the cold, molecular interstellar gas phase.The simulation parameters are calibrated to the $z=0$ galaxy stellar mass function and mass-size relation. The effect of the various parameters and the calibration choices are described in detail in \citet{Crain15}. The initial gas particle mass is $1.6 \times 10^6 \rm{M_{\odot}}$. The maximal gravitational force softening is 700 pc and a pressure floor is implemented for the interstellar medium in order to prevent spurious fragmentation \citep{Schaye08}.

The simulation relies on subgrid physics for unresolved processes at small scales and low temperatures in the interstellar medium. This means that the simulation by design does not give cold thin discs. The minimum resolved scale is about 1 kpc, which means that the simulation is best suited to study bulges at the larger end of the mass-size spectrum and the transformation of disc galaxies to elliptical galaxies. However, in appendix \ref{SectionAppendixA} we show that a comparison of the $\rm (25 \, Mpc)^3$ sized reference run (RefL0025N376) and the recalibrated run at a factor 8 higher mass resolution (RecalL0025N0752) suggests a good convergence of our results for $M_{*}\gtrsim10^{9}{\rm M_{\odot}}$.

In this work we adopt a kinematic definition for a classical bulge as the spheroidal, dispersion dominated component within 5 proper kpc (pkpc). We will study central galaxies at $z=0$ and their main progenitors at higher redshifts (which are expected, but not required to be central galaxies). For satellites additional processes such as ram pressure stripping and strong tidal forces might induce morphological changes, complications that we aim to avoid in this work. 

\section{Kinematic morphology}
\label{SectionKinematicMorphology}

\begin{figure*}
\text{\hskip2em\relax Disc, $S/T=0.18$ \hskip10em\relax Disc+Bulge, $S/T=0.51$ \hskip10em\relax Elliptical, $S/T=0.86$}\\
\includegraphics[width=0.32\textwidth]{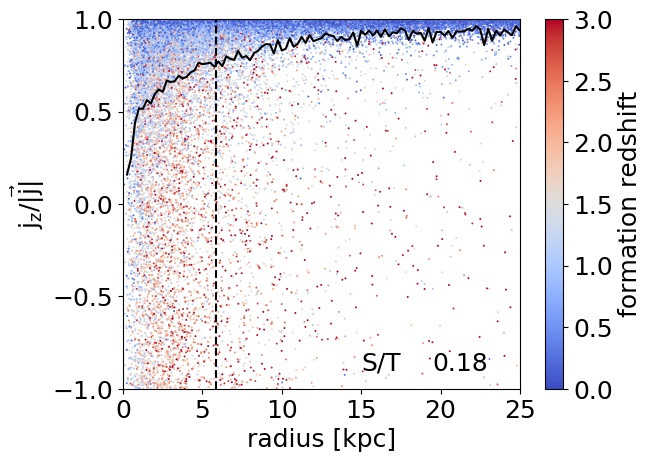}
\includegraphics[width=0.32\textwidth]{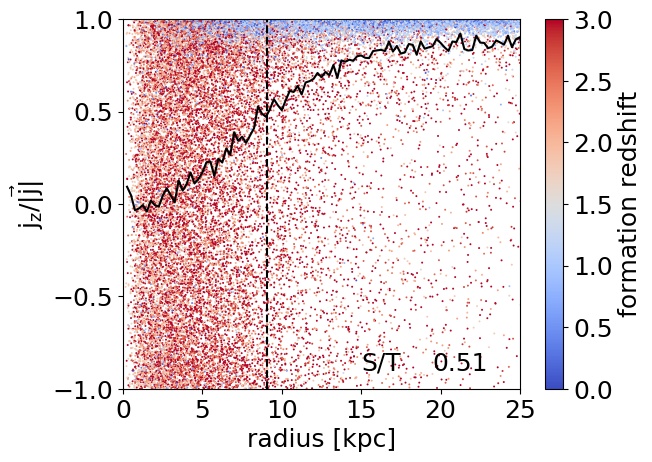}
\includegraphics[width=0.32\textwidth]{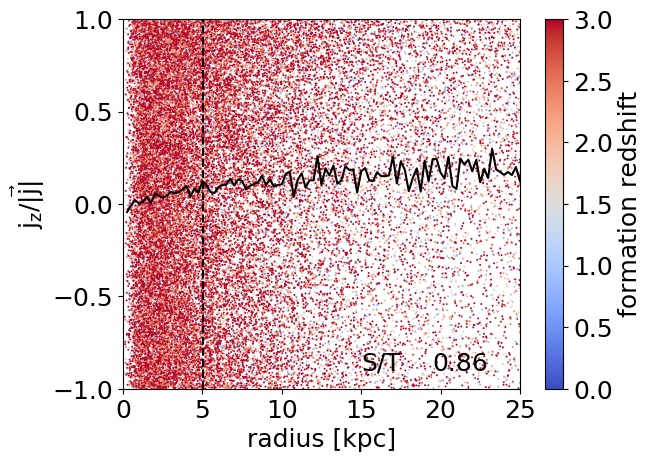}\\
\includegraphics[width=0.32\textwidth]{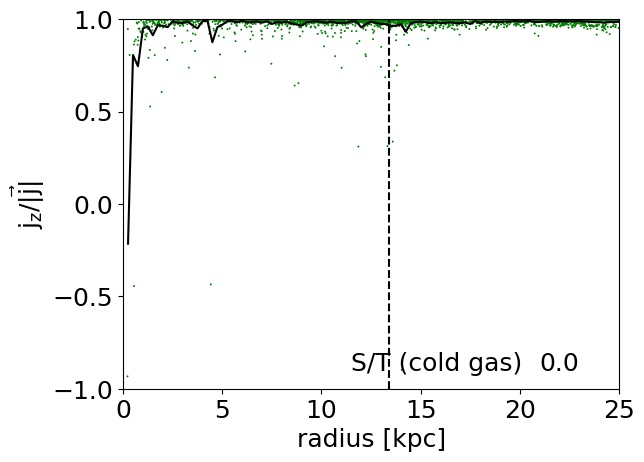}
\includegraphics[width=0.32\textwidth]{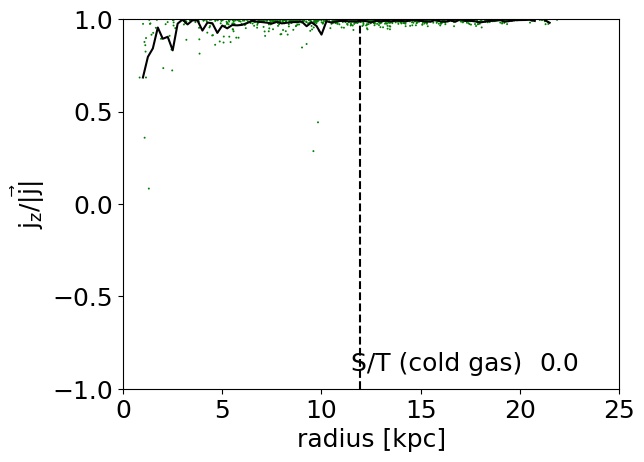}
\includegraphics[width=0.32\textwidth]{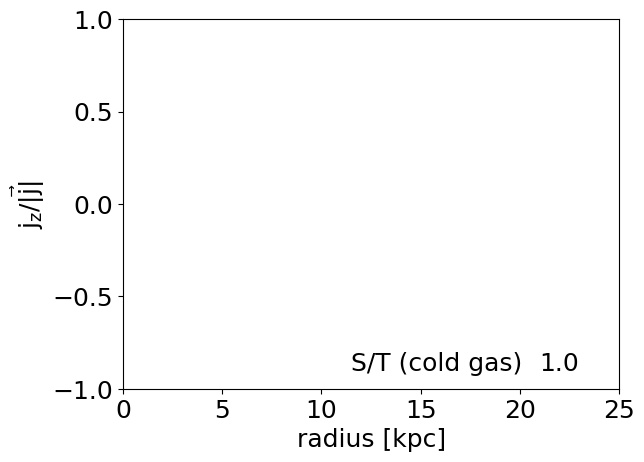}
\caption{The kinematic structure of three typical EAGLE galaxies with different kinematic morphologies ranging from a disc structure (top left panel) to a disc+bulge structure (top middle) and an elliptical structure (top right). All three galaxies (f.l.t.r. GalaxyID 8772511, 8960069, 10645724) are at $z=0$ and have similar stellar masses, $\log_{10}(T/{\rm M_{\odot}})=10.56,10.60,10.70$ respectively. Each dot denotes a stellar particle, colour coded by the redshift at which it was formed. The horizontal axis denotes the 3D distance from the centre. The stellar half-mass radius is denoted by the vertical black dashed line. The vertical axis shows the alignment of the angular momentum of a given stellar particle ($\vec{j}$) with the angular momentum direction of the galaxy ($\hat{Z}$), where $\hat{Z}$ denotes the direction of the total angular momentum of all star particles within the stellar half-mass radius. With this definition purely corotating particles have $j_Z/|\vec{j}|=1$ and purely counterrotating particles have $j_Z/|\vec{j}|=-1$. In this plot, a random distribution of angular momenta would have a uniform distribution of points in the vertical direction. We decompose each galaxy into two components, a `spheroidal' component with mass ($S$) equal to twice the mass of all particles $j_Z/|\vec{j}|<0$ and a `disc component' ($D$) which comprises the rest of the total stellar mass ($T$). In this way the kinematic structure of each galaxy is characterised by a single ratio $S/T$, which is 0.18, 0.51 and 0.86 respectively for these galaxies. The solid black curve in each panel denotes the running average of $j_Z/|\vec{j}|$ as a function of radius. In the top middle panel this goes from 0 at small radii, corresponding to a truly random angular momentum distribution to a value of close to 1 at large radi, corresponding to a pure disc. The bottom panels show the same diagnostics for the star-forming gas particles in the same three galaxies, but using the same direction for $\hat{Z}$ as in the top row. The right galaxy has no star-forming gas.} 
\label{figureExampleGalaxies}
\end{figure*}

In this work we use a kinematic morphology indicator, rather than a photometric one. The kinematic morphology of a galaxy is generally condensed into a single indicator such as a bulge-to-total ratio $(B/T)$, disc-to-total ratio $(D/T)$ or a kinematic morphology parameter $\kappa_{rot}$ \citep[e.g.][]{Scannapieco10,Sales10,Sales12,Zavala15,Bottrell17,Correa17}, with varying prescriptions for each indicator. In this work we use a simple prescription similar to the one applied to the GIMIC simulation by \citet{Crain10} and to the Illustris simulation by \citet{Bottrell17}.

First we determine for each galaxy the direction of total stellar angular momentum of all stellar particles within the stellar half-mass radius, denoted as $\hat{Z}$. Then we project the angular momentum of individual stellar particles $\vec{j}$ onto the $\hat{Z}$-direction and normalise it by the total angular momentum $|\vec{j}|$ of the given particle. The resulting variable $j_Z/|/\vec{j}|$ denotes the amount of corotation for each stellar particle with the central half of the galaxy. Stellar particles that corotate with the stellar disc have $j_Z/|/\vec{j}|=1$, stellar particles that counter-rotate have $j_Z/|/\vec{j}|=-1$ and stellar particles with random directions of angular momentum (a pure non-rotating spheroid) are distributed uniformly between -1 and 1 (which is the reason why we chose this definition). 

Fig. \ref{figureExampleGalaxies} shows the distribution of this `angular momentum alignment' parameter versus radius for three typical galaxies, a disc galaxy (top left panel), a disc+bulge galaxy (top middle panel) and an elliptical galaxy (top right panel). Each point corresponds to a stellar particle and its colour indicates its formation redshift. There is a clear visual distinction between the stellar disc component (stars with $j_Z/|/\vec{j}|\approx 1$), which tends to be younger, and the spheroidal component (uniformly distributed $j_Z/|/\vec{j}|$) which consists of older stars. In order to disentangle both components in a robust way, we define the `spheroidal component' with mass $S$ to be twice the mass of counter-rotating stars (with $j_Z/|/\vec{j}|<0$) \citep{Crain10}. The stellar disc mass, $D$, is defined as the total mass, $T$, minus the spheroidal component $S$. In rare cases where more than half of the stellar mass is counter-rotating we set $S=T, D=0$.

We use the ratio $S/T$ to quantify the stellar morphology of each galaxy. It varies from low to high (specific values are included in the top panels of Fig. \ref{figureExampleGalaxies}) ranging from disc galaxies, via disc+bulge galaxies to elliptical galaxies. We specifically denote this as $S/T$ instead of the more common $B/T$ ratio, because there is no distinction based on radius and the spheroidal component includes both the bulge and halo, although in many cases the spheroidal component is more centrally concentrated than the disc component. The difference with the B/T-ratio from \citet{Bottrell17} lies in the calculation of the $\hat{Z}$ direction. \citet{Bottrell17} use all stellar particles within ten half-mass radii. We use all stellar particles within one half-mass radius. We do this because the total stellar angular momentum can be dominated by structures at large radii (for example due to recent mergers) which could lead to a misclassification of the direction of rotation of the stellar disc. For a significant fraction of galaxies, the direction of the total stellar angular momentum varies with radius. The results thus depend on the choice of radius. The advantage of our prescription with respect to prescriptions based on kinetic energy \citep[e.g.][]{Sales10,Sales12,Correa17} is that the decomposition into a disc component and a spheroidal component is not sensitive to small variations in this $\hat{Z}$-direction. In fact for a hypothetical galaxy with a pure disc component and a purely random spheroidal component, the $S/T$ ratio will remain the same as long as the $\hat{Z}$-direction points to within 90$^{\circ}$ of the disc direction, because all disc stars will have a positive $j_Z/|/\vec{j}|$ and all spheroid stars will remain uniformly distributed.

\begin{figure*}
\text{Complex, $S/T=0.39$ \hskip28em\relax}\\
\includegraphics[width=\columnwidth]{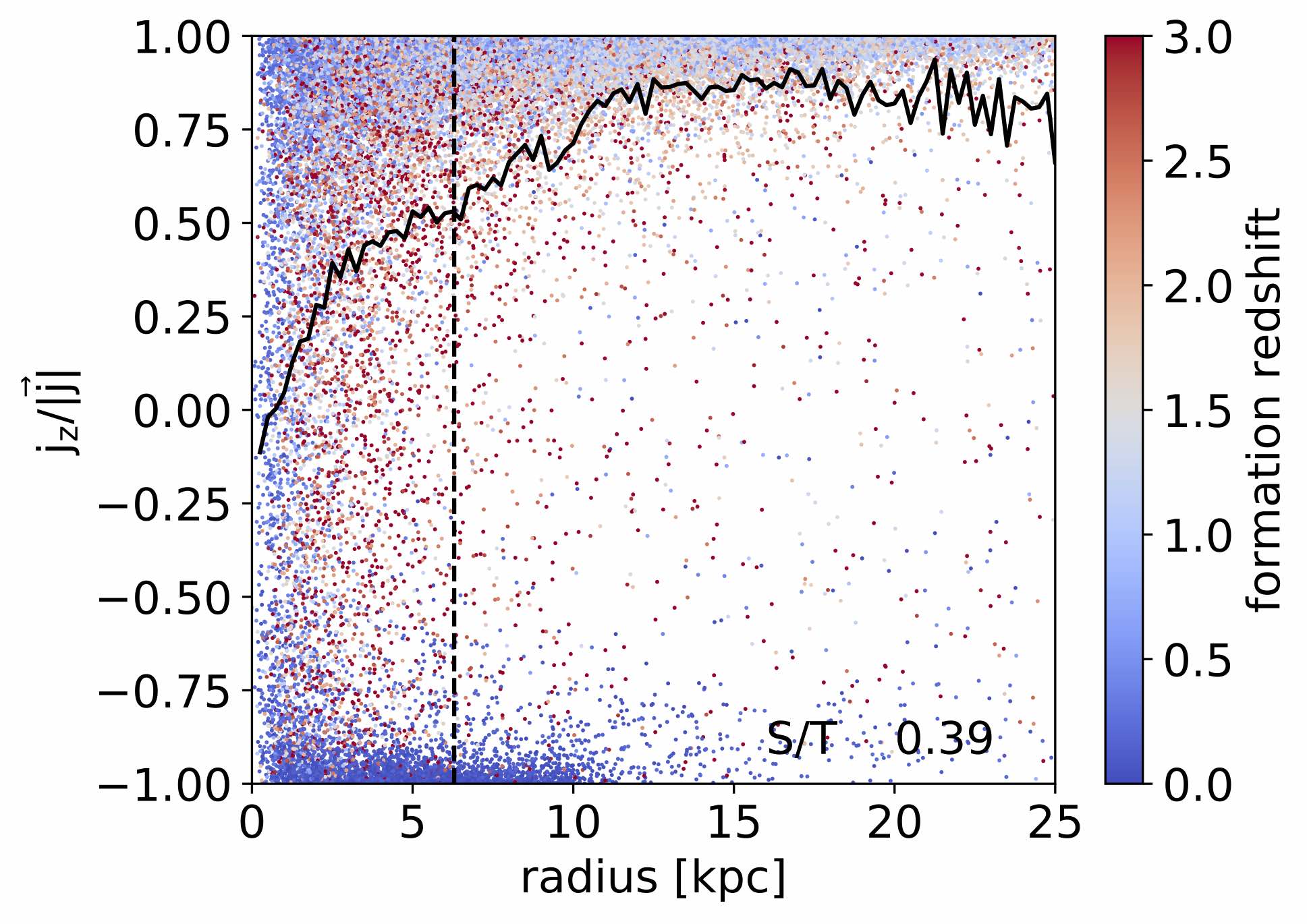}
\includegraphics[width=\columnwidth]{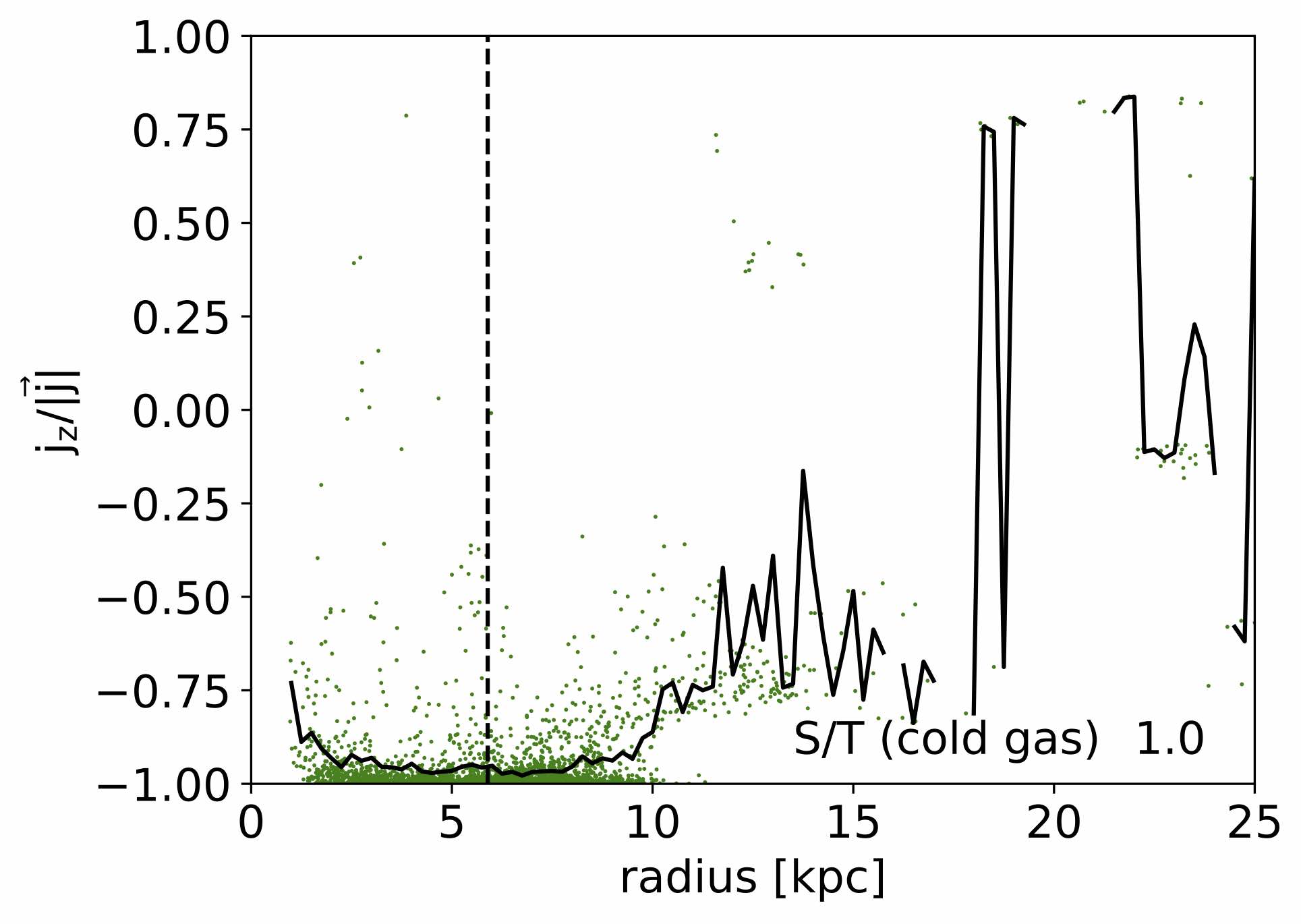}
\caption{The same diagnostics as in Fig. \ref{figureExampleGalaxies} for GalaxyID 18281742 at $z=0$ which has a stellar mass of $\log_{10}(T/{\rm M_{\odot}})=10.71$. This galaxy has a more exotic kinematic structure. The right panel shows a star-forming gas disc which is counterrotating with respect to the stars. In the left panel we see a counterrotating young stellar disc together with an extended corotating disc that consists of stars of varying ages and there is a hint of a bulge. The $S/T$ ratio does not discriminate between a classical bulge, a counterrotating disc or a counterrotating bar, but it does capture the kinematic content of the majority of galaxies that are more akin to the ones in Fig. \ref{figureExampleGalaxies}.}
\label{figureCounterrotatingGalaxy}
\end{figure*}

\begin{figure*}
\text{\hskip1em\relax Disc, $S/T=0.18$ \hskip5em\relax Disc+Bulge, $S/T=0.51$ \hskip4em\relax Elliptical, $S/T=0.86$ \hskip5em\relax Complex, $S/T=0.39$}\\
\vspace{0.5em}
\includegraphics[width=0.24\textwidth]{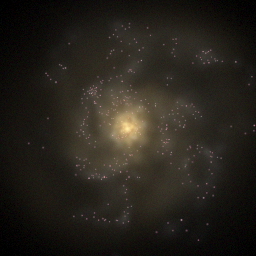}
\includegraphics[width=0.24\textwidth]{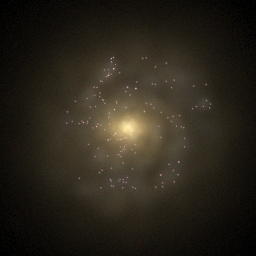}
\includegraphics[width=0.24\textwidth]{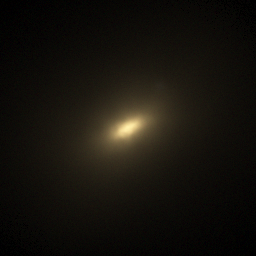}
\includegraphics[width=0.24\textwidth]{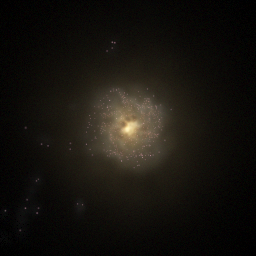}\\
\includegraphics[width=0.24\textwidth]{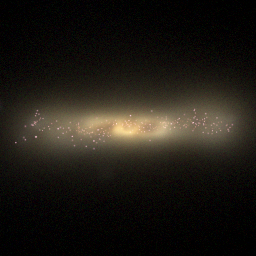}
\includegraphics[width=0.24\textwidth]{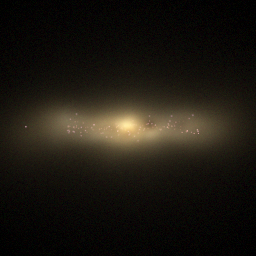}
\includegraphics[width=0.24\textwidth]{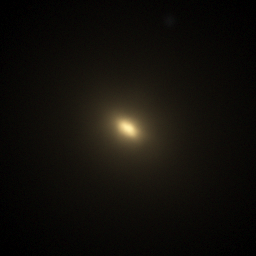}
\includegraphics[width=0.24\textwidth]{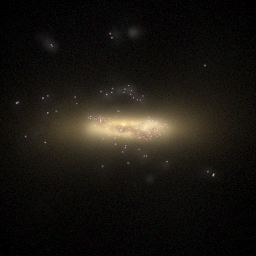}
\caption{Mock $gri$-images for the four galaxies from Figs. \ref{figureExampleGalaxies} and \ref{figureCounterrotatingGalaxy}. The images are 60 by 60 pkpc large. See \citet{Trayford15} and \citet{McAlpine16} for details. The top row shows the face-on views for GalaxyID 8772511, 8960069,10645724 and 18281742 respectively. The bottom row shows the corresponding edge-on views. The images for the first three galaxies agree by eye with the morphology that we deduced from the angular momenta of the stellar particles, representing a disc, disc+bulge and an elliptical galaxy respectively. The fourth galaxy would be classied by eye as a simple disc galaxy. Its counterrotating star-forming gas-disc is not apparent in the image.}
\label{figureGalaxyImages}
\end{figure*}

Of course a good portion of galaxies have more complicated structures than just a disc and a spheroid. When plotting the total angular momenta (instead of just the $\hat{Z}$-component), they show signs of for example bars or misaligned accretion \citep{Sales12}, but our simple decomposition catches the essence of the major kinematic morphology transformations that occur in the EAGLE simulation. Fig. \ref{figureCounterrotatingGalaxy} shows an example of a galaxy with a more complicated structure. The left panel shows that the youngest stars form a disc of 20 kpc diameter that is counter-rotating with respect to main disc, which is composed of older stars. The right panel shows that the star forming gas corresponds to this young counter-rotating disc. This galaxy has an $S/T$ ratio of 0.39, where in reality there is almost no hot spheroidal component, but instead two discs. This shows that in some cases the interpretation is not as simple as suggested by Fig. \ref{figureExampleGalaxies}.

The bottom row of Fig. \ref{figureExampleGalaxies} also shows the distribution of the cold star forming gas for our three example galaxies. Typically the angular momenta of star forming gas particles are very well aligned, starting at small radii (as in the left and middle bottom panels) yielding `star forming gas $S/T$' ratios of $\approx1$. Note that for the gas we still use the same $\hat{Z}$-direction defined by the stars within the half-mass radius. The elliptical galaxy (right bottom panel) has no star forming gas left.

Fig. \ref{figureGalaxyImages} shows mock $gri$-images for the four galaxies from Figs. \ref{figureExampleGalaxies}, \ref{figureCounterrotatingGalaxy}. The visual morphology corresponds well with our classification based on the $S/T$ ratio. Keep in mind that our $S/T$ ratio is mass weighted.  Disc stars are typically younger than spheroid stars, see Fig. \ref{figureExampleGalaxies}. A luminosity-weighted $S/T$ ratio for these galaxies would thus be smaller and the visual impression will thus be discier than suggested by the mass-weighted $S/T$. We understand that for comparison with observations our definition of $S/T$ is not ideal as it would be hard to extract this ratio from observations, for which less detailed kinematic information is available. A direct comparison with observations is not the purpose of this work though. Our aim is to gain physical insight into the formation of speroid and disc components in the simulation. For this, the $S/T$ ratio, which is based on detailed kinematic information, is well suited.

We do not retrieve pure stellar discs with $S/T\approx0.0$ (although this ratio is common for the star forming gas), whereas galaxies with very small bulge-to-disc ratios are thought to be fairly common \citep[e.g.][]{Kormendy10}. At the other end we interpret the elliptical galaxy from Figs. \ref{figureExampleGalaxies}, \ref{figureGalaxyImages} as having a 14\% disc component ($S/T=0.86$), whereas this galaxy would probably be classified photometrically as a pure elliptical galaxy. However, observations of ETG's that include stellar kinematics \citep[e.g.][]{ATLAS3} point towards varying degrees of rotational support for these galaxies. The 14\% surplus of stars over a uniform distribution in Fig. \ref{figureExampleGalaxies} is concentrated at $j_Z/|/\vec{j}|\approx1$.

In addition to the $S/T$ ratio for the entire galaxy, we will use the $S/T$ ratio for stars within 5 pkpc of the galaxy's centre and for stars outside 5 pkpc. This splits the `spheroidal' component into a `stellar bulge' and a `stellar halo' respectively and the disc component into an `inner disc' and an `outer disc'. 

\section{Morphology evolution}
\label{SectionMorphologyEvolution}

\begin{figure}
\includegraphics[width=\columnwidth]{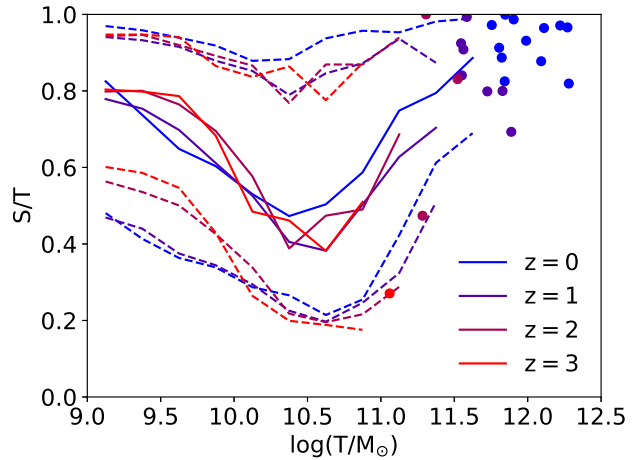}
\caption{The kinematic morphology of central galaxies, specifically the spheroid-to-total stellar mass ratio, as a function of stellar mass for different redshifts (colours). Solid curves denote running medians and dashed curves denote 10\%-90\% ranges. For mass bins with fewer than 10 galaxies, individual galaxies are shown as coloured dots. Although there are minor differences between the different redshifts, the overall picture is very similar. At low and high masses galaxies are mostly spheroidal (high $S/T$). In between, at $T \approx 10^{10.5} {\rm M_{\odot}}$, galaxies are mostly disky (low $S/T$). This trend is slightly stronger at high redshift than at low redshift.}
\label{figureSTpopulation}
\end{figure}

Fig. \ref{figureSTpopulation} shows the relation between the stellar $S/T$ ratio and stellar mass for central galaxies in the RefL0100N1504 simulation at different redshifts. At all redshifts this relation follows a similar trend. Low-mass galaxies ($T\lesssim 10^{9.5}{\rm M_{\odot}}$) are mostly spheroidal. Around the mass of the Milky Way most galaxies are disky and massive galaxies  ($T\gtrsim 10^{11}{\rm M_{\odot}}$) tend to be elliptical. The kinematic morphology of galaxies in EAGLE is primarily a function of stellar mass rather than redshift, although there are minor additional trends with redshift. At low redshifts ($z\lesssim 1$) the mass-morphology relation is a bit less pronounced and there is more scatter towards disky (low $S/T$) galaxies at low masses.

A convergence test of these results is included in appendix \ref{SectionAppendixA}, Fig. \ref{figureSTpopulationConvergence}. In short, these results are well converged in a `weak convergence' sense \citep{Schaye15}, meaning that the results are consistent at higher resolution when the subgrid model is recalibrated to the present-day galaxy stellar mass function and mass-size relation. This recalibration is needed to obtain the same effective efficiency of feedback processes at large scales when the transition between sub- and super-grid physics changes.

Instead of considering galaxy morphology for the whole population, we will now focus on the evolution of galaxy morphology along the merger tree, thus following the main progenitors\footnote{Main progenitors are loosely speaking the most massive progenitors, although in the case of a merger with a mass ratio close to unity, the choice of main progenitor is somewhat arbitrary. We use the prescription of \citet{DeLucia07} to select the progenitor with the `most massive integrated history', see \citet{Qu16}.} of massive galaxies backwards in time. Ultimately our goal is to understand when and why morphological transformations take place. A question best answered by following the evolution of these galaxies directly.

\begin{figure*}
\includegraphics[width=\columnwidth]{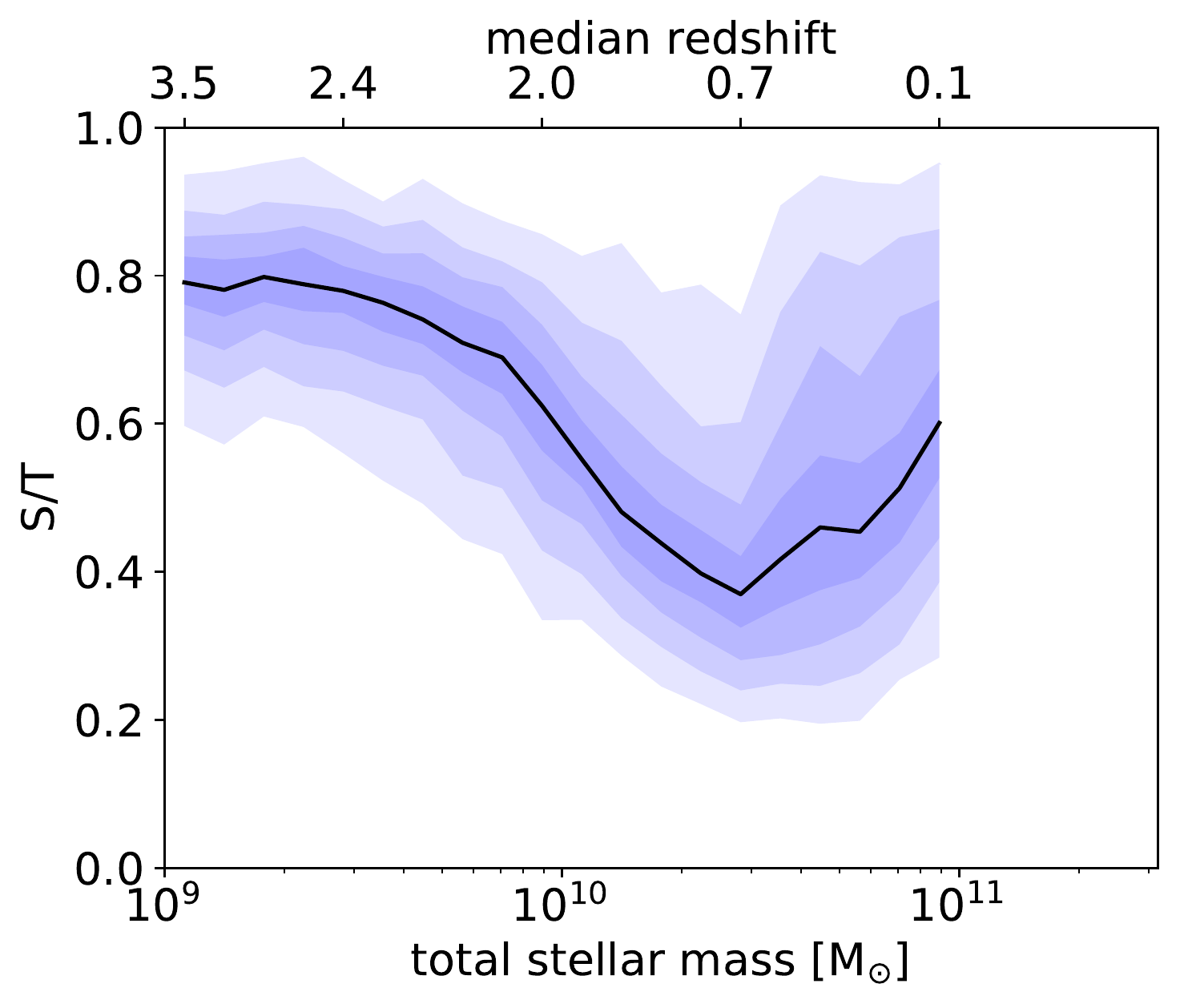}
\includegraphics[width=\columnwidth]{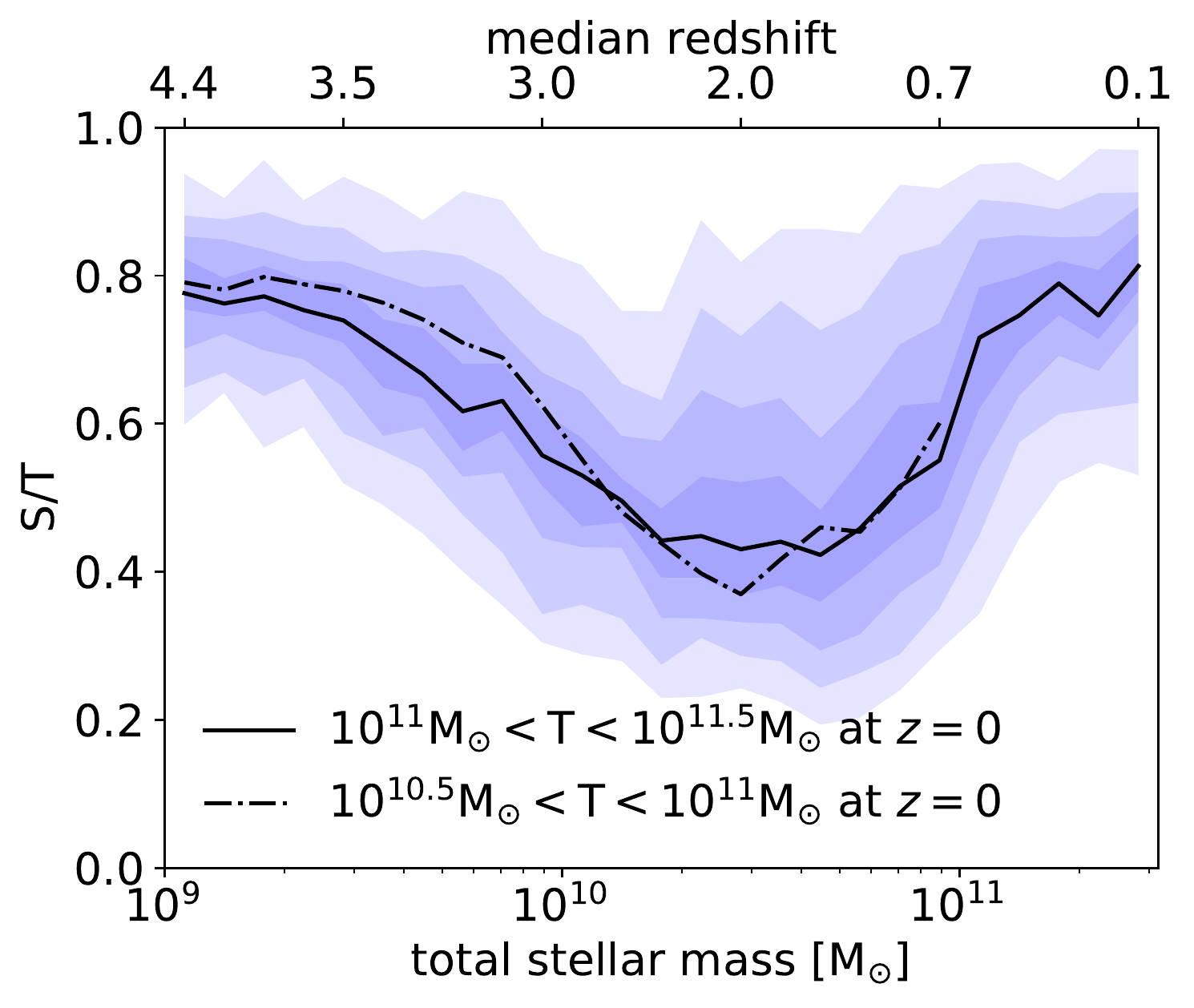}
\caption{The left panel shows the evolution of the stellar kinematics for the main progenitors of central galaxies with a total stellar mass $10.5<\log_{10}(T/{\rm M_{\odot}})<11$ at $z=0$. The vertical axis denotes the mass ratio of the `spheroidal' component with respect to the `total' stellar mass for these progenitors. The horizontal axis denotes the stellar mass of the main progenitors. The median redshift of these progenitors at different masses is indicated by the top horizontal axis. The solid black curve indicates the median of the distribution and the colours represent percentiles in 10\% increments. Most of these galaxies share a common kinematic evolution, starting out at high $S/T$ ratios at $T\lessapprox 3\times 10^9 {\rm M_{\odot}}$, subsequently becoming discier towards $T\approx 3\times 10^{10} {\rm M_{\odot}}$, after which the trend reverses and galaxies become increasingly less disky. The right panel shows the same diagnostics for central galaxies selected to have $11<\log_{10}(T/{\rm M_{\odot}})<11.5$ at $z=0$. The solid curve from the left panel is repeated as a dash-dotted curve for reference. The trend for these galaxies is remarkably similar, although the cosmic timing is very different (compare the top horizontal axes).}
\label{figureHTmass}
\end{figure*}

Fig. \ref{figureHTmass} shows the evolution of the $S/T$ ratio for the main progenitors of galaxies in the $z=0$ mass range $10.5<\log_{10}(T/{\rm M_{\odot}})<11$ (left panel) and $11<\log_{10}(T/{\rm M_{\odot}})<11.5$ (right panel). The median redshift as a function of mass is shown using the top axis. Although these galaxies span an order of magnitude in mass at $z=0$, they follow a very similar trend (compare the black solid and dash dotted curves in the right panel), as expected from the lack of significant evolution found in Fig. \ref{figureSTpopulation}. Galaxies start out with a spheroidal kinematic structure at low masses. In between  $10^{9.5} {\rm M_{\odot}}$ and $10^{10.5} {\rm M_{\odot}}$ they build up a prominent disc, resulting in a decrease of the $S/T$ ratio. At $T>10^{10.5}{\rm M_{\odot}}$ the $S/T$ ratio increases again, indicating a conversion from disky galaxies to spheroidal galaxies.

Perhaps surprisingly, we thus find that that low-mass ($T\lesssim 10^{9.5}{\rm M_{\odot}}$) central galaxies (Fig. \ref{figureSTpopulation}) and the low-mass main progenitors of massive $z=0$ central galaxies (Fig. \ref{figureHTmass}) tend to have a spheroidal (or otherwise non-disky) morphology. One might think that this could be due to the artificial pressure floor which inhibits the formation of cold, thin (i.e. scale height $\ll 1$ kpc) discs. However, we find no direct relation to the galaxy sizes, as would be expected if a puffy gas disc would be the root cause. In fact the median half-mass radius of the main progenitors remains constant over the mass range $10^{9}{\rm M_{\odot}}<T<10^{10.3}{\rm M_{\odot}}$ (not shown), whereas the transformation from elliptical to disc galaxies is practically complete over this mass range. Similarly, the overall mass-size relation in EAGLE is very flat at these masses, see Fig. 9 of \citet{Schaye15}. Also, we find that the star forming gas particles tend to have a disky distribution also at small radii (as is the case for the example galaxies in Fig. \ref{figureExampleGalaxies} but also for many lower mass galaxies), indicating that the cause for the spheroidal morphology is likely not the pressure-floor induced puffiness of the cool gas disc.

Recently \citet{ElBadry17} have found similar results for $z=0$ galaxies in the FIRE-2 simulation. The FIRE-2 simulation has a much higher resolution than EAGLE for low mass galaxies and it includes cooling of the interstellar matter down to 10 K. They found that the HI gas shows much more corotation than the stars for galaxies in the wide stellar mass range $10^{6.3}{\rm M_{\odot}}<T<10^{11.1}{\rm M_{\odot}}$. They also found that the gas fails to form a disc below $10^{8}{\rm M_{\odot}}$ and they furthermore found no signs of stellar discs for 15 out of their 17 galaxies with $T<10^{9.5}{\rm M_{\odot}}$. A similar early phase is found in the VELA simulation suite \citep{Zolotov15,Tacchella16b,Tacchella16a,Tomassetti16}, where galaxies below $T\approx10^{9}{\rm M_{\odot}}$ tend to be triaxial, prolate, and dispersion dominated.

\citet{Simons15} observe a similar transition based on the kinematics determined from nebular emission lines for a morphological blind selection of emission line galaxies at $z<0.375$. They define $10^{9.5}{\rm M_{\odot}}$ as the `mass of disc formation', because above this mass most galaxies are rotation dominated discs, while below this mass a large fraction of galaxies show no kinematic signs of disc rotation. Earlier work by \citet{Conselice06} finds a sharp transition from irregular galaxies to spiral galaxies around the same mass scale. Combined with the recent finding of \citet{Wheeler16} that 80\% of the local volume dwarf galaxies, including dwarf irregulars, have dispersion-supported, rather than rotation-supported, stellar motions, this would lead to a similar mass-dependent transition in kinematic morphology. Moreover, a recent analysis of galaxy morphology by \citet{Zhu17} based on a statistical modelling of stellar orbits of galaxies in the CALIFA survey, reveals the same transition from warm, hot, and counterrotating orbits to cold stellar orbits at $T\approx10^{9.5}{\rm M_{\odot}}$. Their observational analysis is most similar to our treatment, since they determine the fraction of counterrotating orbits, which is the basis of our spheroid definition. However, \citet{Fisher11} find the opposite trend based on a photometric $B/T$ decomposition of the light profiles of galaxies in the local (11 Mpc) Universe. They find an increasing fraction of bulgeless galaxies with decreasing mass. 

\citet{ElBadry17} argue that the reduced rotational support in their low-mass FIRE-2 galaxies is due to stellar feedback driving non-circular motions in the gas, in combination with heating by the UV background which suppresses the accretion of high angular momentum gas. \citet{Zolotov15,Tacchella16b,Tacchella16a,Tomassetti16} argue that the rotational support of low-mass VELA galaxies is reduced during the phase in which dark matter dominates the gravitational budget, also in the centres of galaxies. The transition to the disc-dominated phase is then initialised by a compaction event, which leads to a peak in the central star formation rate and a subsequent quenching of the core. A stellar disc can form at larger radii from freshly accreted high-angular-momentum gas. The compaction event which triggers the transformation shows up as a marked drop in the effective radius \citep[][Fig. 9]{Zolotov15}. Although for individual galaxies in EAGLE a similar compaction event could occur, overall the mass-size relation shows no sign of this \citep[][Fig. 9]{Schaye15}.

We found that the fraction of low-mass galaxies that have a disky morphology decreases somewhat with redshift (see Fig. \ref{figureSTpopulation}). At high redshifts we expect effects from the possibly more violent, disorganised growth of galaxies which we discussed in the introduction: the collapse of primordial gas clouds \citep{Eggen62}, clump migration in violently unstable discs \citep[e.g.][]{Noguchi98,Bournaud07,Elmegreen08,Perez13}, strong gas flows to the centre in marginally unstable discs \citep[e.g.][]{Krumholz17} and misaligned accretion \citep[e.g.][]{Sales12,Aumer13}. The merger rates are much higher at these redshifts \citep[e.g.][]{Genel10,Qu16}. Furthermore, turbulence generated by stellar feedback may well be stronger at higher redshifts as a result of the higher specific star formation rates \citep[e.g.][]{Johnson18}. We note that if feedback produces turbulence but little counterrotation, then it is possible that other measures of kinematic morphology may yield somewhat stronger evolution at fixed mass.

\begin{figure*}
\includegraphics[width=\columnwidth]{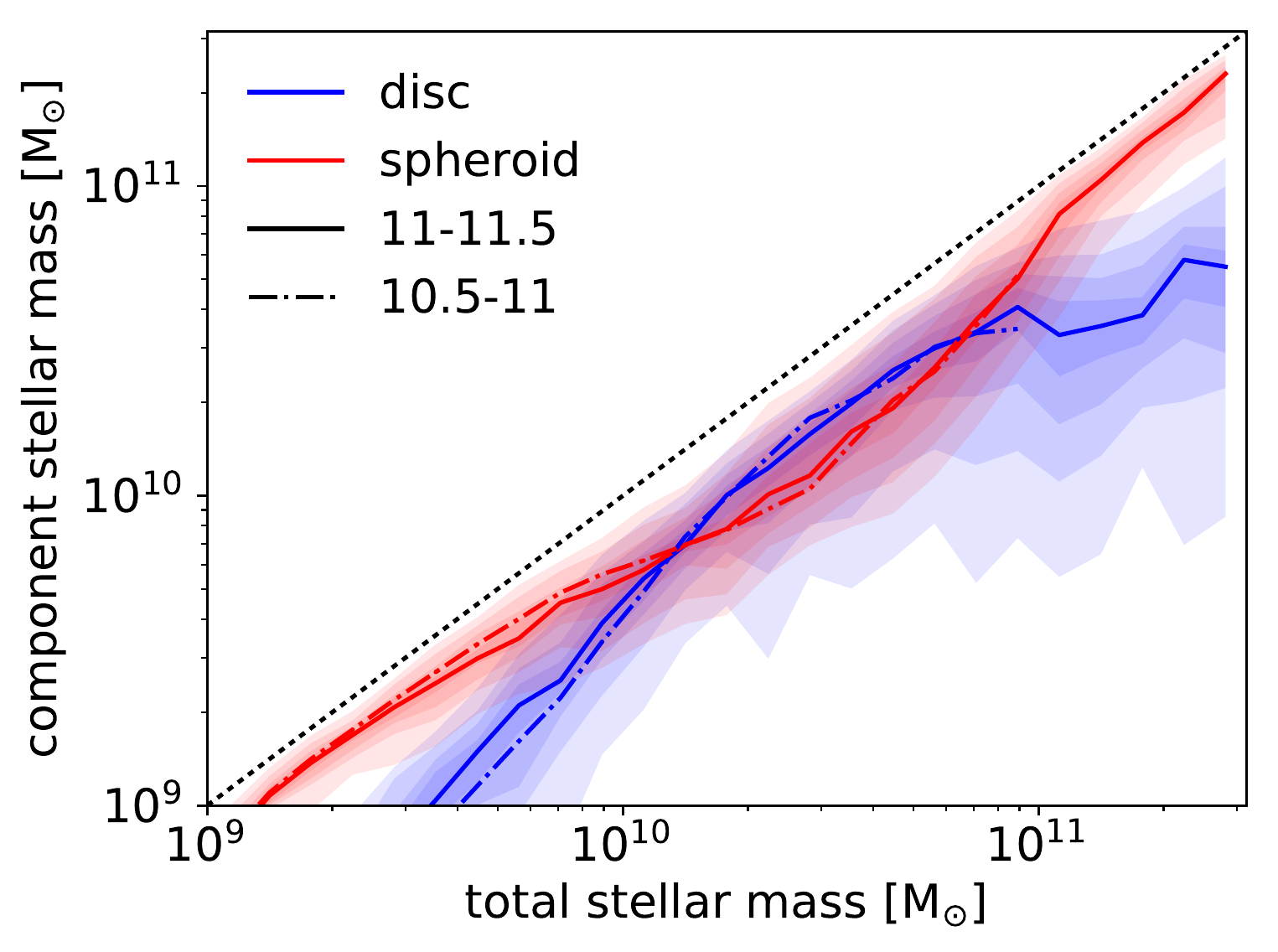}
\includegraphics[width=\columnwidth]{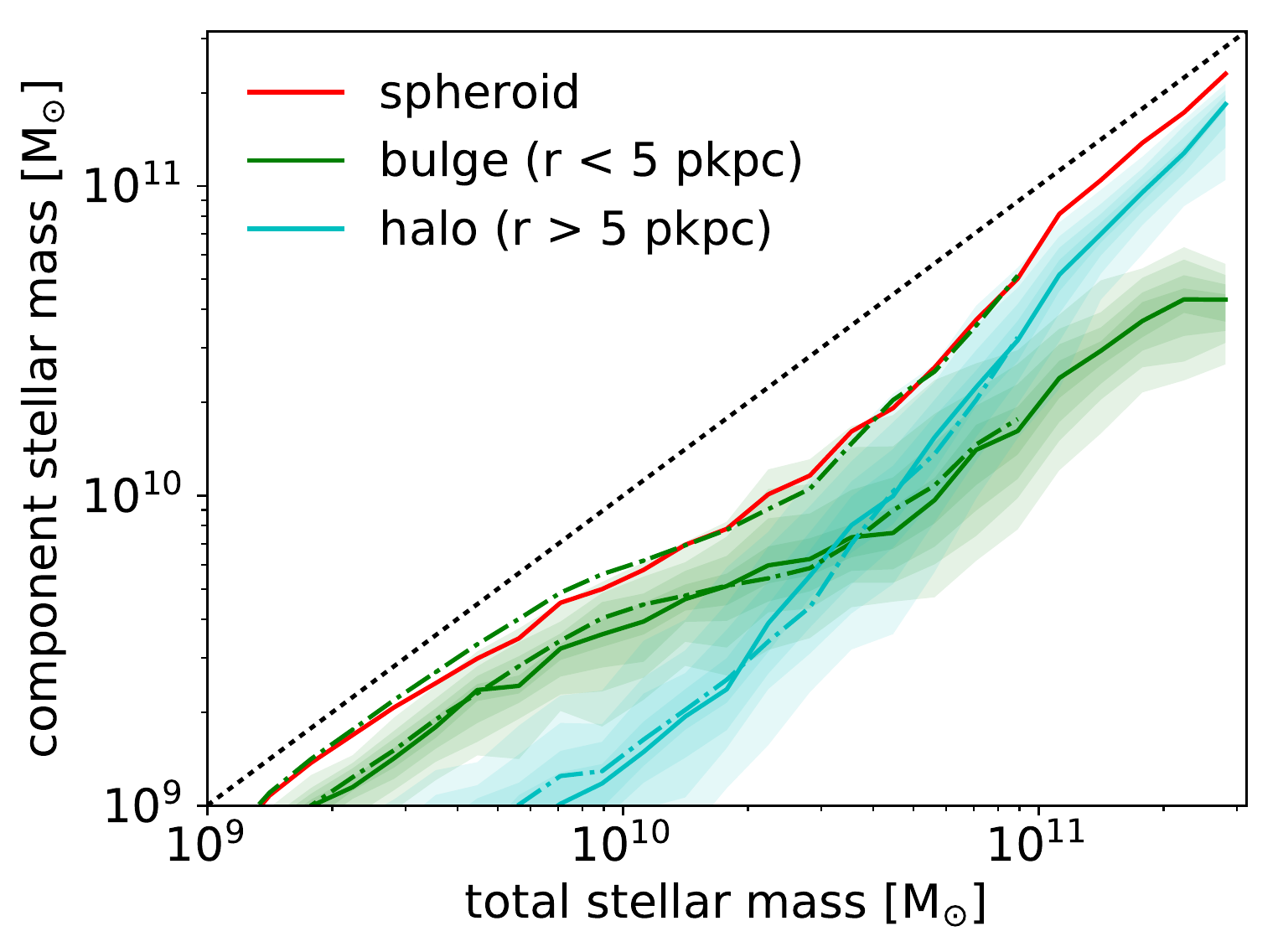}
\caption{The left panel shows the evolution of the `disc' mass component (blue) and the `spheriodal' mass component (red) for the main progenitors of central galaxies. Solid and dash-dotted curves represent the same $z=0$ mass ranges as in the right panel of Fig. \ref{figureHTmass}. The curves show running medians. We have indicated percentile ranges in  10\% shade increments only for the solid curve selection, but the ranges for the dash-dotted selection are very similar. The sum of the `disc' and `spheroid' components by definition equals the total mass (dotted black line). The two selections give very similar results. Galaxies start out witha spheroidal morphology, but the disc component grows fast, overtaking the spheroidal component just above $\rm 10^{10} M_{\odot}$. At large mass scales the spheroidal component catches up and it dominates at $\rm 10^{11.5} M_{\odot}$. The right panel splits the `spheroid' component (repeated in red) into two radial bins. We define the stellar bulge (in green) as the hot component within 5 pkpc of the galaxy centre and we define the stellar halo (in cyan) as the hot component outside 5 pkcp. This distinction demonstrates that the rise in the hot component at large masses is mostly due to the growth of a hot stellar halo at large radii. However, the bulge component keeps increasing over the whole mass range.}
\label{figureComponentMass}
\end{figure*}

Fig. \ref{figureComponentMass} (left panel) shows the evolution of the masses of the disc and spheroid components of the main progenitors. We see that during the period of rapid disc growth ($10^{9.5}{\rm M_{\odot}}\lesssim T \lesssim10^{10.5}{\rm M_{\odot}}$), the spheroidal component does grow in mass, albeit at a reduced rate. At the high-mass end the growth of the disc component flattens out, but the average disc mass still increases slightly. Although on average we do not see a destruction of disc mass, there will certainly be individual massive galaxies for which this is the case. For massive ($\approx10^{11.5}{\rm M_{\odot}}$) galaxies the spheroidal component clearly dominates, with the 10$^{\rm th}$ percentile of the spheroidal component being more massive than the 90$^{\rm th}$ percentile of the disc component. The relative scatter in disc masses is larger than the relative scatter in the spheroid masses.

In the right panel of Fig. \ref{figureComponentMass} we split the spheroidal component into a bulge and halo, i.e. inside and outside 5 pkpc respectively. This shows that the low-mass progenitors are dominated by a bulge, while bulge growth slows down considerably at $T\approx 10^{9.75}{\rm M_{\odot}}$ and makes place for a fast growth of the halo component at $T\gtrsim 10^{10.3}{\rm M_{\odot}}$. However, the mean bulge mass continues to grow during the period of rapid disc growth and subsequent halo growth. Roughly 24\% of the bulge mass of a  $10^{10.5}{\rm M_{\odot}}$ galaxy was on average in place at  $10^{9.5}{\rm M_{\odot}}$, before the epoch of rapid disc growth. At $10^{11}{\rm M_{\odot}}$ this percentage has dropped to 7\%, although a good portion of the bulge growth above $10^{10.5}{\rm M_{\odot}}$ takes place in galaxies with extensive halos, for which the bulge may not be perceived as a separate component. This is certainly the case for the ellipticals at the massive end.

\section{The origin of bulge stars}
\label{SectionOriginBulge}

The stars that make up a present-day galaxy have either been formed in its main progenitor (in-situ) or have been formed in another progenitor (ex-situ) and have subsequently been accreted during a merger. Disc stars are expected to have mainly formed in-situ. For the bulge and the halo components it is less obvious where their stars formed. These components could be the result of:
\begin{enumerate}
\item{various secular processes in the absence of mergers (in-situ),}
\item{the disruption of stellar discs by mergers (in-situ)}
\item{merger induced gas flows and subsequent star formation (in-situ),}
\item{accretion of stars during mergers (ex-situ).} 
\end{enumerate}
In this section we aim to estimate the contribution of process (iv) in the EAGLE simulation: direct bulge/halo formation from accreted stars. In section \ref{SectionMergers2} we will focus on the total merger contribution to bulge/halo formation, processes (ii), (iii) and (iv). Any remaining non-merger related bulge/halo formation will be attributed by definition to process (i) which includes the potential disruption of stellar discs by non-merger induced mechanisms as well as the non-merger induced direct formation of stars in a spheroidal component. An analysis of galaxies in the VELA simulation by \citet{Zolotov15} shows that at high redshift potentially half of the bulge stars are formed in-situ directly in the bulge component, items (i) and (iii).

\begin{figure*}
\includegraphics[width=\columnwidth]{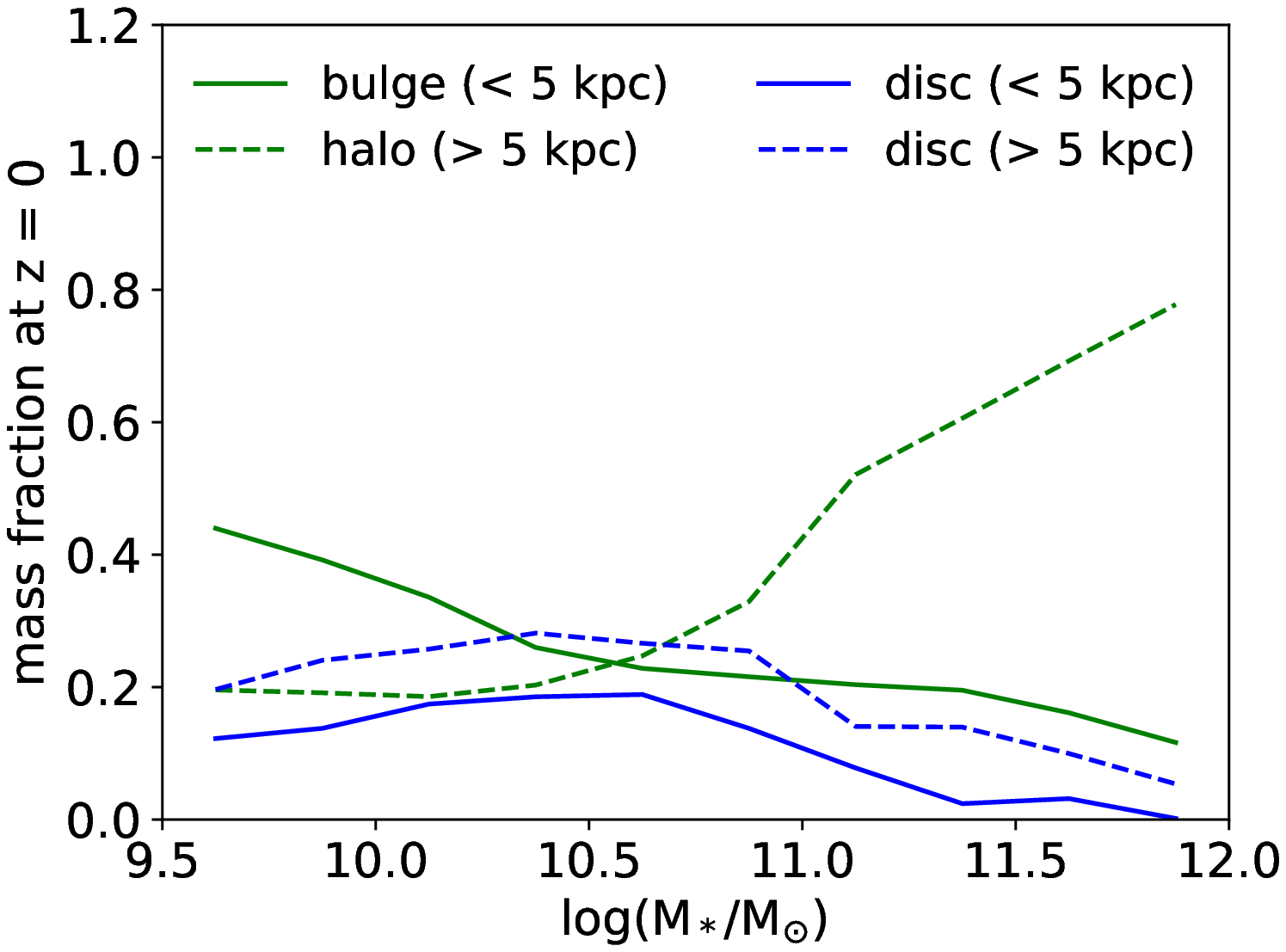}
\includegraphics[width=\columnwidth]{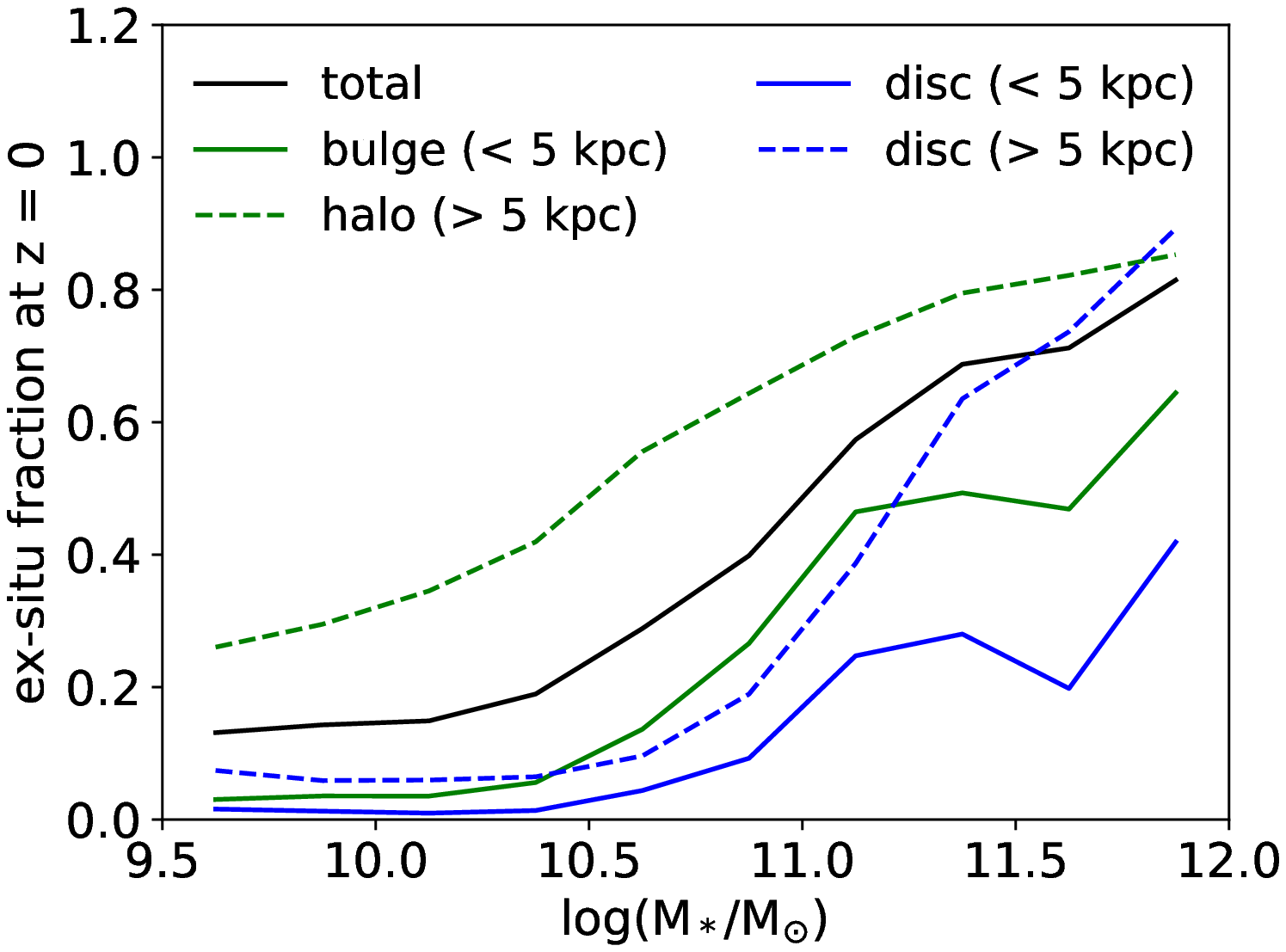}
\caption{The left panel shows the median stellar mass fractions of four kinematic stellar components of central galaxies at $z=0$. The spheroidal component is split into a  bulge' (solid green curve) and a `halo' (dashed green curve) as in the right panel of Fig. \ref{figureComponentMass}. The disc is similarly split at a radius of 5 kpc, giving an `inner disc' component (solid blue curve) and an `outer disc' component (dashed blue curve). Note that the horizontal axes in Figs. \ref{figureHTmass} and \ref{figureComponentMass} corresponded to the mass of the main progenitors, which corresponded to high redshifts for low masses, whereas in this figure the horizontal axis corresponds to $z=0$ only. The picture is however qualitatively similar. At low masses the bulge dominates, at high masses the halo dominates and in between the disc has its largest contribution. The right panel shows, for each component separately, the median mass fraction of stars belonging to that component that has been accreted (rather than formed in-situ). The black solid curve gives the median ex-situ mass fraction for the total galaxy. For $T\lesssim 10^{10.5}{\rm M_{\odot}}$ the disc, as well as the bulge, are almost entirely made up of in-situ formed stars, whereas the halo has a large contribution from ex-situ formed stars. At larger masses also the bulge and disc components contain more ex-situ formed stars.}
\label{figureExsituRedshiftZero}
\end{figure*}

The left panel of fig. \ref{figureExsituRedshiftZero} shows the makeup of $z=0$ galaxies as a function of mass in terms of bulge, halo and disc components. The disc components are most prominent around and below the knee of the galaxy stellar mass function, $T\lesssim10^{10.75} {\rm M_{\odot}}$ (where most of the stellar mass in the universe resides). At higher masses the halo component dominates while at $T\lesssim10^{10} {\rm M_{\odot}}$ the bulge component dominates the mass budget. This is all in qualitative agreement with the trend we saw for the main progenitors at high redshift in Fig. \ref{figureComponentMass}.

We now aim to calculate the fraction of stars for all of those morphological components that have an ex-situ origin. Remember that our decomposition into a hot/disc component is statistical in the sense that stellar particles with $j_Z/|\vec{j}|>0$ are not uniquely assigned to be in either component. It is therefore not possible to trace the provenance of the stars in each component directly. We can, however, circumvent this problem by first doing an $S/T$ decomposition for the in-situ and ex-situ formed stars separately (both inside and outside 5 pkpc). We then obtain masses for eight components (combinations of in-situ/ex-situ, spheroid/disc, inside/outside 5 kpc) from which we can calculate the ex-situ fractions. The right panel of Fig. \ref{figureExsituRedshiftZero} shows the medians of these mass fractions for all central $z=0$ galaxies.

For $T\lesssim 10^{10.5} {M_{\odot}}$ the contribution from ex-situ formed stars to the bulge is very small ($\lesssim10\%$) (as it is for the disc). This means that these bulges were not formed directly from stars that were accreted during mergers, process (iv). The halo does have a prominent contribution from ex-situ stars, even for low-mass systems. At the massive end ($T\gtrsim 10^{11} {M_{\odot}}$) where the overall ex-situ content of galaxies rises (solid black curve), all components contain a larger fraction of ex-situ formed stars. For the disc components we should not overinterpret this finding though, because these are ex-situ fractions for components that themselves constitute only a minor fraction of the total stellar mass budget of these massive galaxies, as is evident from the left panel of Fig. \ref{figureExsituRedshiftZero}.  The sharp transition from in-situ dominated galaxies to ex-situ dominated galaxies at the massive end agrees well with the inference from close pair counts in GAMA \citep[Fig. 17 of][]{Robotham14}.

\section{The effects of star formation and mergers on morphology}
\label{SectionMergers}

In the previous section we investigated the importance of the direct formation of bulges and halos from stars accreted during mergers. This does not include the indirect effect that mergers might have in triggering morphological changes. In this section we first investigate the effect of mergers and in-situ star formation on the overall kinematic morphology $S/T$, before isolating the effect on the buildup of the individual morphological components in section \ref{SectionMergers2}.

We investigate the changes in kinematic morphology between consecutive snapshots along the merger tree and relate those to the merger activity and in-situ star formation. We use all main progenitors of central galaxies in the mass range $10^{10.5}{\rm M_{\odot}}<T<10^{12}{\rm M_{\odot}}$ at $z=0$. The time resolution of this analysis is roughly 0.7 Gyr, although the time between consecutive snapshots is not completely constant. This is a convenient time step, because it is small compared to the ages of the galaxies, but long enough to capture the main effect of a merger on the morphology of a galaxy (except for cases where the merger happens close to the snapshot time). In principle we use all snapshots, although at very high redshifts few main progenitors will be in the mass range under consideration.

\begin{figure*}
\includegraphics[height=0.21\textwidth]{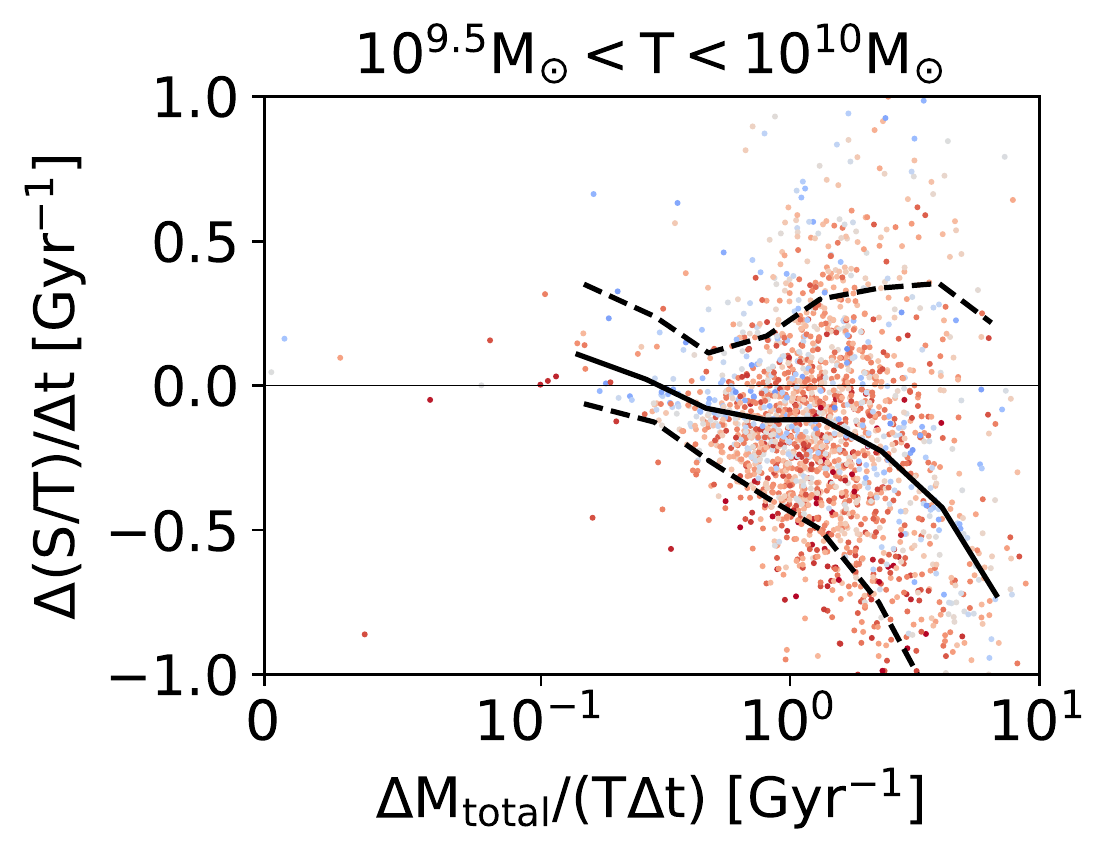}
\includegraphics[height=0.21\textwidth]{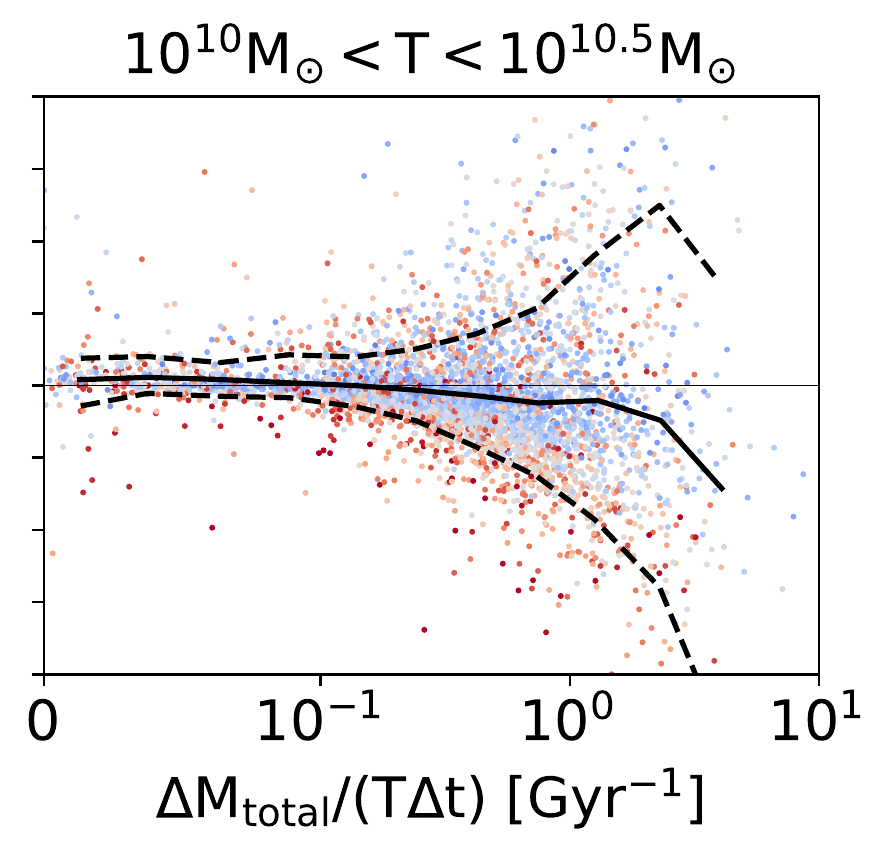}
\includegraphics[height=0.21\textwidth]{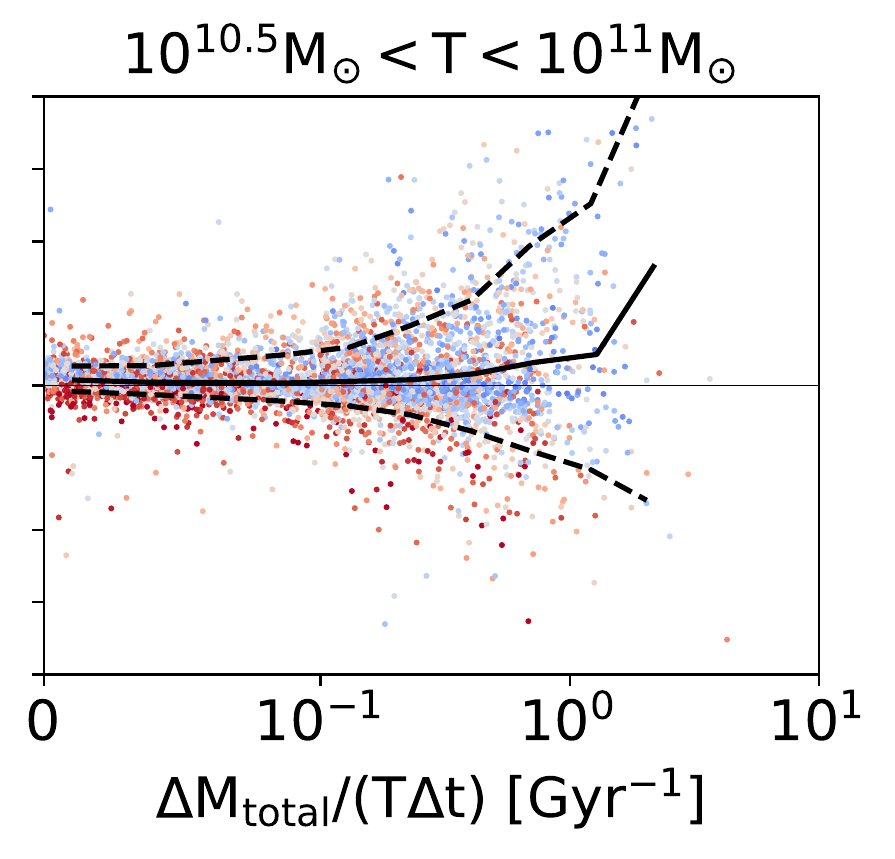}
\includegraphics[height=0.21\textwidth]{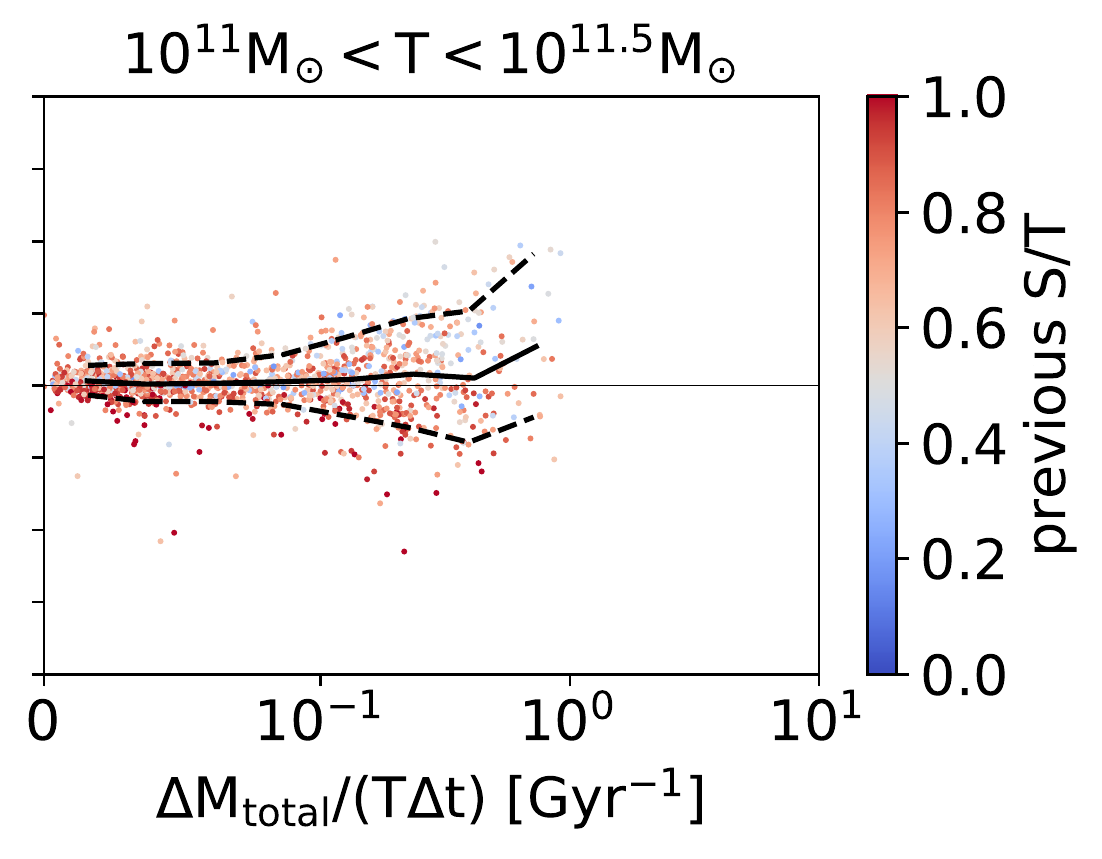}
\includegraphics[height=0.21\textwidth]{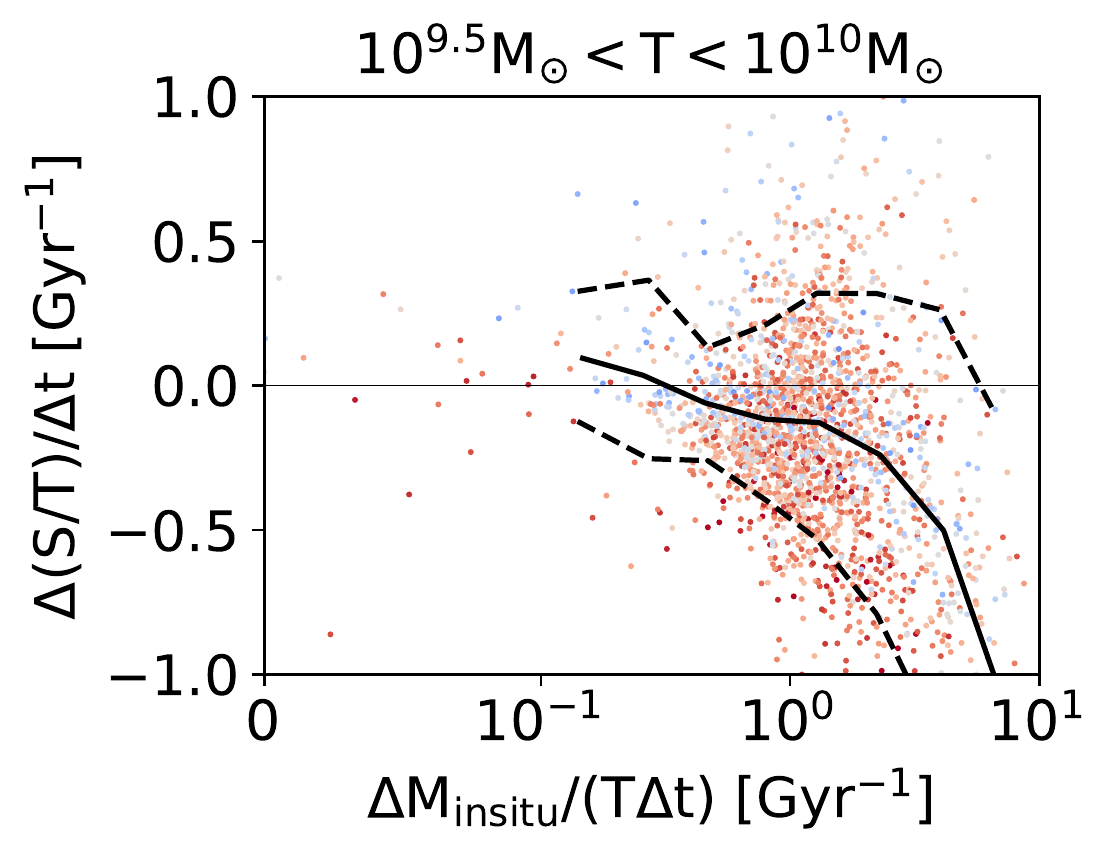}
\includegraphics[height=0.21\textwidth]{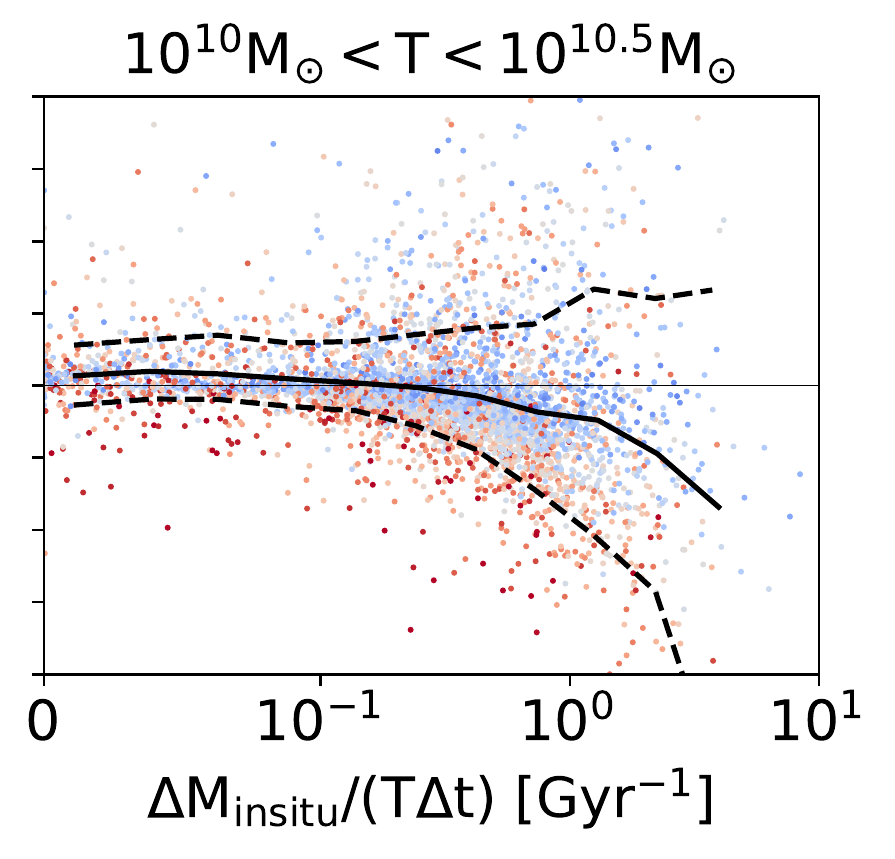}
\includegraphics[height=0.21\textwidth]{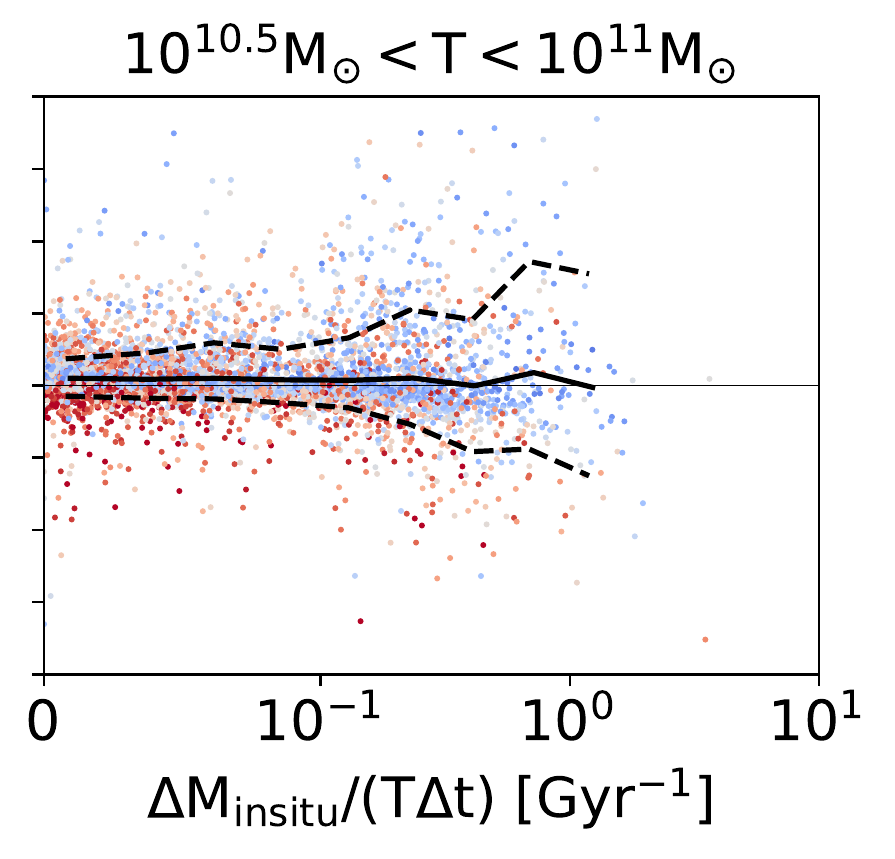}
\includegraphics[height=0.21\textwidth]{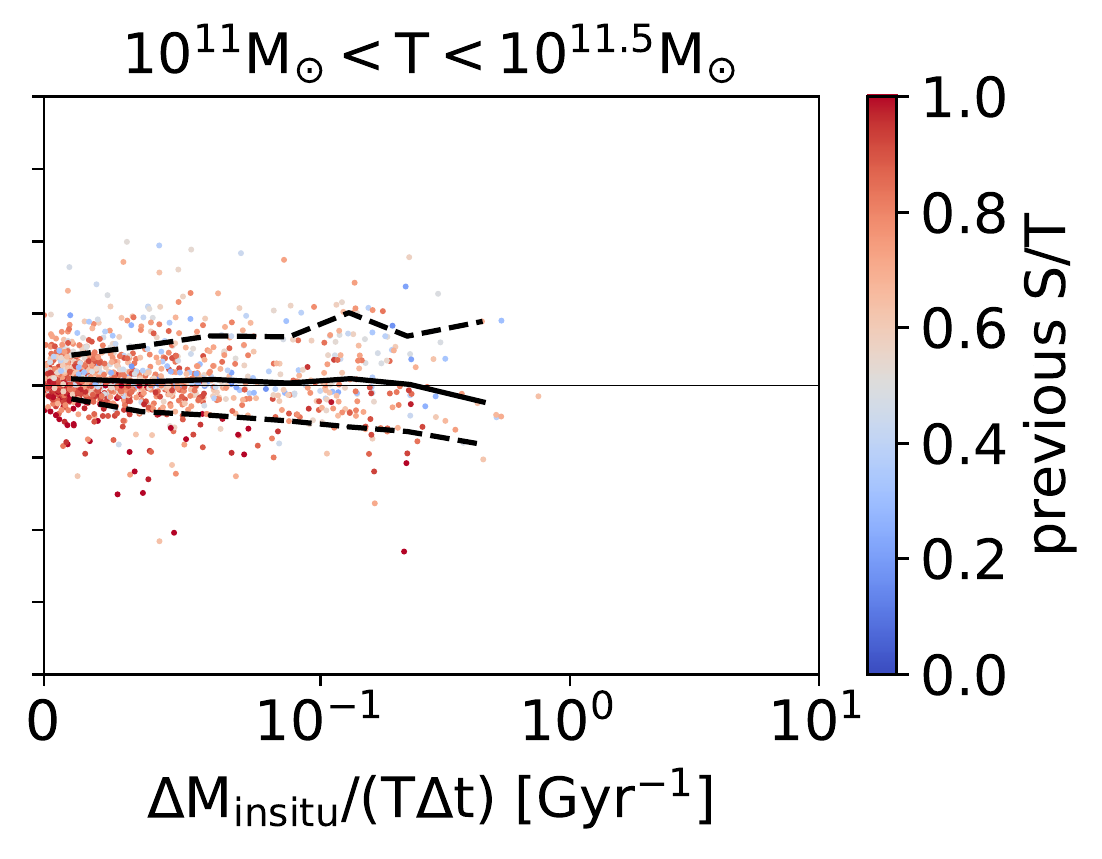}
\includegraphics[height=0.21\textwidth]{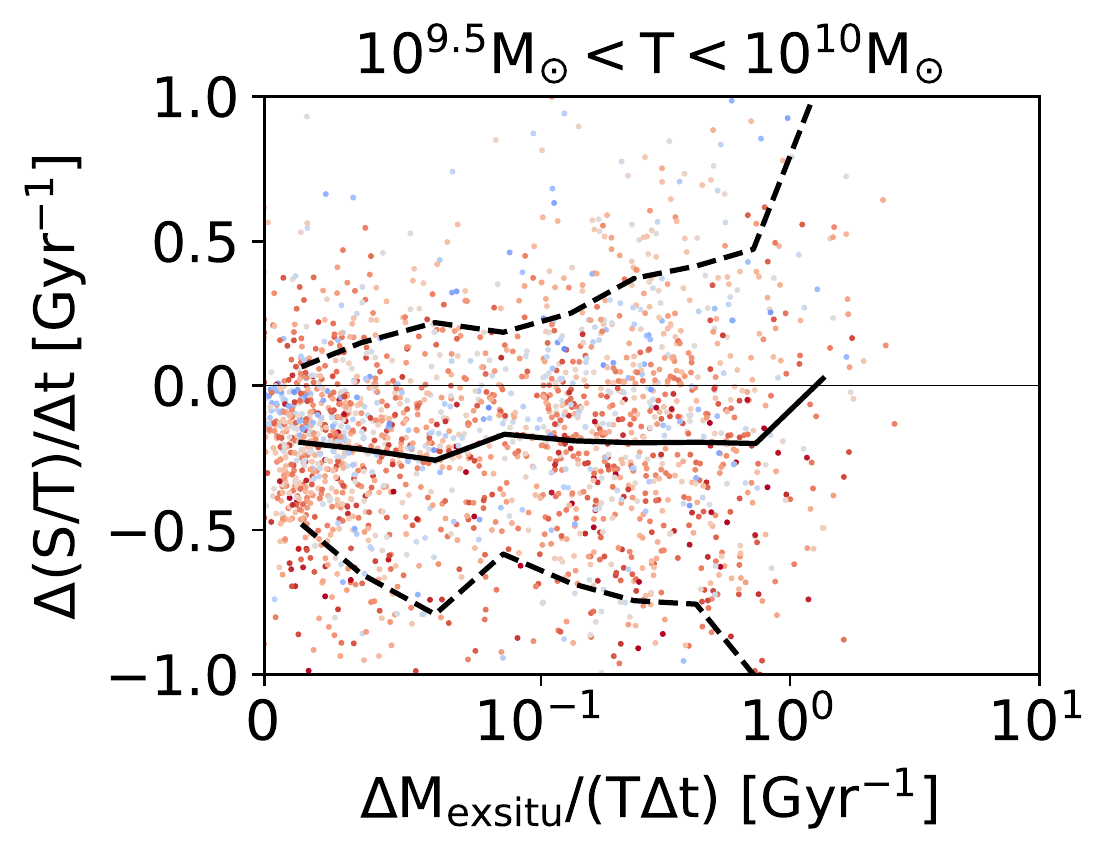}
\includegraphics[height=0.21\textwidth]{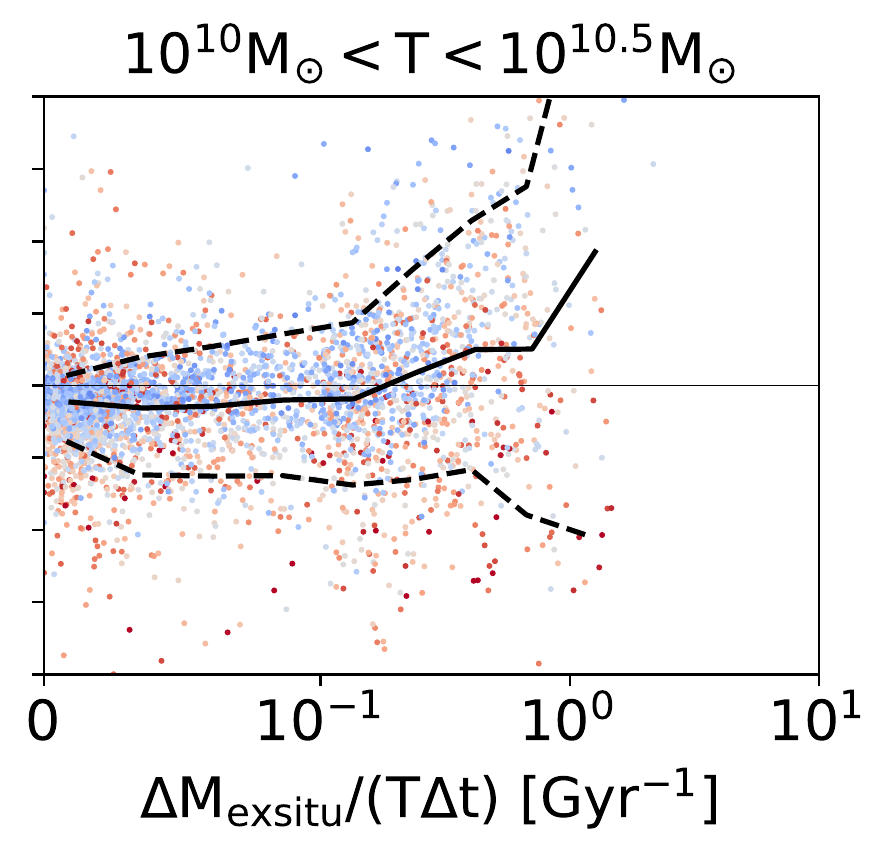}
\includegraphics[height=0.21\textwidth]{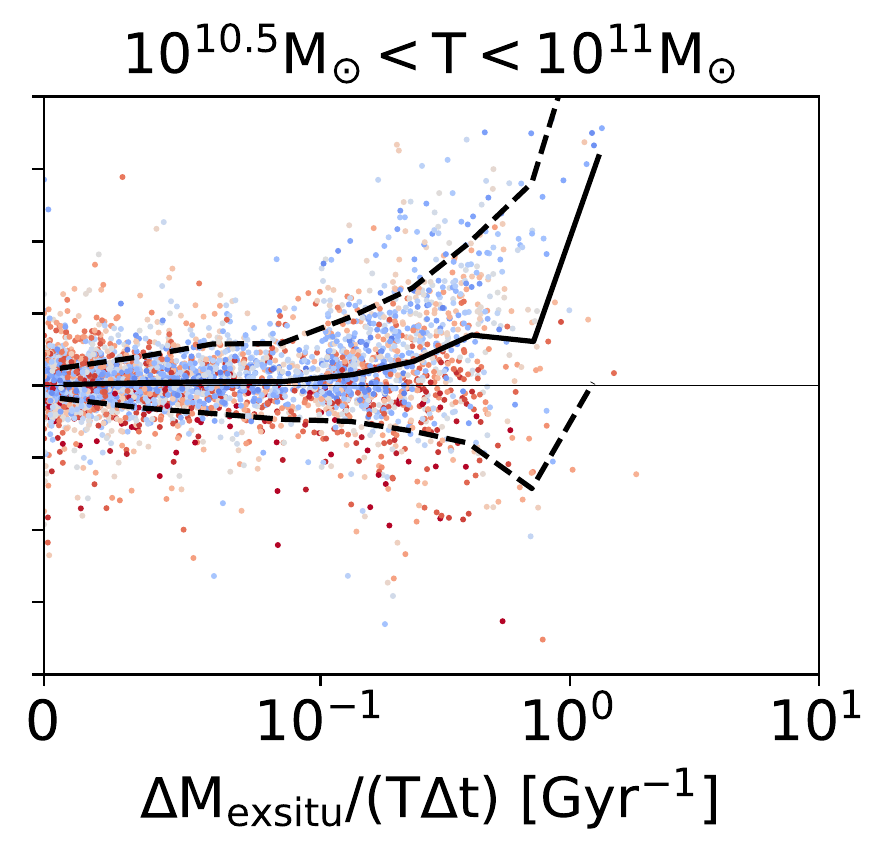}
\includegraphics[height=0.21\textwidth]{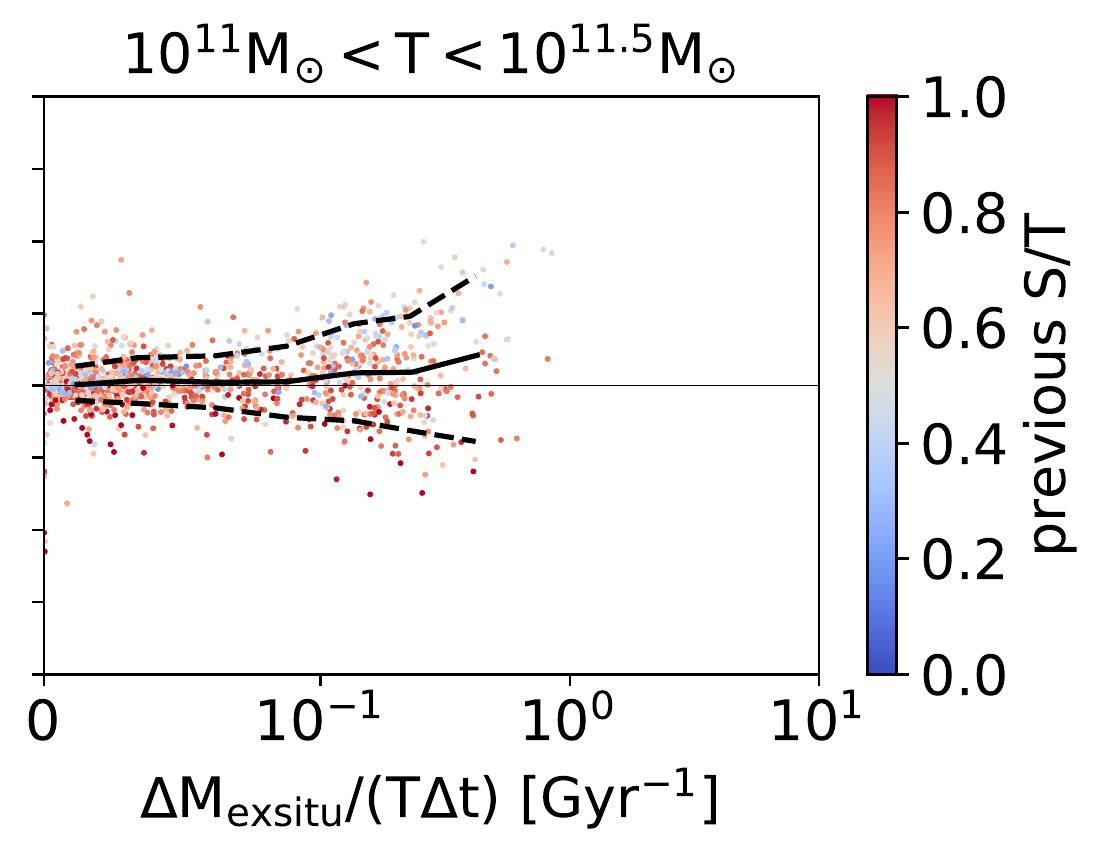}
\caption{The change in kinematic structure between consecutive snapshots, denoted by the change in the $S/T$ ratio per Gyr, as a function of respectively the relative stellar mass growth per Gyr (top row), the mass growth through in-situ star formation (middle row) and the mass growth through accretion of stars (bottom row). Each column corresponds to a different main progenitor mass range (evaluated at the earliest snapshot) as indicated above each panel. This figure contains all main progenitors of central galaxies at $z=0$ in the mass range $10^{10.5}{\rm M_{\odot}}<T<10^{12}{\rm M_{\odot}}$. Each main progenitor appears multiple times over multiple panels (once for each snapshot for which it falls in the assigned mass range). The galaxies are colour-coded by the $S/T$ ratio at the earliest snapshot. In each panel the running average is denoted by a solid curve and the 10\%-90\% range by dashed curves. The horizontal axis is linear below $10^{-1}$ and logarithmic above that. The top row shows that  in the mass range $10^{9.5} {\rm M_{\odot}} \lesssim T\lesssim 10^{10.5}  {\rm M_{\odot}}$ mass growth leads on average to a more disky kinematic structure (decreasing $S/T$), while in the  range $10^{10.5} {\rm M_{\odot}} \lesssim T\lesssim 10^{11.5}  {\rm M_{\odot}}$ mass growth leads to a more spheroidal kinematic structure (increasing $S/T$). The middle row shows that the strong trend for growing galaxies to become more disky below $\approx 10^{10.5} {\rm M_{\odot}}$ is a direct result of the in-situ star formation activity. The bottom row shows that merger activity on average leads to a more spheroidal kinematic structure.}
\label{figureSnapshotsCombined}
\end{figure*}

Fig. \ref{figureSnapshotsCombined} shows how the rates of kinematic morphology changes, $\Delta(S/T)/\Delta t$, relate to the stellar mass growth rates of galaxies (top row), to the mass growth rates through in-situ star formation ($\Delta M_{\rm insitu}/(T\Delta t)$  middle row) and to the mass growth rate through accretion of ex-situ formed stars ($\Delta M_{\rm exsitu}/(T\Delta t)$) which we use as a proxy for merger activity (bottom row). For each time step we define $\Delta M_{\rm exsitu}/T$ as the fraction of stellar mass at the later snapshot that has been accreted after the earlier snapshot. We normalise this by the time difference, $\Delta t$, between the two snapshots to obtain a rate per Gyr. In this calculation, the mass of the star particles, which is not constant due to stellar mass loss, is evaluated at the later snapshot (both for $\Delta M$ and for $T$). The in-situ mass fraction is calculated in a similar way. It includes all stars that have been formed since the earlier snapshot, thus also the stars that formed during a merger\footnote{Technically it also includes  stars that formed in a merger companion after the earlier snapshot and just before accretion. These should ideally be classified as ex-situ stars. This happens due to the finite time resolution but constitutes an insignificant fraction of the total $\Delta M_{\rm insitu}$ budget.}. We have split the sample into mass bins (columns) that represent the main progenitor stellar mass at the earliest of the two consecutive snapshots. This gives a much clearer picture than splitting by redshift (not shown).

Below $10^{10.5}{\rm M_{\odot}}$ galaxies tend to become more disky when they experience fast mass growth (downward trend in the first two panels of the top row), which is consistent with Fig. \ref{figureHTmass}. This push towards a disky kinematic structure is clearly caused by the in-situ star formation, as is evident from the strong downward trend in the first two panels of the middle row of Fig. \ref{figureSnapshotsCombined}, although mergers try to push the galaxies in the opposite direction towards a spheroidal kinematic structure (mostly the second panel of the bottom row).

Above $10^{10.5}{\rm M_{\odot}}$ the trend is reversed. Galaxies tend to become more spheroidal as they grow in mass (upward trend in the last two panels of the top row). The trend weakens at the highest masses because these galaxies are already mostly spheroidal. This transformation is driven by merger activity (upward trend in the last two panels of the bottom row) with a negligible contribution to the morphology changes by in-situ star formation (negligible trend in the last two panels of the middle row). The lack of a pronounced trend with the in-situ mass growth above $10^{10.5} {\rm M_{\odot}}$ could be due in part to the fact that the relative growth rate through in-situ star formation at these masses does not reach the high values that are responsible for most of the trend at lower masses. The importance of in-situ and ex-situ growth for morphology change thus shows a strong dependence on the mass of the main progenitor\footnote{The same probably holds for central galaxies that are not main progenitors of $z=0$ galaxies. We have specifically investigated main progenitors, because we are interested in long-lasting changes in morphology that are not wiped out by the disappearance of galaxies during mergers.}.

The reason that morphological changes can be decomposed into changes induced by mergers and by in-situ star formation, is that the in-situ and ex-situ mass growth of galaxies is mostly unrelated. They are positively correlated, meaning that galaxies of a given mass with a higher merger activity tend to have a higher in-situ star formation rate, but this is a small effect. The Spearman ${\rm R^2}$ coefficient between $\Delta(M_{\rm insitu}/(T\Delta t))$ and  $\Delta(M_{\rm exsitu}/(T\Delta t)$ varies from 0.13 to 0.17 for the different 0.5 dex wide mass bins, which means, loosely speaking, that they are for 85\% unaware of each other's existence and peak at different ($\approx$ 0.7 Gyr) time steps. This is in qualitative agreement with observations from CANDELS at $z\approx2$ which indicate that only 3\% of the star formation budget in $T>10^{10}{\rm M_{\odot}}$ galaxies is triggered by major mergers \citep{Lofthouse16} and with observations from GAMA that only show enhanced star formation in primary merger galaxies for short duration ($<0.1$ Gyr) star formation indicators and find a reduced star formation rate in secondary galaxies \citep{Davies15}.

We show a figure analogous to Fig. \ref{figureSnapshotsCombined} in Appendix \ref{SectionAppendixB} (Fig. \ref{figureSnapshotsCombinedInside}), but for $S/T$ changes within 5 pkpc, thus relating to bulge formation. The trends are the same as for Fig. \ref{figureSnapshotsCombined}. Below $10^{10.5}{\rm M_{\odot}}$ in-situ star formation builds up a central disc, above this mass mergers dominate and push the central region towards a bulge structure.

\section{The merger contribution to spheroid and disc formation rates}
\label{SectionMergers2}

\begin{figure*}
\includegraphics[height=0.21\textwidth]{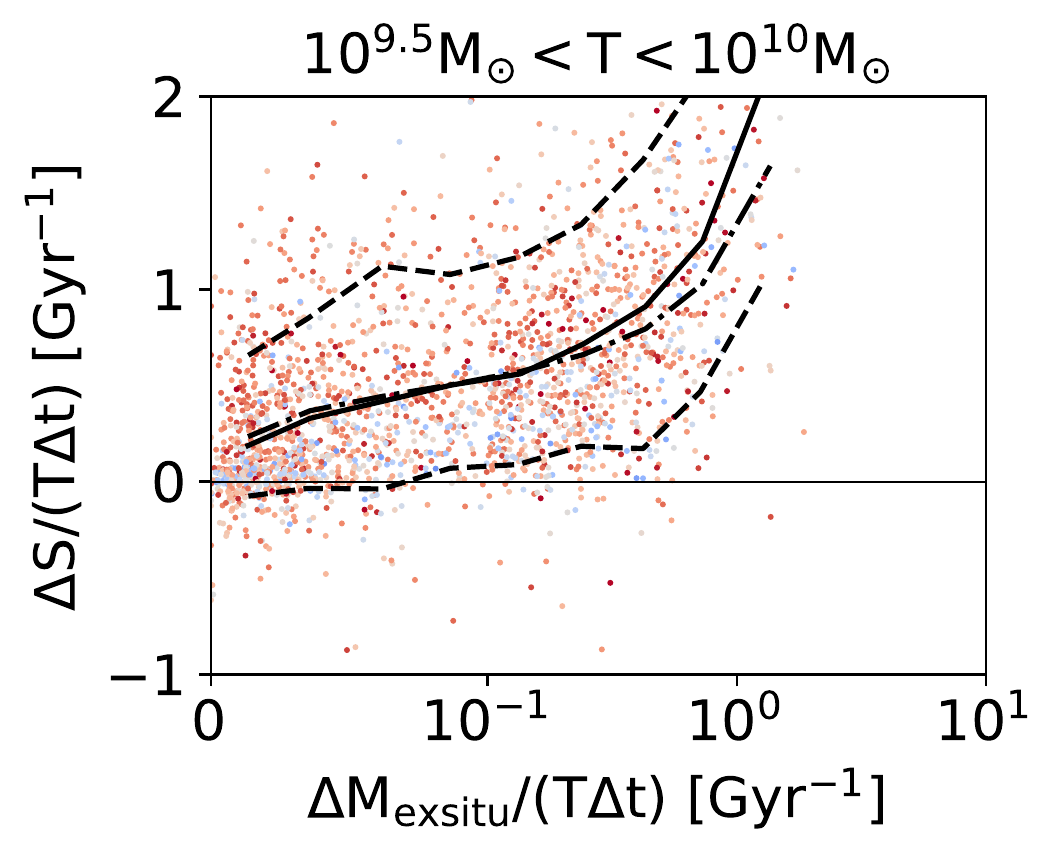}
\includegraphics[height=0.21\textwidth]{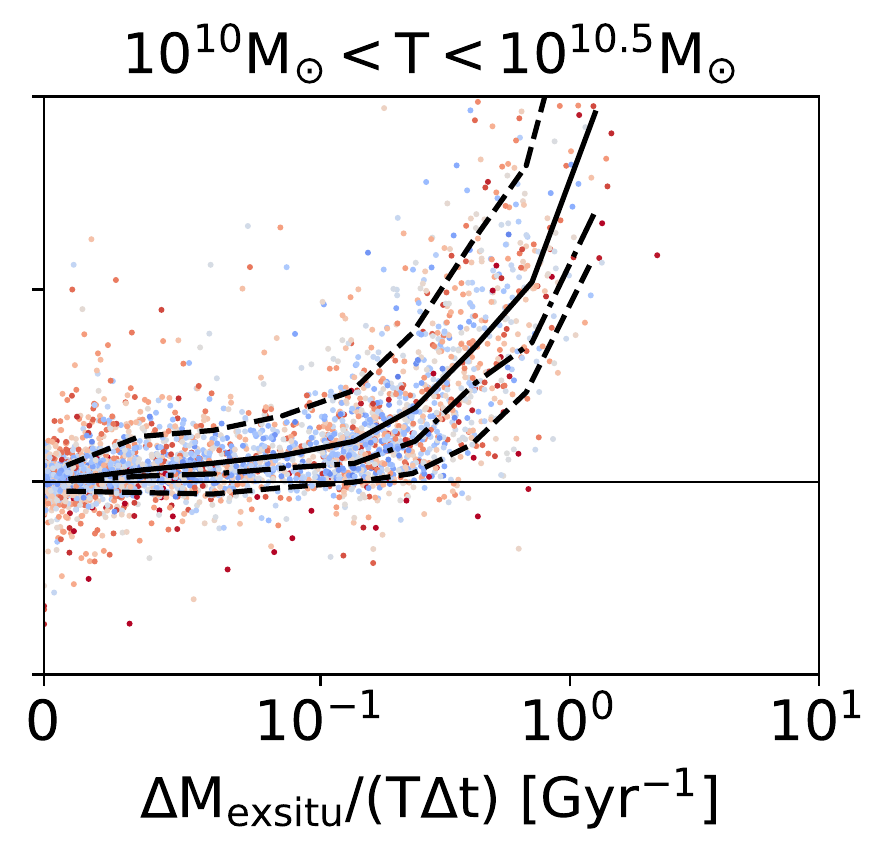}
\includegraphics[height=0.21\textwidth]{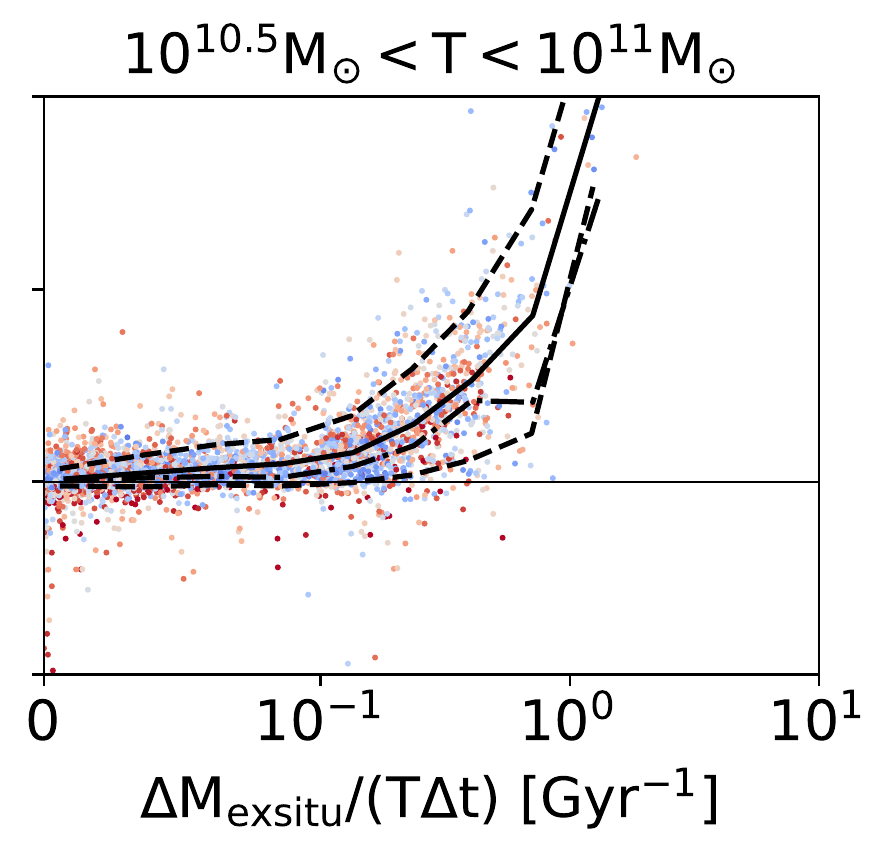}
\includegraphics[height=0.21\textwidth]{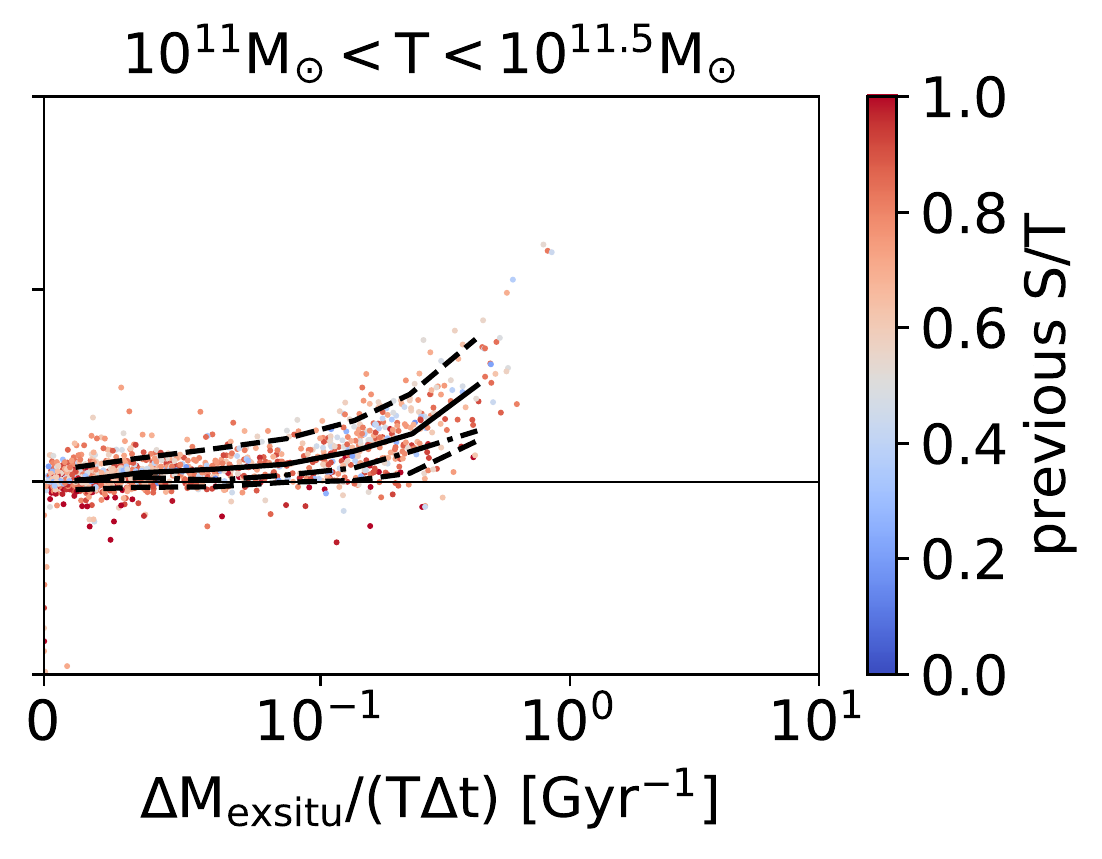}\\
\includegraphics[height=0.21\textwidth]{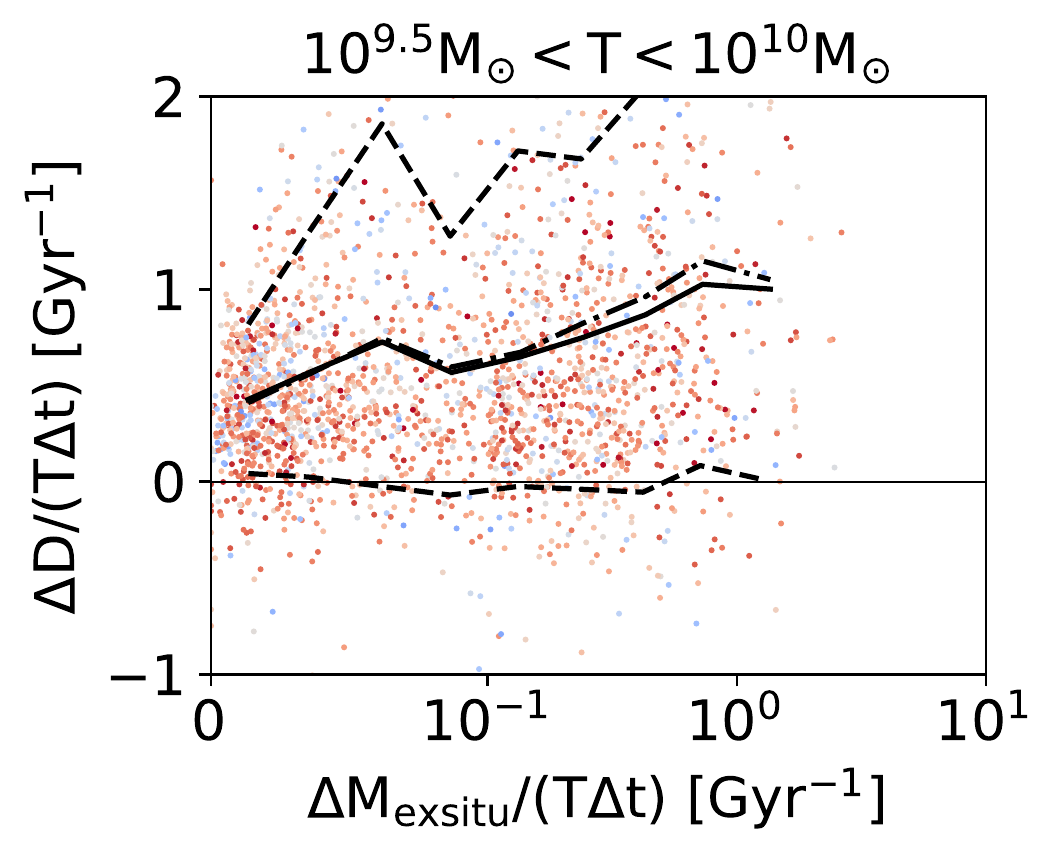}
\includegraphics[height=0.21\textwidth]{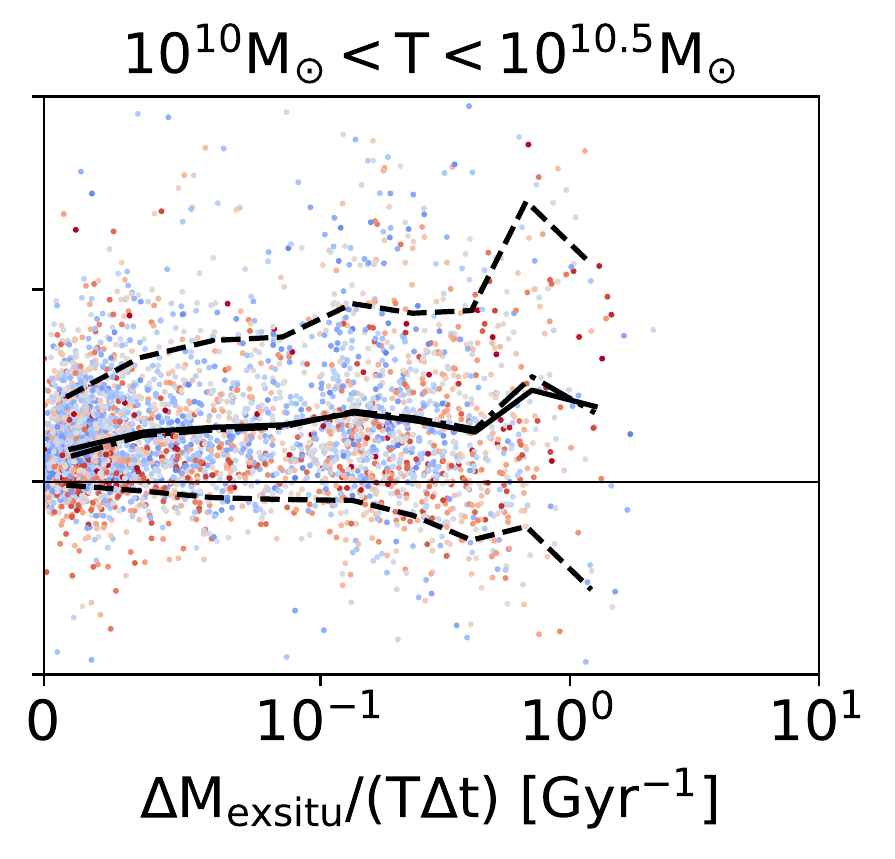}
\includegraphics[height=0.21\textwidth]{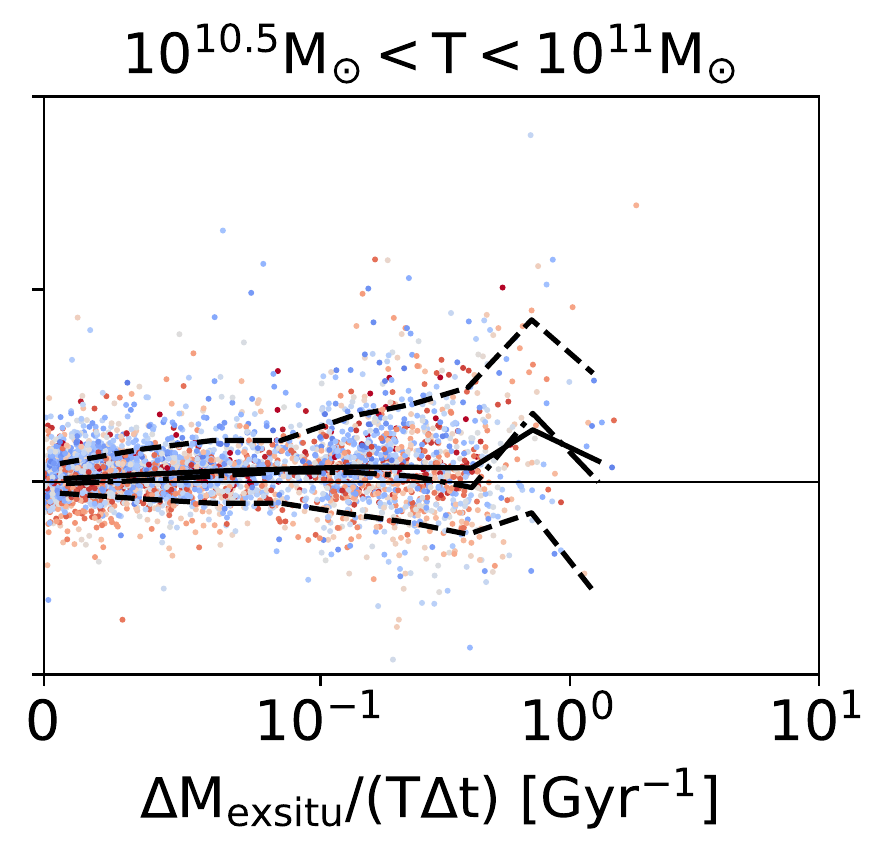}
\includegraphics[height=0.21\textwidth]{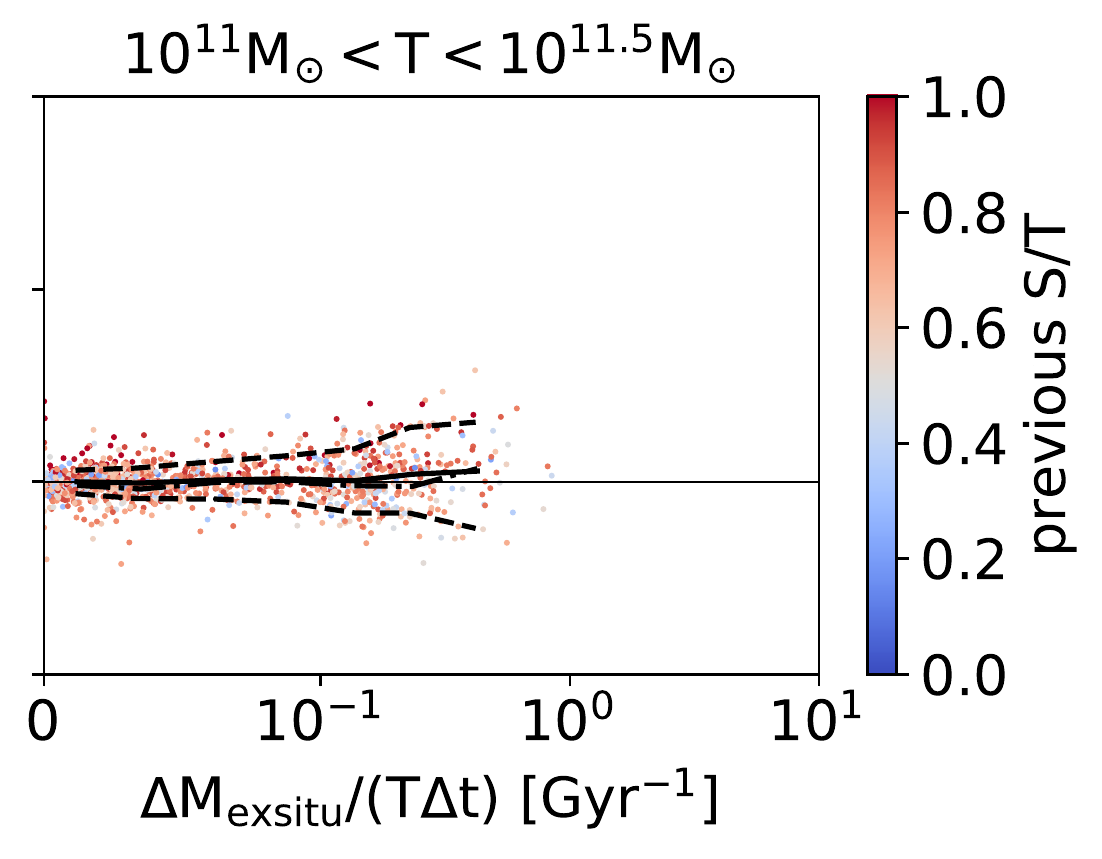}
\caption{The dependence of spheroid growth (top row) and disc growth (bottom row) on merger activity. The top row is the same as the bottom row of Fig. \ref{figureSnapshotsCombined}, but now the vertical axis denotes $\Delta S/(T\Delta t)$ instead of $\Delta(S/T)/\Delta t$. This isolates the growth rate of the spheroid component instead of the morphological change rate. Note that the horizontal axes are linear up to 0.1 and logarithmic beyond that. The solid curves denote running averages, dashed curves denote the 10\%-90\% range and the dash-dotted curves denote the running averages for the bulge growth rate $(\Delta S/T)_{\rm{<5kpc}}/\Delta t$ (the underlying distribution inside 5 kpc is not shown, but is very similar). In all top panels the relative growth rate of the spheroid depends strongly on merger activity. Time steps with little to no accretion of stars, on average show little to no growth of the spheroid (solid curve) or the bulge (dash-dotted curve). The bottom row shows the same diagnostics, but then for the growth rate of the disc (solid curves denote the running averages of the underlying distribution, dashed curves denote the 10\%-90\% range and the dash-dotted curves denote the running averages for the disc within 5 kpc). Disc growth shows a small dependency on merger activity. The curves are rising in the left part of the panels, where the accreted stellar mass rates are very small. It could well be that these `tiny' mergers trace the smooth accretion of gas. On average we do not see evidence for the destruction of discs by mergers (the solid and dash-dotted curves in the bottom panels are not declining towards the right).}
\label{figureMergerSpheroidFormation}
\end{figure*}

In this section we look at the total effect that mergers have on spheroid formation. Fig. \ref{figureMergerSpheroidFormation} (top row) shows the dependence of the spheroid growth rate, $\Delta S/(T\Delta t)$, on merger activity\footnote{In the calculation of $\Delta S/T$, we take for $T$ the average $T$ of both snapshots. This is done because in rare cases during a merger $T$ can be artificially low, due to a misidentification of which stellar particle belongs to which subhalo. If $T$ is very small, $\Delta S/T$ blows up. Furthermore we reject time steps for which $T$ drops by more than two-thirds. This only alters the percentages in Table \ref{TableSecular} by at most $2\%$.}, $\Delta M_{\rm exsitu}/(T \Delta t)$. This measure for merger activity includes mergers of all resolved mass ratios. For all mass ranges (columns) the average growth rate of the spheroid (solid black curve) increases strongly with merger activity and approaches zero during periods of low merger activity. This means that most of the spheroid formation is triggered by mergers.

We use the total ex-situ mass accretion rate as our proxy for merger activity instead of the more commonly used merger ratio and classification into minor and major mergers, because we expect the growth rate of the spheroid to not only depend on the merger ratio of the most prominent merger, but also on the number of mergers that occur during a $\approx0.7$ Gyr time step. Nevertheless, the horizontal axis in Fig. \ref{figureMergerSpheroidFormation} can be roughly translated into a merger ratio. $\Delta M_{\rm exsitu}/(T \Delta t) \approx 0.1\; {\rm Gyr^{-1}}$ is equivalent to a single merger with a mass ratio 1:13 within 0.7 Gyr. Similarly a rate of 0.3 Gyr$^{-1}$ corresponds to a single merger with a mass ratio 1:3.7 within 0.7 Gyr. The contribution at $\Delta M_{\rm exsitu}/(T \Delta t) > 0.3\; {\rm Gyr^{-1}}$ can thus roughly be attributed to major mergers. The contribution at $0.1\; {\rm Gyr^{-1}}< \Delta M_{\rm exsitu}/(T \Delta t) < 0.3\; {\rm Gyr^{-1}}$ can roughly be attributed to minor mergers and the contribution at $\Delta M_{\rm exsitu}/(T \Delta t) < 0.1\; {\rm Gyr^{-1}}$ can be attributed to `tiny' mergers, which in some works is referred to as the `smooth accretion' of stars. The average spheroid growth rates (solid curves in the top row) mostly rise in response to stellar accretion rates in the `minor and major' merger regime, especially above $\rm{10^{10} M_{\odot}}$. The same is true for the bulge growth rates (dash-dotted curves in the top row).

\begin{table*}
\begin{center}
\caption{The estimated contribution of mergers (of any mass ratio) to the rate of spheroid formation (3\textsuperscript{rd} column) and  bulge formation (5\textsuperscript{th} column) in different main progenitor mass bins (1\textsuperscript{st} column). We include a mass bin 0.5 dex smaller than in Figs. \ref{figureSnapshotsCombined} and \ref{figureMergerSpheroidFormation}. The 2\textsuperscript{nd} column gives the average rate of relative spheroid mass growth in the different mass bins. See the main text for an explanation of how we use this, together with the solid curves in the top row of Fig. \ref{figureMergerSpheroidFormation}, to estimate the total merger contribution to spheroid formation (3\textsuperscript{rd} column) and the approximate contribution of `minor + major' and `major' mergers (3\textsuperscript{rd} column, in parentheses).   The 4\textsuperscript{th} and 5\textsuperscript{th} columns repeat the same procedure for the bulge (i.e. the spheroid inside 5 kpc; corresponding to the dash-dotted curves in the top row of Fig. \ref{figureMergerSpheroidFormation}). Overall we see that the merger contributions to spheroid and bulge growth are very large, especially above $10^{10}{\rm M_{\odot}}$. Major, minor and tiny mergers all contribute to a similar degree.}
\begin{tabular}{llllll}
\hline
$\log_{10}(T/{\rm M_{\odot}})$ & $\left<\Delta S/(T\Delta t)\right>$  & merger contribution to & $\left<\Delta S/(T\Delta t)\right>_{\rm <5kpc}$  & merger  contribution to \\ 
 &   &  spheroid formation rate &  & bulge formation rate \\ 
 & & & & \\ 
 & Gyr$^{-1}$ & all (minor+major, major)& Gyr$^{-1}$ & all (minor+major, major)  \\ \hline
  9-9.5& 1.39 & >51\% ($\sim$39\%, $\sim$25\%) & 1.43 & >47\% ($\sim$33\%, $\sim$19\%)  \\ 
 9.5-10& 0.57 & >67\% ($\sim$46\%, $\sim$28\%) & 0.55 & >57\% ($\sim$36\%, $\sim$21\%)  \\ 
 10-10.5& 0.16 & >91\% ($\sim$76\%, $\sim$46\%) & 0.10 &  >95\% ($\sim$82\%, $\sim$55\%)  \\ 
 10.5-11& 0.08 & >82\% ($\sim$65\%, $\sim$33\%) & 0.05 &  >82\% ($\sim$74\%, $\sim$41\%)  \\ 
  11-11.5& 0.08 & >92\% ($\sim$64\%, $\sim$21\%) & 0.04 & >76\% ($\sim$64\%, $\sim$20\%) \\ \hline
\end{tabular}
\label{TableSecular}
\end{center}
\end{table*}

From the trends in the top row of  Fig. \ref{figureMergerSpheroidFormation} we can estimate the percentages of the spheroid- and bulge formation rates that are associated with mergers. This represents the combined effect of items (ii), (iii) and (iv) from section \ref{SectionOriginBulge}. First we estimate the secular contribution to spheroid formation, item (i) from section \ref{SectionOriginBulge}, by dividing the spheroid growth rate in the absence of mergers by the average growth rate: $\left<\Delta S/(T\Delta t)\right>_{\Delta M_{\rm exsitu}/(T\Delta t) <0.025 {\rm \; Gyr}^{-1}}$/$\left<\Delta S/(T\Delta t)\right>$. The denominator of this fraction is given by the 2\textsuperscript{nd} column of Table \ref{TableSecular} and the numerator is given by the left-most points of the solid curves in the top row of Fig. \ref{figureMergerSpheroidFormation}. The merger contribution to spheroid formation is then simply defined as 1 minus the secular contribution and is listed in the 3\textsuperscript{rd} column of Table \ref{TableSecular}. This merger contribution includes growth due to ex-situ formed (i.e. accreted) stars, stars formed in-situ during merger events and stars displaced from the disc to the spheroid component. It includes `tiny' mergers with very small mass ratios. We also estimate the approximate contribution of `minor plus major' and `major' mergers (3\textsuperscript{rd} column, in parentheses). For these estimates we use a cut at  ${\Delta M_{\rm exsitu}/(T\Delta t) <0.1{\rm \; Gyr}^{-1}}$ and ${\Delta M_{\rm exsitu}/(T\Delta t) <0.3{\rm \; Gyr}^{-1}}$ respectively in the numerator. The listed merger contributions are rough estimates. On the one hand they could be biased low, because in cases where the merger happens close to the snapshot time, the merger-triggered growth might be spread out over three consecutive snapshots, in which case we would miss part of it. On the other hand the estimates for the contributions of `minor+major' and `major' mergers could be biased high if multiple mergers occur between consecutive snapshots. We use the same procedure in an aperture of 5 pkpc to estimate the merger contribution to bulge formation (using the 4\textsuperscript{th} column of Table \ref{TableSecular} and the dash-dotted curves in Fig. \ref{figureMergerSpheroidFormation}, resulting in the percentages listed in the 5\textsuperscript{th} column of Table \ref{TableSecular}).

The lower limits on the merger contribution to bulge and spheroid (i.e. bulge+halo) formation are quite similar. Above $10^{10}{\rm M_{\odot}}$, $\gtrsim80\%$ of the bulge- or spheroid formation rate is associated with mergers (of any mass ratio).  Major mergers contribute $\sim20\%-55\%$, minor mergers  $25\%-45\%$ and `tiny' mergers $5\%-30\%$. Below $10^{10}{\rm M_{\odot}}$ the total merger contribution drops, but it is still in the $50\%$ range. Comparing this to the fraction of bulge stars that have an ex-situ origin (right panel of Fig. \ref{figureExsituRedshiftZero}), we find that a large part of the bulge forms from either (iii) messy, merger induced episodes of central star formation or from (ii) the disruption of stellar discs by mergers. We thus see that mergers, although not responsible for the direct supply of bulge stars, do trigger the formation of bulges and dominate the transition to elliptical morphologies at high masses in the EAGLE simulation.

\begin{table*}
\begin{center}
\caption{The estimated contribution of mergers (of any mass ratio) to the rate of disc formation (3\textsuperscript{rd} column) and inner disc formation within 5 kpc (5\textsuperscript{th} column) in different main progenitor mass bins (1\textsuperscript{st} column). The diagnostics are the same as for Table \ref{TableSecular}, but for the disc component, $D$, instead of the spheroidal component, $S$ (thus using the bottom row of Fig. \ref{figureSnapshotsCombined}). For the three mass bins with negligible disc growth/destruction no merger contribution percentage is given. During the main period of disc growth ($10^{9.5}{\rm M_{\odot}}\gtrsim T \gtrsim 10^{10.5}{\rm M_{\odot}}$) the disc grows mostly independently from merger activity, but on average mergers (mostly tiny mergers) do have a slight positive effect on the disc growth rate. For $T\lesssim 10^{9.5} {\rm M_{\odot}}$ the disc grows preferentially during mergers. The same is true for  $T\gtrsim 10^{10.5} {\rm M_{\odot}}$, although in this case there is almost no disc growth. The inner disc behaves similarly to the total disc. The main difference is the slight destruction on average of the inner disc for $T \gtrsim 10^{11}{\rm M_{\odot}}$}
\begin{tabular}{llllll}
\hline
$\log_{10}(T/{\rm M_{\odot}})$ & $\left<\Delta D/(T\Delta t)\right>$  & merger contribution to & $\left<\Delta D/(T\Delta t)\right>_{\rm <5kpc}$  & merger  contribution to \\ 
 &   &  disc formation rate &  & inner disc formation rate \\ 
 & & & & \\ 
 & Gyr$^{-1}$ & all (minor+major, major)& Gyr$^{-1}$ & all (minor+major, major)  \\ \hline
  9-9.5&  0.66 & >65\% ($\sim$43\%, $\sim$22\%) & $\;$0.68 &  >69\% ($\sim$46\%, $\sim$25\%) \\ 
 9.5-10& 0.66 & >35\% ($\sim$18\%, $\sim$11\%)  & $\;$0.69 & >41\% ($\sim$22\%, $\sim$12\%) \\ 
 10-10.5&  0.24 & >30\% ($\sim$15\%, $\sim$5\%)  & $\;$0.23 & >42\% ($\sim$22\%, $\sim$6\%) \\ 
 10.5-11& 0.04 & >53\% ($\sim$29\%, $\sim$10\%)  & $\;$0.008  & - \\ 
  11-11.5& 0.005 & -  &  -0.02 &  - \\ \hline
\end{tabular}
\label{TableSecular2}
\end{center}
\end{table*}

The bottom row of Fig. \ref{figureMergerSpheroidFormation} shows the effect that mergers have on the disc formation rate. Overall the trend is upward, but small, indicating that disc formation is on average slightly enhanced during periods of merger activity. The bottom-left panel shows the largest upward trend, hinting that in the lower mass range (corresponding to higher redshifts) the rate of disc formation is enhanced during periods of merger activity. Table \ref{TableSecular2} gives the merger contributions to the disc formation rate (calculated in the same way as the merger contributions to spheroid formation). We see that for $10^9 {\rm M_{\odot}}<T<10^{9.5} {\rm M_{\odot}}$ (which does not have a panel in Fig. \ref{figureMergerSpheroidFormation}) the disc growth rate rises strongly during merger activity (as does the spheroid growth rate from Table \ref{TableSecular}, which is much larger in this mass bin). This indicates that galaxy growth in this main progenitor mass range does not occur in an orderly fashion, but is a rather messy affair. Roughly half of the mass growth is associated with mergers and most of it ends up in the spheroidal component. Note that in our definition of $S$, the spheroid is not necessarily a smooth elliptical structure, but can also be a more complex clumpy structure, as long as it does not have a very well-defined sense of rotation.

In the mass range $10^{9.5} {\rm M_{\odot}}< T < 10^{10.5} {\rm M_{\odot}}$, the mass range in which discs come to dominate (see Figs. \ref{figureHTmass} and \ref{figureComponentMass}), the discs grows in a more orderly fashion, mostly independently from mergers. From the 3\textsuperscript{th} and 5\textsuperscript{th} columns of Table \ref{TableSecular} we see that  $\gtrsim35\%$ of the disc growth in this mass range can be attributed to mergers, of which half is due to `tiny' mergers or the associated smooth accretion of gas.

For $T> 10^{10.5} {\rm M_{\odot}}$ the disc formation rate drops dramatically (see the 2\textsuperscript{nd} and 4\textsuperscript{th} column of Table \ref{TableSecular2}). The sporadic disc formation occurs on large radii and becomes more correlated with merger activity. Perhaps surprisingly, the average effect of mergers on disc growth is positive, indicating that on average mergers do not result in the net destruction of stellar discs. In fact, if we look at the right parts of the solid curves in the bottom row of Fig. \ref{figureMergerSpheroidFormation}, major mergers on average do not result in negative values of $\Delta D$ in any mass bin. For $T>10^{11}{\rm M_{\odot}}$ the inner discs are on average slightly destroyed, but this does not seem to be connected to merger activity. The morphological transformation of massive galaxies in EAGLE is thus more driven by the buildup of spheroids than by the destruction of discs.

The strong trend of morphology with mass at the massive end, which is present in the overall galaxy population at $z\lesssim 2$ (Fig. \ref{figureSTpopulation}) and in the evolution of the progenitors of today's massive galaxies (Fig. \ref{figureHTmass}), is thus caused by the strong reduction of in-situ star formation rates around the knee of the galaxy stellar mass function (i.e. $T\approx10^{10.7} {\rm M_{\odot}}$). In the absence of significant in-situ star formation, galaxies mainly grow through mergers, causing a transformation towards elliptical morphologies. This morphological transformation is thus a direct result of the quenching of star formation in massive galaxies. \citet{Bower17} find that in EAGLE the strong quenching around the knee of the galaxy stellar mass function is caused by feedback from the central black hole. For  $T\lesssim 10^{10.5} {\rm M_{\odot}}$ stellar feedback causes bouyant outflows of hot gas. However, for $T\gtrsim 10^{10.5} {\rm M_{\odot}}$ the hydrostatic gas corona becomes so hot that the gas heated by stellar feedback is no longer buoyant. The subsequent buildup of gas in the centre triggers rapid growth of the central black hole, which eventually, disrupts the supply of cold gas and quenches the star formation. Any other quenching mechanism that kicks in at these masses (as is required by the observed galaxy stellar mass function) would presumably have a similar effect on galaxy morphologies, when combined with the effect of mergers, unless the quenching mechanism itself has a strong direct effect on stellar orbits. 

\section{Conclusions}
\label{SectionConclusions}

We have investigated the kinematic morphological evolution of the stellar component of central galaxies in the EAGLE cosmological simulation. We use a simple prescription based on the angular momenta of the stellar particles to separate each galaxy into a `spheroidal'  and a `disc' component (see Figs. \ref{figureExampleGalaxies} and \ref{figureGalaxyImages}), where the mass of the former is taken to be twice the mass of counterrotating stars. The morphology of each galaxy is characterised by the ratio of the mass in the `spheroidal' component ($S$) and the total stellar mass ($T\equiv M_*$). Note that this mass-weighted $S/T$ ratio is generally higher than a luminosity-weighted ratio (which corresponds to the visual appearance), since stars in the `disc' component tend to be younger than stars in the `spheroidal' component. We separate the `spheroidal' component into a `stellar bulge' (within 5 pkpc) and a `stellar halo' (outside 5 pkpc). We study the evolution of these components for the overall population of central galaxies with $M_{*}>10^{9} {\rm M_{\odot}}$ and we follow the evolution along the merger tree, for the main progenitors of central galaxies in the $z=0$ mass range $10^{10.5}{\rm M_{\odot}}<M_{*}<10^{12}{\rm M_{\odot}}$. We draw the following conclusions:

\begin{itemize}
\item{The kinematic morphologies of central galaxies depend strongly on stellar mass, with little additional dependence on redshift (Fig. \ref{figureSTpopulation}). This mass dependence is the same for the main progenitors of $z=0$ central galaxies (Fig. \ref{figureHTmass}). These galaxies follow a similar kinematic evolution, quite independently from their $z=0$ descendant mass. Galaxies tend to start out with a high $S/T$ ratio at $M_{*}\lesssim{\rm 10^{9.5} M_{\odot}}$, build up a stellar disc an display a decreasing $S/T$ ratio towards $M_{*}\approx 10^{10.5} M_{\odot}$, after which the $S/T$ ratio starts to rise again. The redshift at which galaxies go through these phases  depends strongly on their $z=0$ mass.}
\item{Throughout the whole evolution, the average stellar bulge component keeps growing in mass. Approximately a quarter of the bulge mass of a $10^{10.5}{\rm M_{\odot}}$ galaxy was in place at $M_{*}=10^{9.5}{\rm M_{\odot}}$, before the epoch of rapid disc growth (Fig. \ref{figureComponentMass}).}
\item{The mass growth at high masses ($M_{*}\gtrsim10^{10.5}{\rm M_{\odot}}$) is dominated by the growth of the stellar halo (Fig. \ref{figureComponentMass}).}
\item{The stellar bulges of $z=0$ galaxies with mass $M_{*}\lesssim 10^{10.5} {\rm M_{\odot}}$ consist almost entirely of stars that were formed in-situ. The stellar halo, on the other hand, has a large contribution from stars that were accreted during mergers (Fig.  \ref{figureExsituRedshiftZero}).}
\item{Morphological changes are mainly caused by in-situ star formation for galaxies in the mass range $10^{9.5}{\rm M_{\odot}}\lesssim M_{*} \lesssim 10^{10.5}{\rm M_{\odot}}$ (at the time of star formation) and are mainly associated with merger activity for $M_{*} \gtrsim 10^{10.5}{\rm M_{\odot}}$ (Fig. \ref{figureSnapshotsCombined}).}
\item{For $M_{*} > 10^{10}{\rm M_{\odot}}$ mergers (including all mass ratios) contribute $\gtrsim 80\%$ to the formation rate of bulges (Table \ref{TableSecular}, top row of Fig. \ref{figureMergerSpheroidFormation}). This percentage represents the combined effect of the accretion of stars formed ex-situ, the disruption of stellar discs and merger-triggered star formation in a spheroidal component. We estimate that  $20\%-55\%$ is due to major mergers,  $25\%-45\%$ is due to minor mergers and $5\%-15\%$ is due to `tiny' mergers with very small merger ratios. The merger contribution to bulge formation, especially the contribution from major mergers, is largest in the $10^{10}{\rm M_{\odot}}<M_*<10^{10.5}{\rm M_{\odot}}$ mass bin and becomes a bit smaller towards higher masses.}
\item{For $M_{*} > 10^{10}{\rm M_{\odot}}$ mergers of all mass ratios contribute $\gtrsim 80\%$ to the formation of spheroids (i.e. bulges+halos), of which $20\%-50\%$ is due to major mergers, $30\%-45\%$ due to minor mergers and $15\%-30\%$ due to 'tiny' mergers (Table \ref{TableSecular}, top row of Fig. \ref{figureMergerSpheroidFormation}).}
\item{Most of the mass of the disc component is formed independently from mergers, but mergers do have a slight net positive effect on the disc growth rate (Table \ref{TableSecular2}, bottom row of Fig. \ref{figureMergerSpheroidFormation}). On average mergers thus do not destroy discs. The morphological transformation of massive galaxies is mainly due to the formation of spheroids. Note, however, that our definition of a disc is purely kinematic: a spheroidal galaxy with net rotation could have a substantial `disc' component.}
\item{For $M_{*}\lesssim 10^{9.5} {\rm M_{\odot}}$ the main progenitor galaxies grow preferentially via in-situ star formation during episodes of enhanced merger activity and form mainly a spheroidal, or more complex non-rotationally supported, structure (Tables \ref{TableSecular} and \ref{TableSecular2}, Fig. \ref{figureComponentMass}).}
\end{itemize}

In conclusion, we find that galaxy formation in EAGLE can be classified into three phases, based on galaxy stellar mass. First, an early phase ($M_{*}\lesssim 10^{9.5} {\rm M_{\odot}}$) of disorganised in-situ star formation associated with merger activity, which results in a spheroidal (or more complex, non-rotationally supported) morphology. Second, a phase ($10^{9.5} {\rm M_{\odot}}\lesssim M_{*}\lesssim 10^{10.5} {\rm M_{\odot}}$) of organised in-situ star formation, resulting in a disky morphology. Third, a late phase ($M_{*}\gtrsim 10^{10.5} {\rm M_{\odot}}$) in which mergers trigger the transformation from disc-dominated galaxies to bulge-dominated or elliptical galaxies. The last phase is increasingly driven by the accretion of stars formed ex-situ.

These three phases roughly correspond to irregulars, disks and ellipticals. The main difference with the `two phases of galaxy formation' as presented by \citet{Oser10} is the inclusion of an early/low-mass phase in which galaxies are dispersion dominated. A similar early phase has been reported by \citet{Zolotov15,Tacchella16b,Tacchella16a} for the VELA simulation suite.

Testing this three phase picture observationally is beyond the scope of this work. In order to investigate whether real galaxies go through similar phases as EAGLE galaxies, one could compare to slit-spectroscopy or IFU surveys, applying the same selection criteria and using virtual observations.

\section*{Acknowledgements}

We thank Camila Correa and Scott Trager for reading the manuscript and providing comments. This work was supported by the Netherlands Organisation for Scientific Research (NWO), through VICI grant 639.043.409. We made use of he DiRAC Data Centric system at Durham University, operated by the Institute for Computational Cosmology on behalf of the STFC DiRAC HPC Facility (www.dirac.ac.uk). This equipment was funded by BIS National E-infrastructure capital grant ST/K00042X/1, STFC capital grants ST/H008519/1 and ST/K00087X/1, STFC DiRAC Operations grant ST/K003267/1 and Durham University. DiRAC is part of the National E-Infrastructure. RGB acknowledges the support of STFC consolidated grant ST/L00075X/1.

\bibliographystyle{mn2e} 
\bibliography{Bibliography}

\appendix{}
\section{}
\label{SectionAppendixA}

\begin{figure*}
\includegraphics[width=\columnwidth]{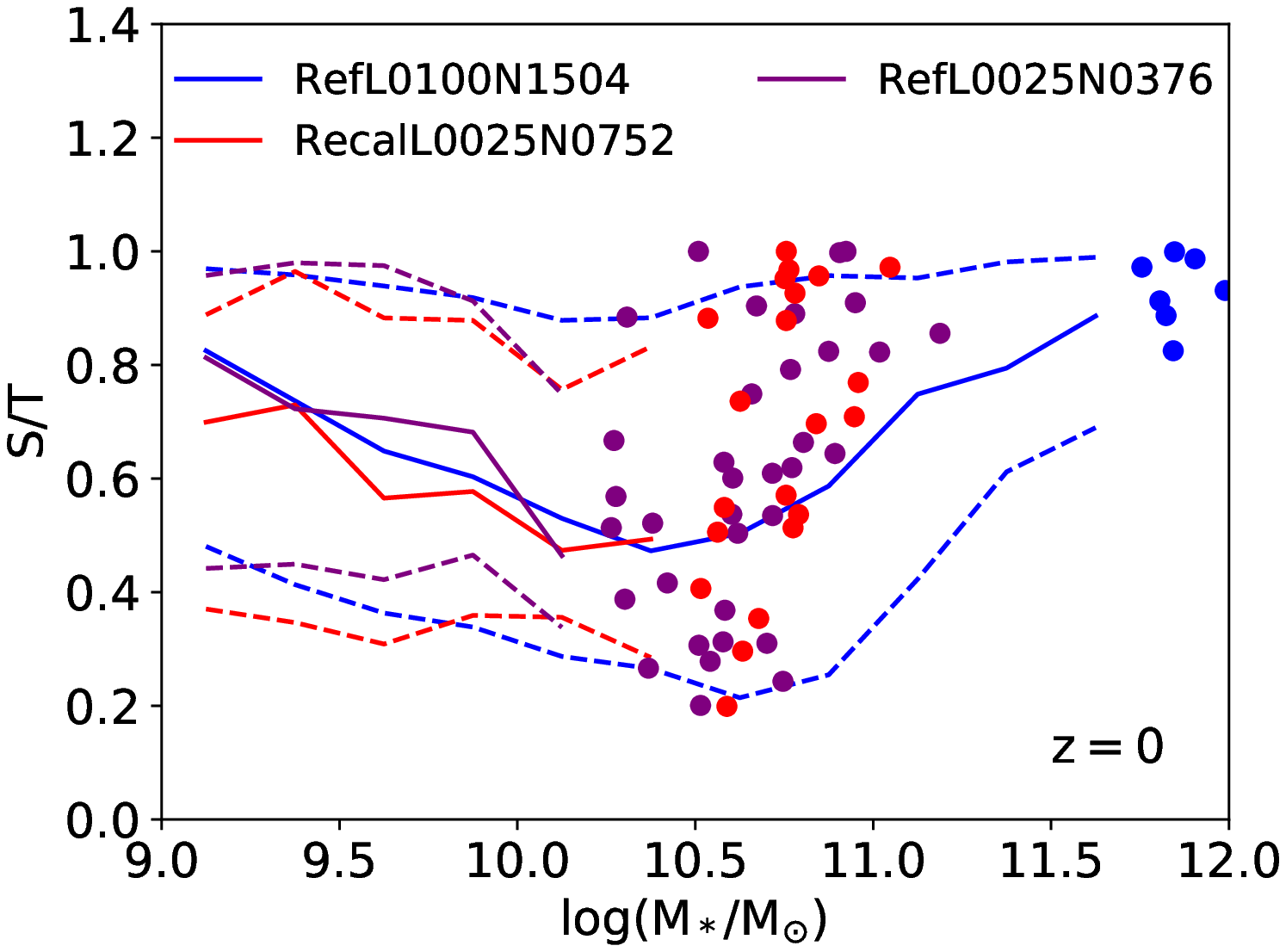}
\includegraphics[width=\columnwidth]{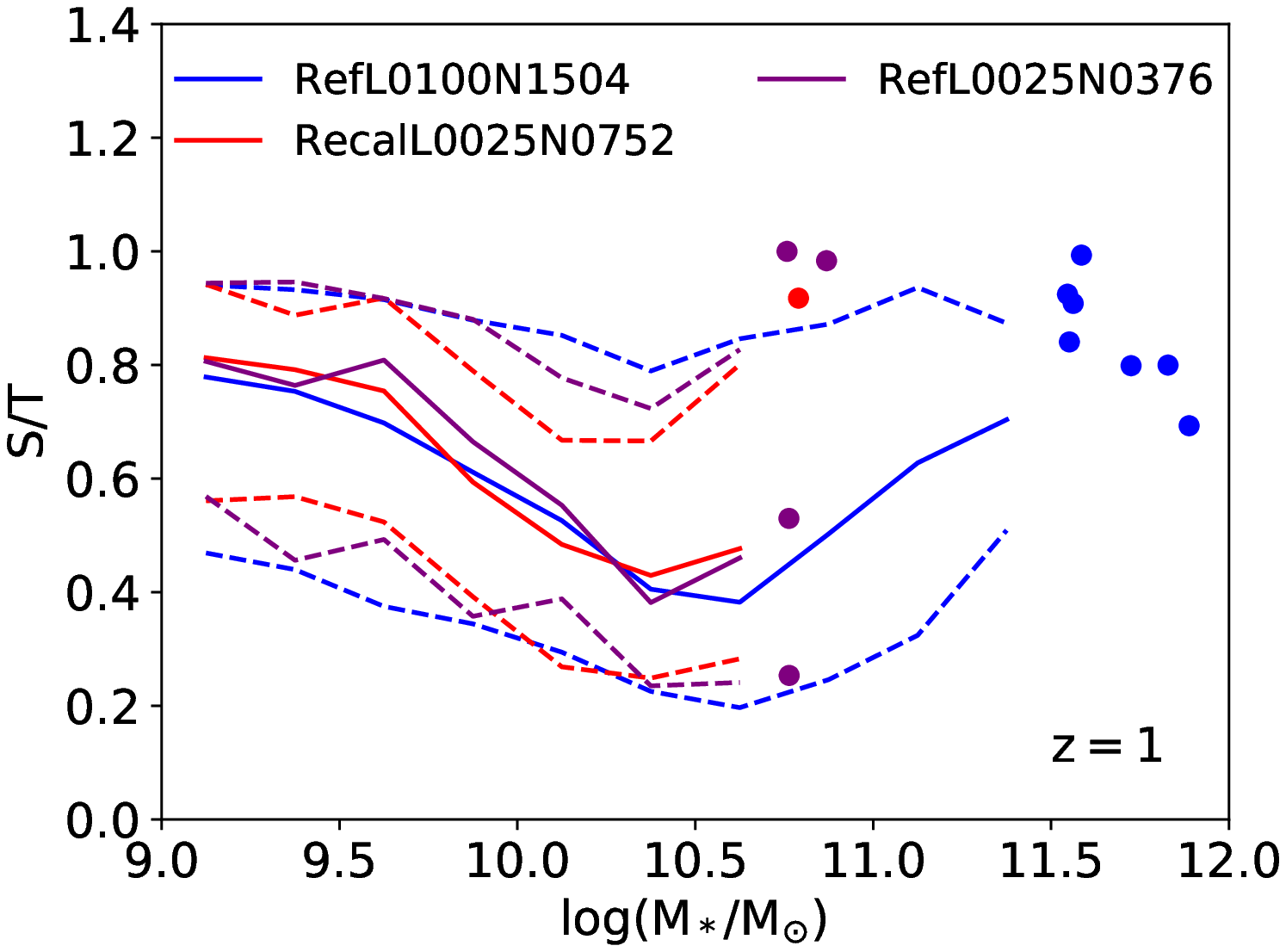}\\
\includegraphics[width=\columnwidth]{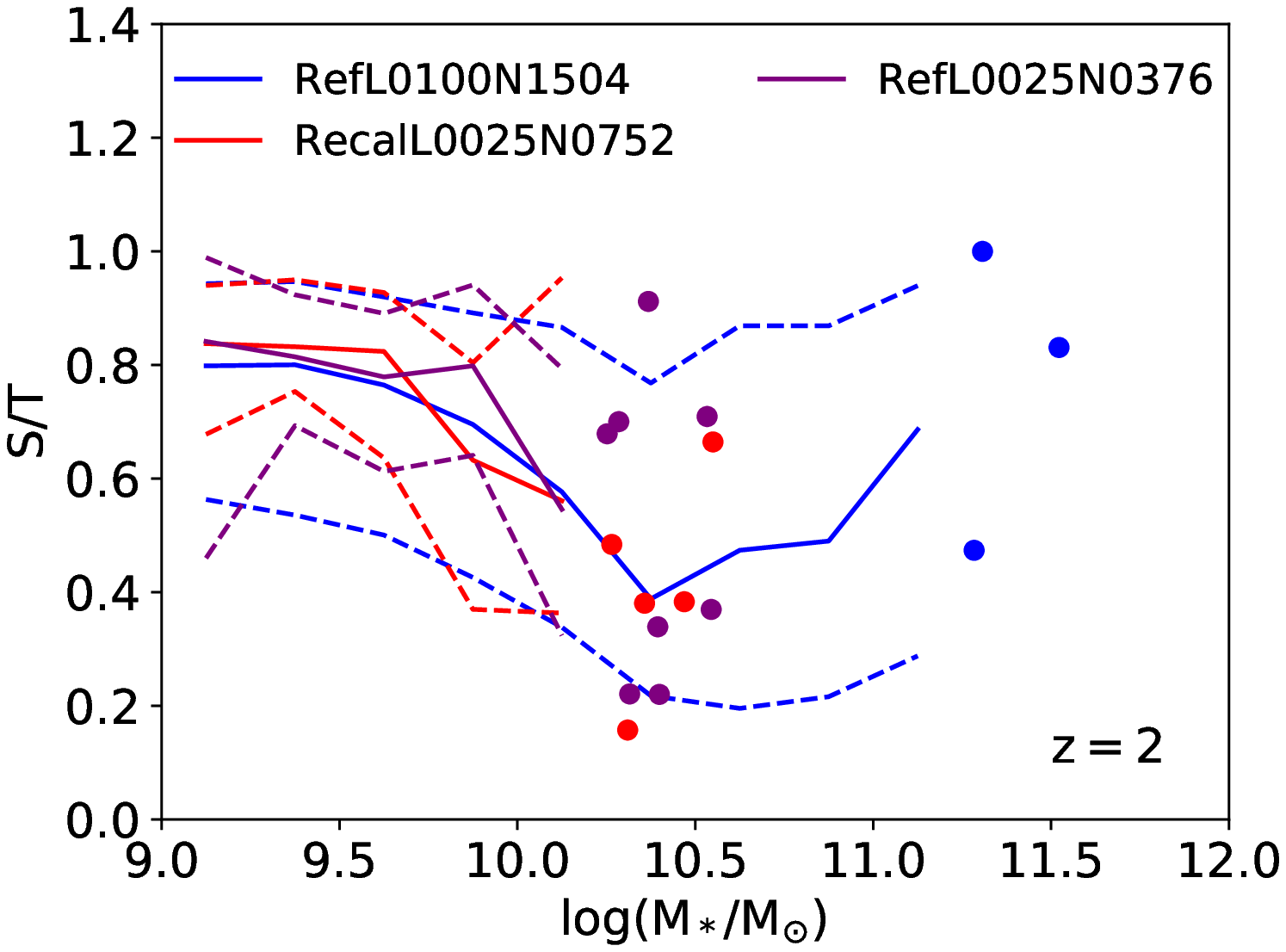}
\includegraphics[width=\columnwidth]{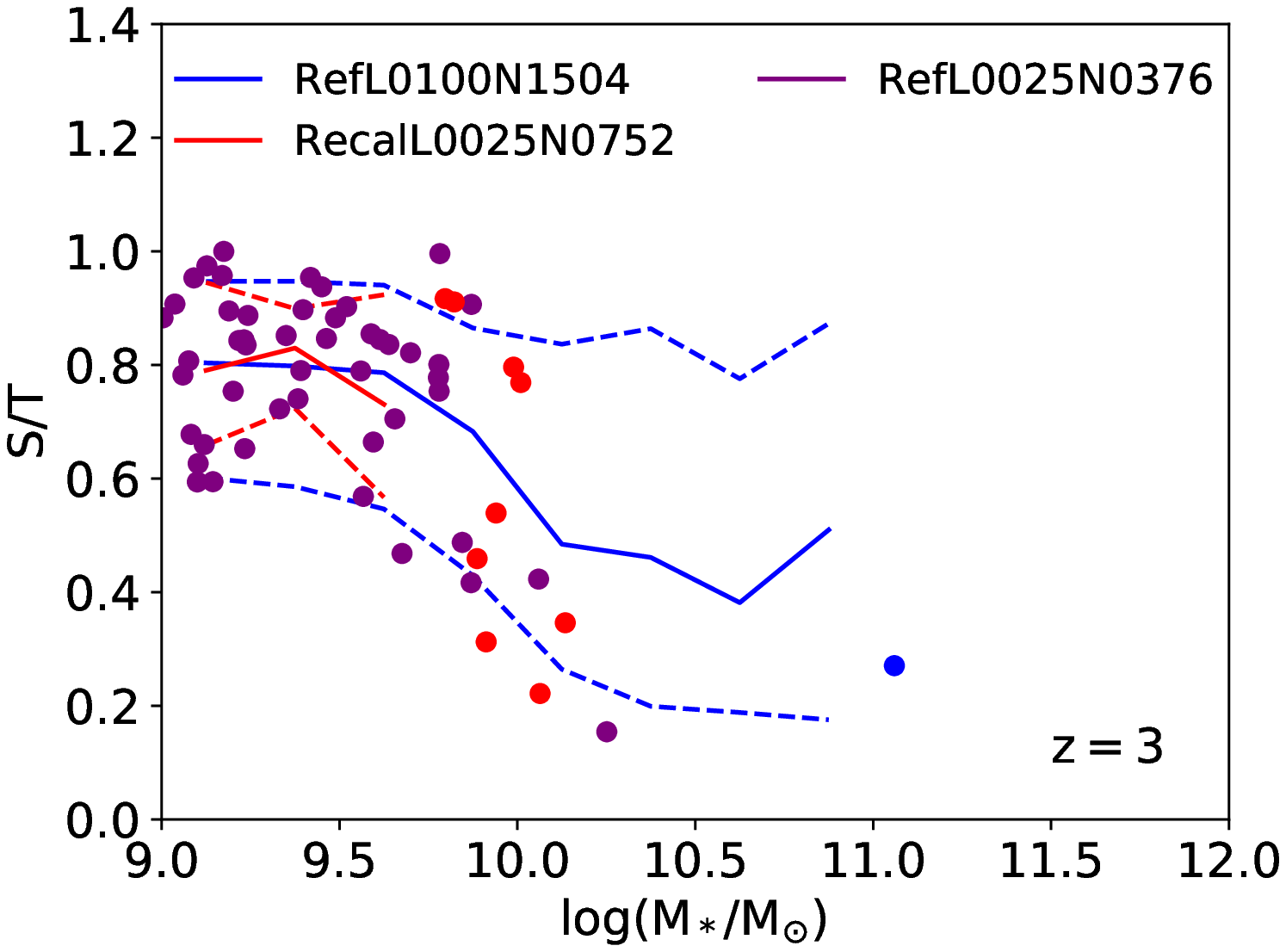}
\caption{Weak convergence test of the mass dependence of the $S/T$ ratio for the population of central galaxies at different redshifts. Different panels show different redshifts. In each panel the running median (solid curve) and 10\%-90\% range (dashed curves) is shown for three different simulation runs (colours). Blue corresponds to the original $\rm (100\;Mpc)^3$ reference run, as shown in Fig. \ref{figureSTpopulation}. Purple and red correspond respectively to the $\rm (25\;Mpc)^3$  reference run and the $\rm (25\;Mpc)^3$ recalibrated run at 8 times higher mass resolution and 2 times higher spatial resolution. individual galaxies are shown as coloured dots for mass bins that contain fewer than 10 galaxies. A comparison of the blue and purple curves mainly tests cosmic variance. These boxes differ by a factor 64 in volume, but use the same resolution. A comparison of the red and purple curves is a test of `weak convergence' with the numerical resolution \citep[see][for a discussion]{Schaye15}. For all redshifts the convergence is excellent, although at the lowest masses and lowest redshifts there is a tendency for galaxies in the higher-resolution RecalL0025N0752 (red) to be slightly more disky.}
\label{figureSTpopulationConvergence}
\end{figure*}

Figure \ref{figureSTpopulationConvergence} shows the convergence with the numerical resolution of the $S/T$ ratio for the population of central galaxies as a function of stellar mass, Fig. \ref{figureSTpopulation}. We compare results from the ${\rm (25\; Mpc)^{3}}$ sized reference run (RefL0025N0376), which has the same resolution as the ${\rm (100\; Mpc)^{3}}$ sized main simulation run (RefL0100N1504), with the ${\rm (25\; Mpc)^{3}}$ sized recalibrated run (RecalL0025N0376), which has an 8 times higher mass resolution (or 2 times higher spatial resolution). This is a test of weak-convergence \citep{Schaye15} as the parameters of the subgrid physics have been recalibrated to the present-day galaxy stellar mass function and mass-size relation. A recalibration is needed because a change in the resolution also affects the division between sub- and super-grid physics. The purpose of recalibrating is to make the large-scale effects of feedback processes the same at the higher resolution. The convergence in Fig. \ref{figureSTpopulationConvergence} is good. Results for the morphology evolution of the main progenitors of massive galaxies can not be tested for convergence in the same way, because the $\rm (25\;Mpc)^3$ sized simulation box does not contain enough massive galaxies. However, the good convergence of our morphology measure for the overall population suggests that results can also be trusted for these main progenitors. One should keep in mind though that the resolution of EAGLE is still too low to treat star formation and the generation of galactic winds without the help of $10^2 - 10^3$~pc-scale subgrid prescriptions. Disks in EAGLE may be artificially puffy because dense gas is not allowed to cool below $10^4$~K.

\section{}
\label{SectionAppendixB}

\begin{figure*}
\includegraphics[height=0.21\textwidth]{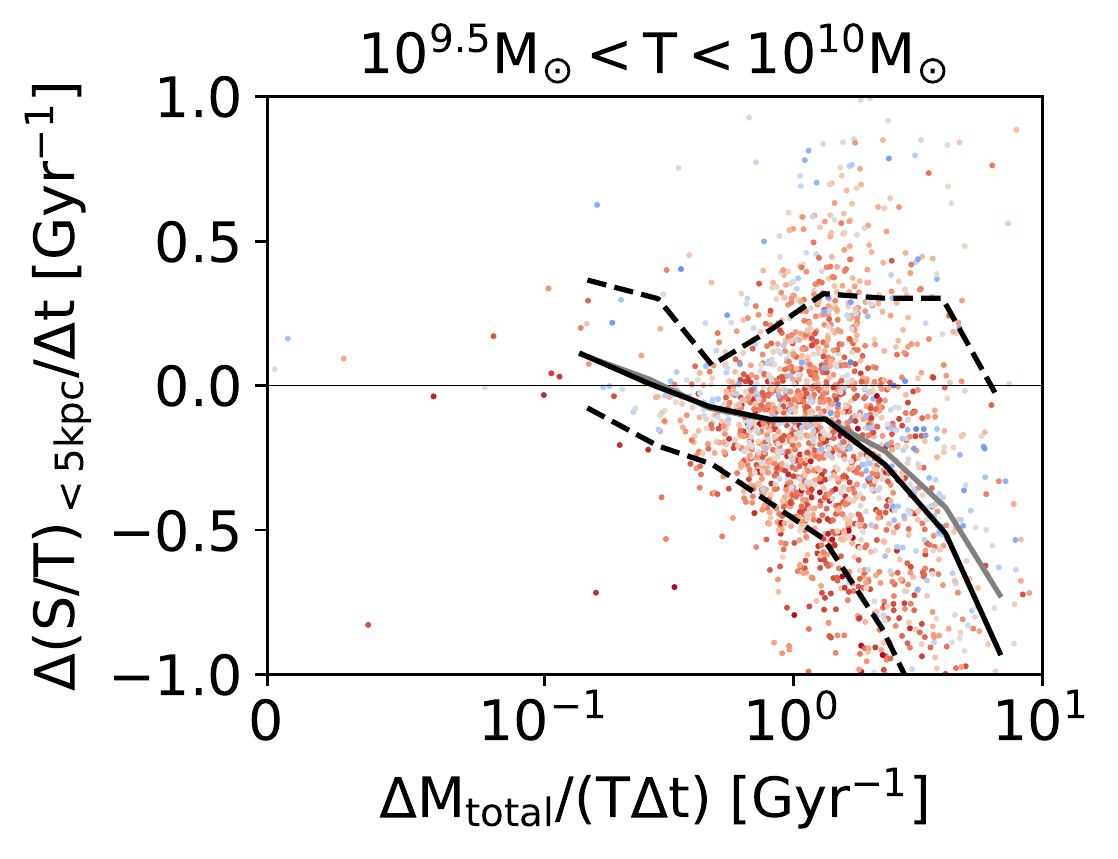}
\includegraphics[height=0.21\textwidth]{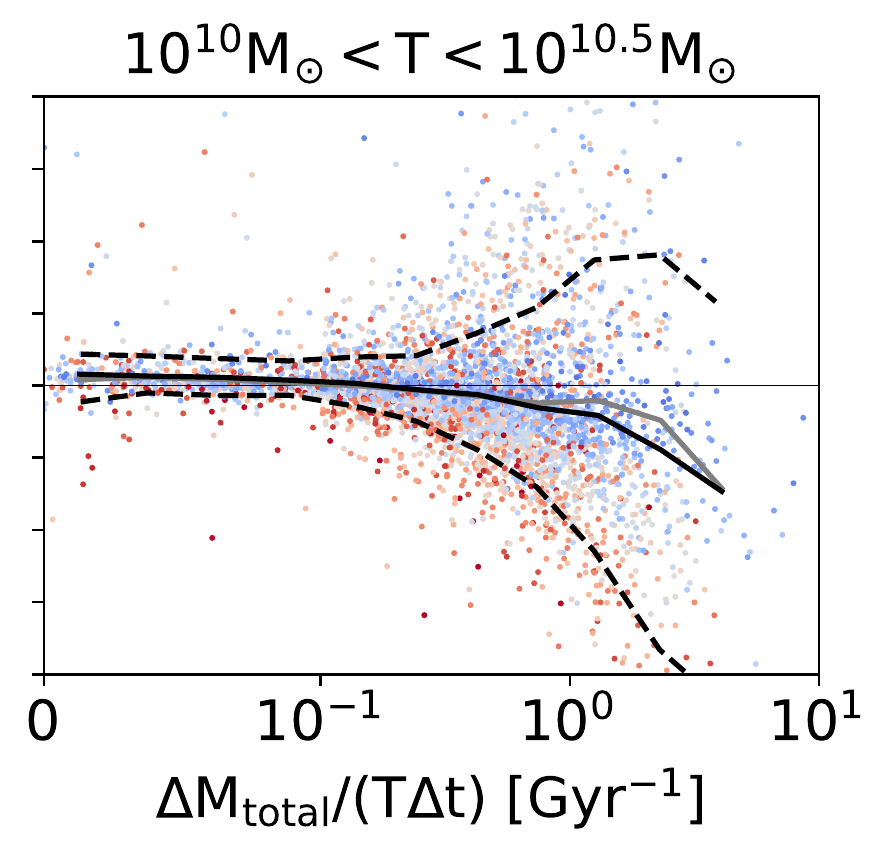}
\includegraphics[height=0.21\textwidth]{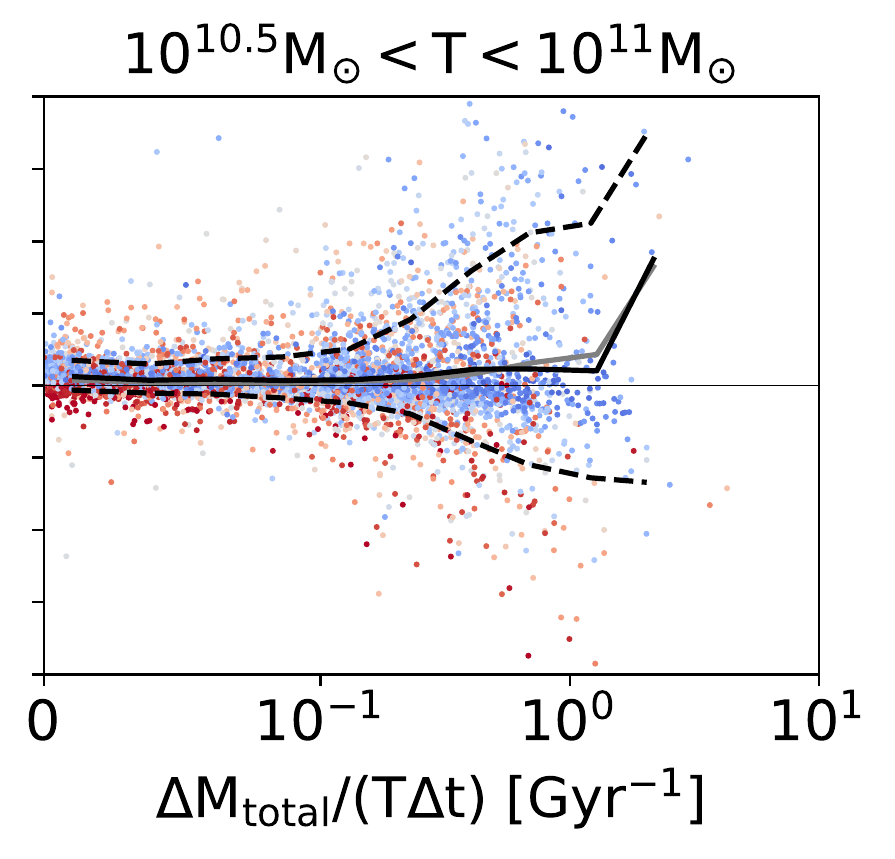}
\includegraphics[height=0.21\textwidth]{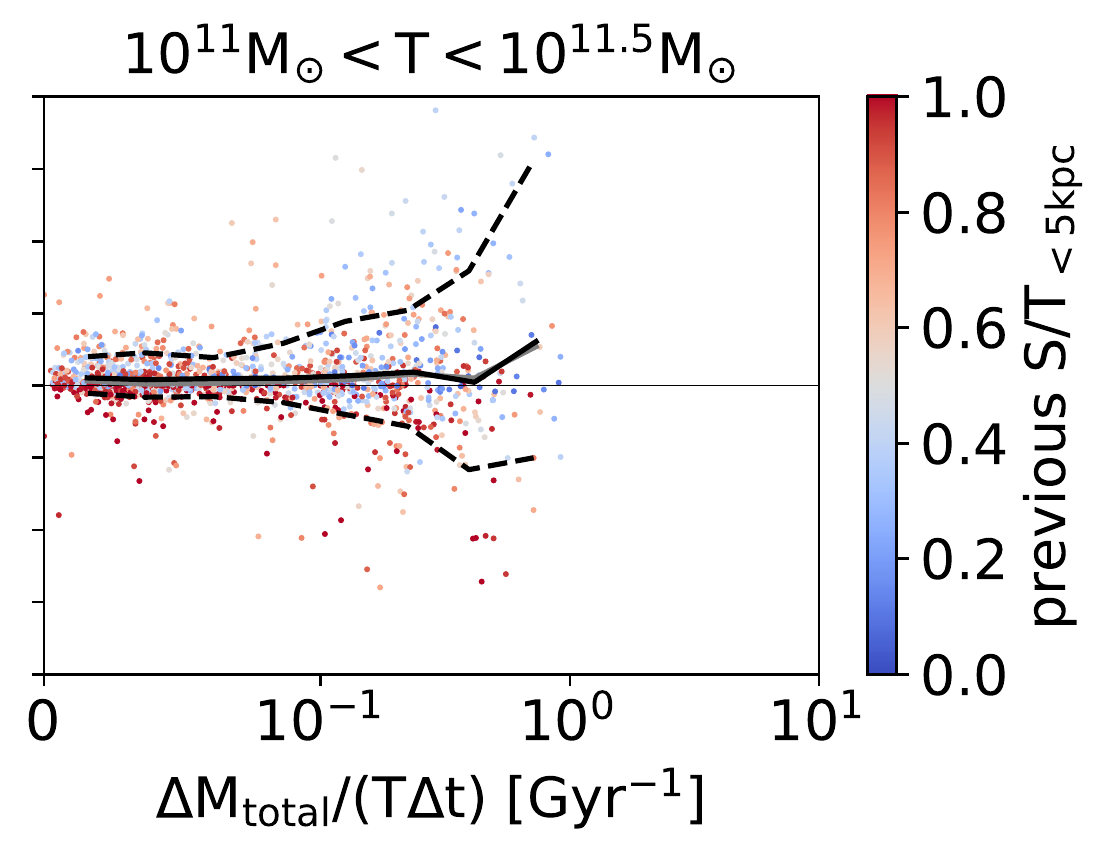}
\includegraphics[height=0.21\textwidth]{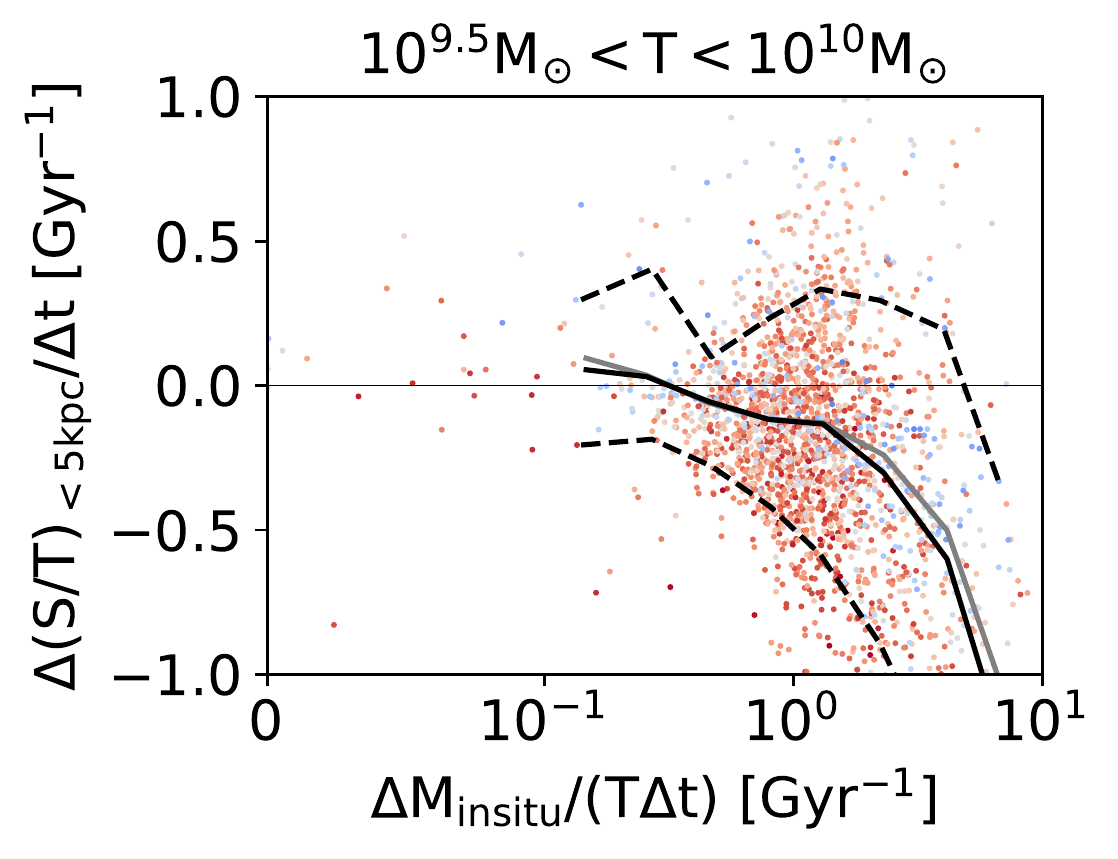}
\includegraphics[height=0.21\textwidth]{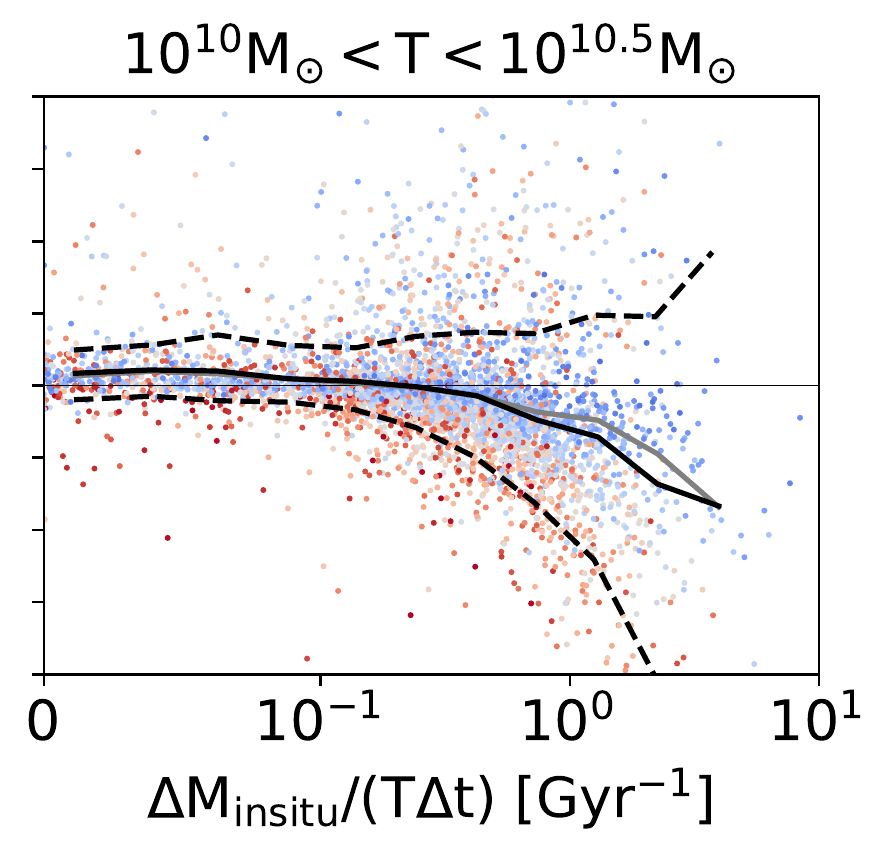}
\includegraphics[height=0.21\textwidth]{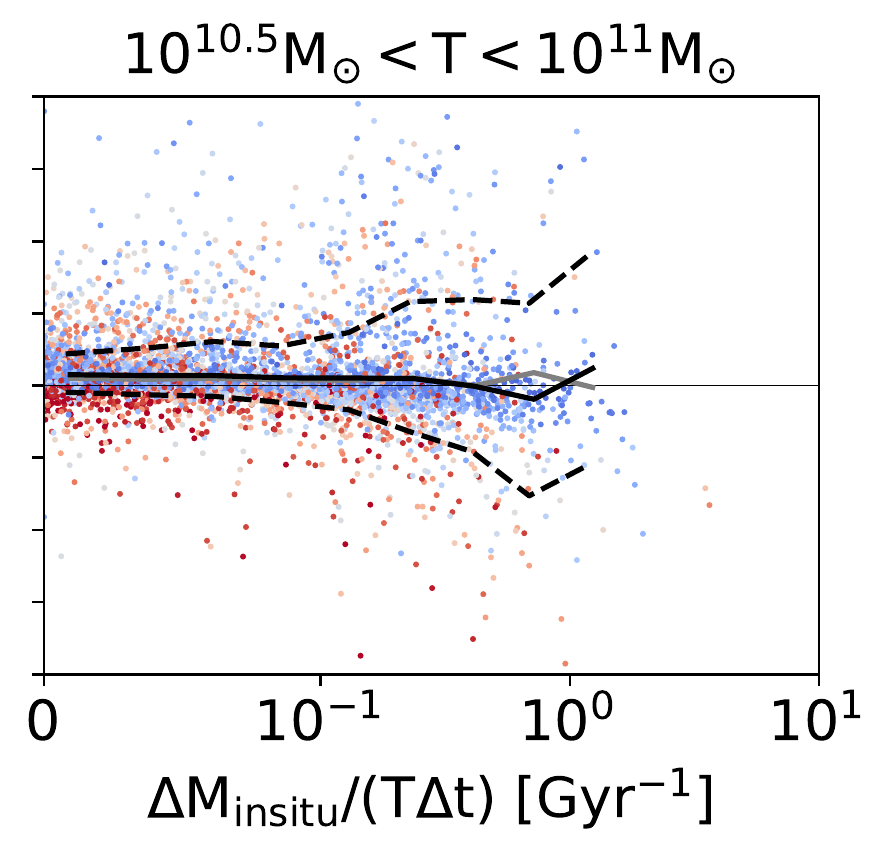}
\includegraphics[height=0.21\textwidth]{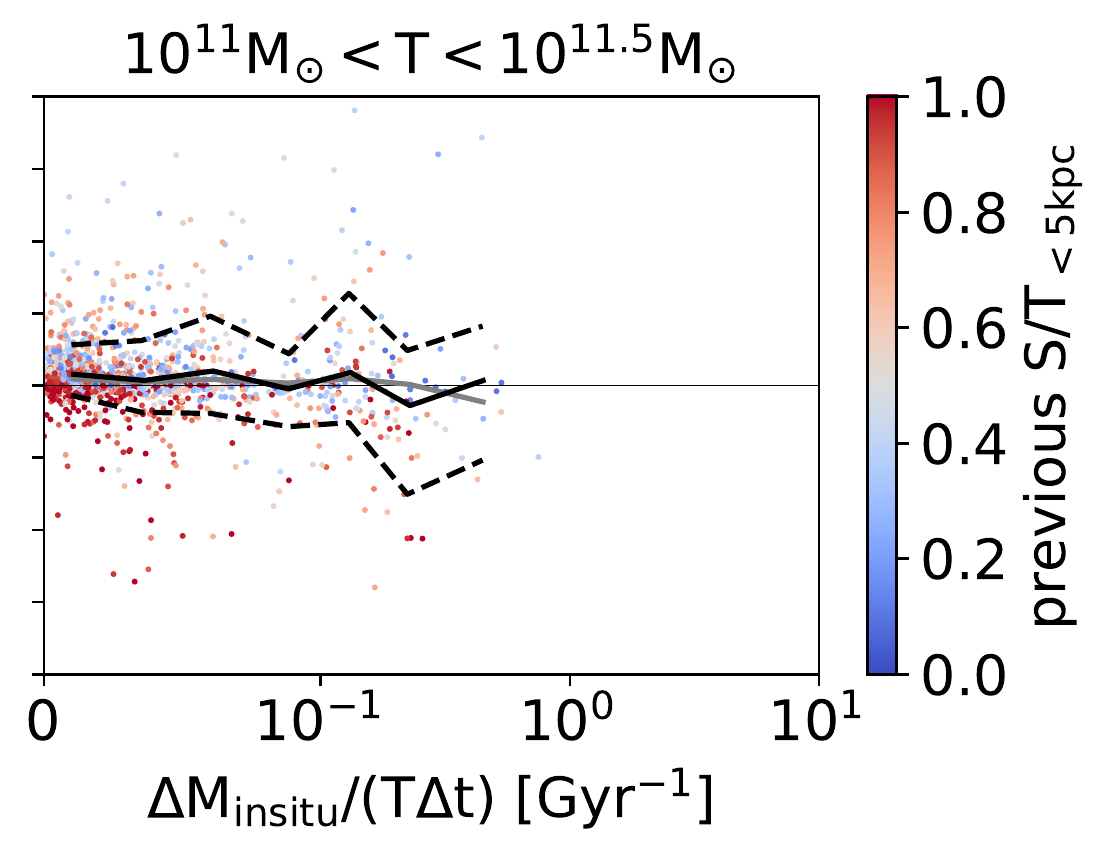}
\includegraphics[height=0.21\textwidth]{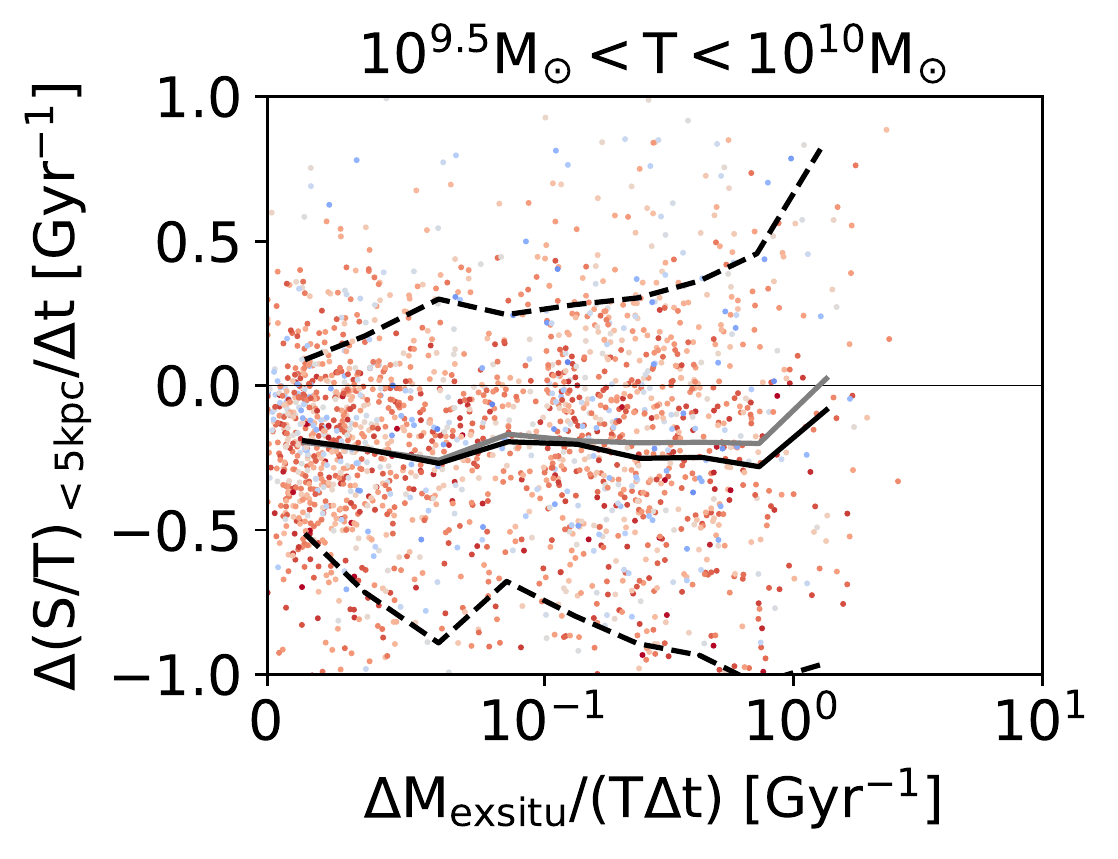}
\includegraphics[height=0.21\textwidth]{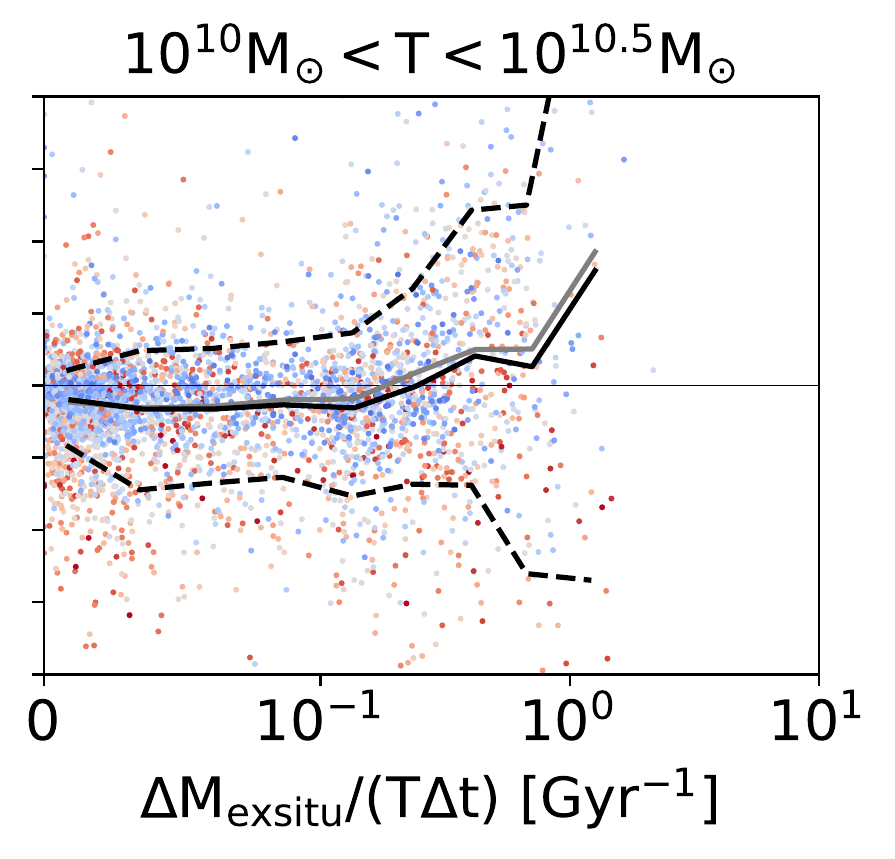}
\includegraphics[height=0.21\textwidth]{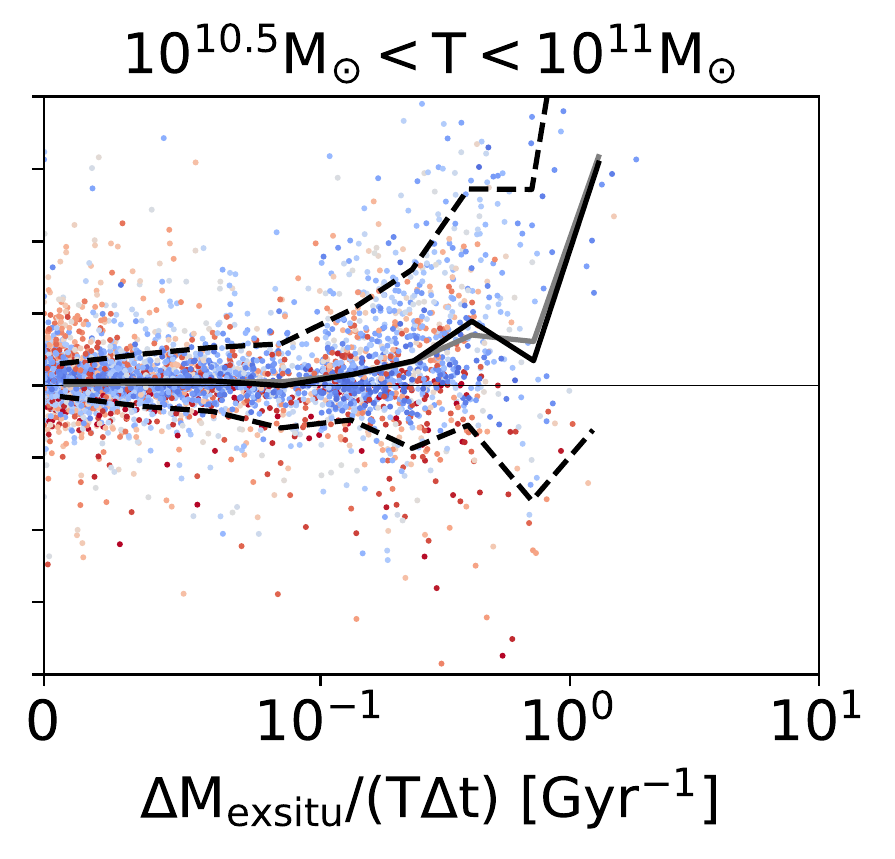}
\includegraphics[height=0.21\textwidth]{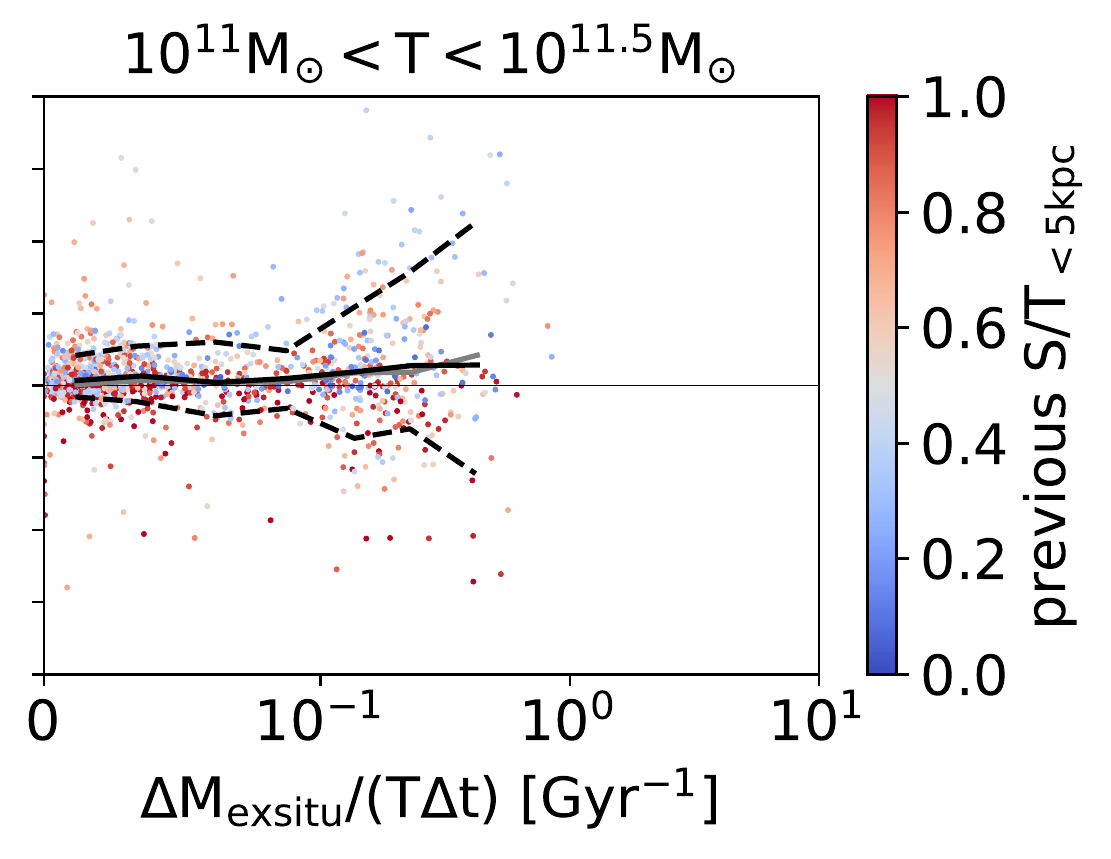}
\caption{As Fig. \ref{figureSnapshotsCombined} for the kinematic changes in the inner 5 pkpc (which separates bulge formation from bulge+halo formation). The running averages from Fig. \ref{figureSnapshotsCombined} are repeated as solid grey curves. In all panels there is a good agreement between the kinematic changes within 5 kpc (solid black curves) and those for the whole galaxy (solid grey curves). This means that the dependence of the inner kinematic changes on merger activity are very similar to that for the whole galaxy.}
\label{figureSnapshotsCombinedInside}
\end{figure*}

Fig. \ref{figureSnapshotsCombinedInside} shows the same diagnostics as Fig. \ref{figureSnapshotsCombined} for the inner 5 pkpc. We include it here instead of in the main text, because these figures turn out to be very similar. This means that the effects of mergers and in-situ star formation on the evolution of the kinematic morphology in the centres of galaxies ($<5\;{\rm pkpc}$) are very similar to the effects they have on the galaxies as a whole.

\label{lastpage}
\end{document}